\documentclass[entropy,article,accept,moreauthors,pdftex,10pt,a4paper]{mdpi} 
\usepackage{booktabs}
\usepackage{multirow}
\usepackage{soul}
\usepackage{microtype}

\usepackage{amsmath,amsfonts,amssymb,amsthm,mathtools,thmtools,etextools} 
 \usepackage{cancel}
 \usepackage{bm}
 \usepackage{dsfont}
 \usepackage{graphicx}
 \usepackage{subfig}
\usepackage{epstopdf}
\usepackage[percent]{overpic}
\usepackage[capitalise]{cleveref}
\usepackage{tikz,pgfplots}
\usepgfplotslibrary{groupplots}
\usetikzlibrary{calc,arrows}
\usetikzlibrary{patterns}
\usetikzlibrary{shapes.misc}
\tikzstyle{line}=[draw]
\tikzstyle{arrow}=[draw, -latex] 
\newcommand\Ccancel[2][black]{\renewcommand\CancelColor{\color{#1}}\xcancel{#2}}
\DeclarePairedDelimiter{\ceil}{\lceil}{\rceil}
\DeclareMathOperator*{\argmax}{arg\,max}
\newenvironment{brsm}{
 \bigl[ \begin{smallmatrix} }{%
 \end{smallmatrix} \bigr]}

\makeatletter
\tikzset{
  every picture/.style={
   execute at begin picture={
     \let\ref\@refstar
   }
  }
}
\makeatother

\articlenumber{00}
\pubvolume{20}
\pubyear{2018}
\copyrightyear{2018}
\history{Received: 15 May 2018; Accepted: 12 June 2018; Published: date}
\issuenum{0}
\updates{yes}

\pdfoutput=1

\Title{Strong Secrecy on a Class of Degraded Broadcast Channels Using Polar Codes}

\Author{Jaume del Olmo Alos*\orcidA{} and Javier Rodríguez Fonollosa\orcidB{}}

\AuthorNames{Jaume del Olmo Alos and Javier Rodríguez Fonollosa}

\address[1]{Departament de Teoria del Senyal i Communications (TSC), Universitat 
Politècnica de Catalunya, Barcelona 08034, Spain; 
javier.fonollosa@upc.edu}

\corres{Correspondence: jaume.del.olmo@upc.edu}

\abstract{Asymptotic secrecy-capacity achieving polar coding schemes are proposed for the memoryless degraded broadcast channel under different reliability and secrecy requirements: layered decoding or layered secrecy. In these settings, the transmitter wishes to send multiple messages to a~set of legitimate receivers keeping them masked from a set of eavesdroppers. The layered decoding structure requires receivers with better channel quality to reliably decode more messages, while~the layered secrecy structure requires eavesdroppers with worse channel quality to be kept ignorant of more messages. Practical constructions for the proposed polar coding schemes are discussed and their performance evaluated by means of simulations.}


\begin{document}


\section{Introduction}
Information-theoretic security over noisy channels was introduced by Wyner in \cite{6772207}, which~characterized the (secrecy-)capacity of the degraded wiretap channel.~Later, Csisz\'{a}r and K{\"o}rner in \cite{1055892} generalized Wyner's results to the general wiretap channel.~In these settings, one~transmitter wishes to reliably send one message to a legitimate receiver, while keeping it secret from an eavesdropper, where~secrecy is defined based on a condition on some information-theoretic measure that is fully quantifiable.~One of these measures is the {information leakage}, defined as the mutual information $I(W;Z^n)$ between a~uniformly-distributed random message $W$ and the channel observations $Z^n$ at the eavesdropper, $n$~being the number of uses of the channel.~Based on this measure, the most common secrecy conditions required to be satisfied by channel codes are the {weak secrecy}, which requires $\lim_{n \rightarrow \infty} \frac{1}{n} I(W;Z^n) = 0$, and the {strong secrecy}, requiring $\lim_{n \rightarrow \infty} I(W;Z^n) = 0$. Although the second notion of security is stronger, surprisingly, both secrecy conditions result in the same secrecy-capacity region \cite{maurer2000information}.

In the last decade, information-theoretic security has been extended to a large variety of contexts, and this paper focuses on two different classes of discrete memoryless Degraded Broadcast Channels (DBC) surveyed in \cite{7264977}: (a) with Non-Layered Decoding and Layered Secrecy (DBC-NLD-LS) and (b) with Layered Decoding and Non-Layered Secrecy (DBC-LD-NLS). In these models, the transmitter wishes to send a set of messages through the DBC, and each message must be reliably decoded by a~particular set of receivers and kept masked from a particular set of eavesdroppers. The degradedness condition of the channel implies that individual channels can be ordered based on the quality of their received signals. The layered decoding structure requires receivers with better channel quality to reliably decode more messages, while the layered secrecy requires eavesdroppers with worse channel quality to be kept ignorant of more messages. 

The capacity region of these models was first characterized in \cite{7264977,6687232,Ekrem1661312}. However, the achievable schemes used by these works rely on random coding arguments that are nonconstructive in practice. In this sense, the purpose of this paper is to provide coding schemes based on polar codes, which were originally proposed by Arikan \cite{arikan2009channel} to achieve the capacity of binary-input, symmetric, point-to-point channels under Successive Cancellation (SC) decoding. Capacity achieving polar codes for the binary symmetric degraded wiretap channel were introduced in \cite{6034749,6620400}, satisfying the weak and the strong secrecy condition, respectively. Recently, polar coding has been extended to the general wiretap channel in \cite{renes2013efficient, 7346401,cihad2014achieving,7426806}. Indeed, \cite{cihad2014achieving,7426806} generalize their results providing polar coding schemes for the broadcast channel with confidential messages, and \cite{7346401} also proposes polar coding strategies to achieve the best-known inner bounds on the secrecy-capacity region of some multi-user settings.

Although recent literature has proven the existence of different secrecy-capacity achieving polar coding schemes for multi-user scenarios (for instance, see \cite{6975233,cihad2014achieving,7426806,7346401,7217814,6874845,7541446,7370934}), polar codes for the two models on which this paper is focused have, as far as we know, not been analyzed yet. As mentioned in \cite{7264977}, these~settings capture practical scenarios in wireless systems, in which channels can be ordered based on the quality of the received signals (for example, Gaussian channels are degraded). Hence, the ultimate goal of this work is not only to prove the existence of two asymptotic secrecy-capacity achieving polar coding schemes for these models under the strong secrecy condition, but also to discuss their practical construction and evaluate their performance for a finite blocklength by means of simulations.

 \subsection{Relation to Prior Work}
A good overview of the similarities and differences between the polar codes proposed in \cite{renes2013efficient, 7346401,cihad2014achieving,7426806} for the general wiretap channel can be found in \cite{7426806} (Figure 1). The polar coding schemes proposed in this paper are based mainly on those introduced by \cite{7426806} because of the following reasons:
\begin{itemize}[leftmargin=*,labelsep=5.8mm]
\item {To provide strong secrecy}.~Despite both weak and strong secrecy conditions resulting in the same secrecy-capacity region, the weak secrecy requirement in practical applications can result in important system vulnerabilities \cite{bloch2011physical} ({Section 3.3}).
\item {To provide polar coding schemes that are implementable in practice}. Notice in \cite{7426806} ({Figure 1}) that the coding scheme presented in \cite{renes2013efficient} relies on a construction for which no efficient code is presently known. Moreover, the polar coding scheme in \cite{cihad2014achieving} relies on the existence, through averaging, of certain deterministic mappings for the encoding/decoding process. 
\end{itemize}

As in \cite{7426806}, our polar coding schemes are totally explicit. However, to provide strong secrecy and reliability simultaneously, the transmitter and the legitimate receivers need to share a secret key of negligible size in terms of rate, and the distribution induced by the encoder must be close in terms of statistical distance to the original one considered for the code construction.~Moreover, we adapt the deterministic SC encoder of \cite{7447169} to our channel models, and we show that it can perform well in practice.~As concluded in \cite{7447169}, this deterministic SC encoder will avoid the need to draw large sequences according to specific distributions at the encoder, which can be useful in communication systems requiring low complexity at the transmitter.

{ In \cite{7426806} ({Remark~3}), the authors highlight the connection between polar code constructions and random binning proofs that allows them to apply their designs to different problems in network information theory. Nevertheless, in our polar coding schemes, the chaining construction used in \cite{7426806} is not needed because of the degradedness condition of the channels, and consequently, we can introduce small changes in the design in order to make our proposed coding schemes more practical. In this sense, we~assume that a source of common randomness is accessible to all parties, which allows the transmitter to send secret information in just one block of size $n$ by only using a secret key with negligible size in terms of rate. Despite this common randomness being available to the eavesdroppers, no information will be leaked about the messages. Moreover, if we consider a communication system requiring transmissions over several blocks of size $n$, the same realization of this source of common randomness can be used at each block without compromising the strong secrecy condition.}


 \subsection{Overview of Novel Contributions}
The main novelties of this paper can be summarized as follows:
\begin{enumerate}[leftmargin=*,labelsep=4.9mm]
\item {Scenario}.~This paper focuses on two different models of the DBC with an arbitrary number of legitimate receivers and an arbitrary number of eavesdroppers for which polar codes have not yet been proposed. These two models arise very commonly in wireless communications.
\item {Existence of the polar coding schemes}.~We prove the existence for sufficiently large $n$ of two secrecy-capacity achieving polar coding schemes under the strong secrecy condition. 
\item {Practical implementation.} We provide polar codes that are implementable in real communication systems, and we discuss further how to construct them in practice. As far as we know, although the construction of polar codes has been covered in a large number of references (for instance, see~\cite{tal2013construct,vangala2015comparative,6601656}), they only focus on polar code constructions under reliability constraints. 
\item {Performance evaluation.} Simulations results are provided in order to evaluate the reliability and secrecy performance of the polar coding schemes. The performance is evaluated according to different design parameters of the practical code construction.~As far as we know, this paper is the first to evaluate the secrecy performance in terms of the strong secrecy, which is done by upper-bounding the information leakage at the eavesdroppers.
\end{enumerate}

\vspace{-0.2cm}
\subsection{Notation}
Through this paper, let $[n] = \{1,\dots,n\}$ for $n \in \mathbb{Z}^{+}$, $a^n$ denote a row vector $(a(1), \dots, a(n))$. We~write $a^{1:j}$ for $j \in [n]$ to denote the subvector $(a(1),\dots,a(j))$.~Let $\mathcal{A} \subset [n]$, then we write $a[\mathcal{A}]$ to denote the sequence $\{a(j)\}_{j\in \mathcal{A}}$, and we use $\mathcal{A}^{\text{C}}$ to denote the set complement with respect to the universal set $[n]$, that is $\mathcal{A}^{\text{C}} = [n] \setminus \mathcal{A}$.~If $\mathcal{A}$ denotes an event, then $\mathcal{A}^{\text{C}}$ also denotes its complement. We use $\ln$ to denote the natural logarithm, whereas $\log$ denotes the logarithm base two.~Let $X$ be a~random variable taking values in $\mathcal{X}$, and let $q_x$ and $p_x$ be two different distributions with support $\mathcal{X}$, then $\mathbb{D}(q_x,p_x)$ and $\mathbb{V}(q_x,p_x)$ denote the Kullback-Leibler divergence and the total variation distance, respectively. Finally, $h_2(p)$ denotes the binary entropy function, i.e., \mbox{$h_2(p) = -p \log p - (1-p) \log (1-p)$}, and we define the indicator function $\mathds{1} \{ u \}$ such that it equals one if the predicate $u$ is true and zero~otherwise. 
 
\subsection{Organization}
The remainder of this paper is organized as follows. In Section~\ref{sec:SM}, the channel models DBC-NLD-LS and DBC-LD-NLS are introduced formally, and their secrecy-capacity regions are characterized. In~Section~\ref{sec:rev}, the fundamentals theorems of polar codes are revisited. In Sections~\ref{sec:PCS_dbcnldls}~and~\ref{sec:PCS_dbcldnls}, two polar coding schemes are proposed for the DBC-NLD-LS and DBC-LD-NLS, respectively, and we prove that both are asymptotic secrecy-capacity achieving. In Section~\ref{sec:results}, practical polar code constructions are discussed for both models, and the performances of the polar codes are evaluated by means of simulations. Finally, the concluding remarks are presented in Section~\ref{sec:conclusions}.

\section{System Model and Secrecy-Capacity Region}\label{sec:SM}
Formally, a DBC $( \mathcal{X}, p_{Y_K\dots Y_1 Z_M \dots Z_1|X}, \mathcal{Y}_K \times \cdots \times \mathcal{Y}_1 \times \mathcal{Z}_M \times \cdots \times \mathcal{Z}_1 )$ with $K$ legitimate receivers and $M$ eavesdroppers is characterized by the probability transition function $p_{Y_{K} \dots Y_1 Z_M \dots Z_1| X}$, \mbox{where~$X \in \mathcal{X}$} denotes the channel input, $Y_k \in \mathcal{Y}_k$ denotes the channel output corresponding to the legitimate receiver $k \in [1,K]$ and $Z_m \in \mathcal{Z}_m$ denotes the channel output corresponding to the eavesdropper $m \in [1,M]$. The broadcast channel is assumed to gradually degrade in such a way that each legitimate receiver has a better channel than any eavesdropper, that is: 
\begin{align}
X-Y_K-\dots-Y_1-Z_M-\dots-Z_1
\label{eq:degch}
\end{align}
forms a Markov chain.~Although we consider physically degradation, the polar coding schemes proposed in this paper are also suitable for stochastically degraded channels (see Remark~\ref{re:subp}). 

\subsection{Degraded Broadcast Channel with Non-Layered Decoding and Layered Secrecy}\label{sec:SM_dbcnldls}
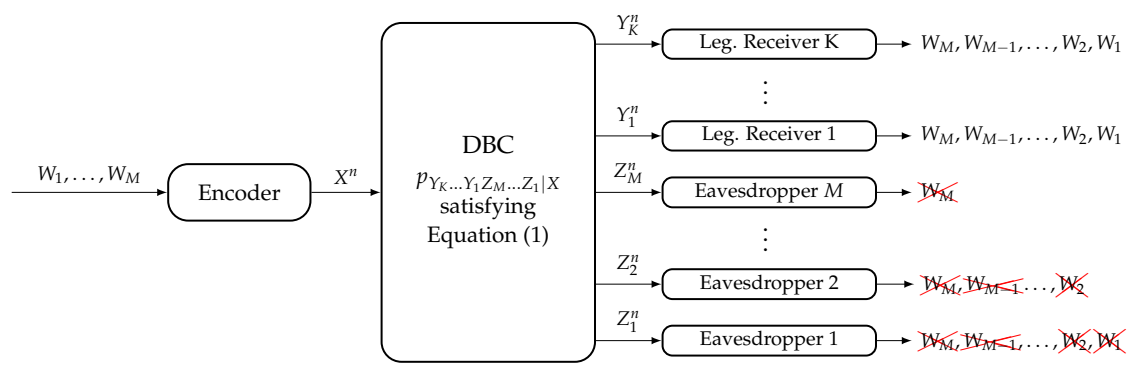
\begin{figure}[H]
 \resizebox{0.95\linewidth}{!}{%
\centering
\begin{tikzpicture}
\node (E) at (-4.7,0) [shape=rectangle,thick,rounded corners=8pt,inner sep=0pt,minimum width=2cm, minimum height=0.8cm, draw] {\small Encoder};
\node (C) at (-1.2,0) [shape=rectangle,thick,rounded corners=8pt,inner sep=0pt,minimum width=3cm, minimum height=4.8cm,align=center,draw] {DBC \\ {\small $p_{Y_K \dots Y_1 Z_M \dots Z_1|X}$} \\ {\small satisfying} \\ {\small Equation~\eqref{eq:degch}} };
\node (E1) at (2.75,-2.1) [shape=rectangle,thick,rounded corners=4pt,minimum width=3cm, inner xsep=6pt, inner ysep=2pt, draw] {\footnotesize Eavesdropper 1};
\node (E2) at (2.75,-1.3) [shape=rectangle,thick,rounded corners=4pt,minimum width=3cm, inner xsep=6pt, inner ysep=2pt, draw] {\footnotesize Eavesdropper 2};
\node (EM) at (2.75,0) [shape=rectangle,thick,rounded corners=4pt,minimum width=3cm, inner xsep=6pt, inner ysep=2pt, draw] {\footnotesize Eavesdropper $M$};
\node (RK) at (2.75,2.1) [shape=rectangle,thick,rounded corners=4pt,minimum width=3cm, inner xsep=6pt, inner ysep=2pt, draw] {\footnotesize Leg. Receiver K};
\node (R1) at (2.75,0.8) [shape=rectangle,thick,rounded corners=4pt,minimum width=3cm, inner xsep=6pt, inner ysep=2pt, draw] {\footnotesize Leg. Receiver 1};
\node (Wt) at (2.7,1.52) [] {$\vdots$};
\node (Wt) at (2.7,-0.56) [] {$\vdots$};
\node (E1t) at (6.3,-2.1) [] {\footnotesize $\Ccancel[red]{W_M}, \Ccancel[red]{W_{M-1}}, \dots, \Ccancel[red]{W_{2}},\Ccancel[red]{W_1}$};
\node (E2t) at (6.3,-1.3) [] {\footnotesize $\Ccancel[red]{W_M}, \Ccancel[red]{W_{M-1}}\dots,\Ccancel[red]{W_2}$\color{white}{$,W_1$}};
\node (EMt) at (6.3,0) [] {\footnotesize $\Ccancel[red]{W_M}$\color{white}{$,W_{M-1},\dots, W_2, W_1$}};
\node (RKt) at (6.3,2.1) [] {\footnotesize $W_	M,W_{M-1},\dots, W_2, W_1$};
\node (R1t) at (6.3,0.8) [] {\footnotesize $W_	M,W_{M-1},\dots, W_2, W_1$};
\draw [arrow] (-7.9,0) -- (E) node [midway, above] (TextNode) {\footnotesize $W_1,\dots,W_M$};
\draw [arrow] (E) -- (C) node [midway, above] (TextNode) {\footnotesize $X^n$};
\draw [arrow] (0.3,2.1) -- (RK) node [midway, above] (TextNode) {\footnotesize $Y_K^n$};
\draw [arrow] (0.3,0.8) -- (R1) node [midway, above] (TextNode) {\footnotesize $Y_1^n$};
\draw [arrow] (0.3,0) -- (EM) node [midway, above] (TextNode) {\footnotesize $Z_M^n$};
\draw [arrow] (0.3,-1.3) -- (E2) node [midway, above] (TextNode) {\footnotesize $Z_2^n$};
\draw [arrow] (0.3,-2.1) -- (E1) node [midway, above] (TextNode) {\footnotesize $Z_1^n$};
\draw [arrow] (RK) -- (RKt) node [midway, above] (TextNode) {};
\draw [arrow] (R1) -- (R1t) node [midway, above] (TextNode) {};
\draw [arrow] (EM) -- (EMt) node [midway, above] (TextNode) {};
\draw [arrow] (E2) -- (E2t) node [midway, above] (TextNode) {};
\draw [arrow] (E1) -- (E1t) node [midway, above] (TextNode) {};
\end{tikzpicture}
}
\caption{DBC with Non-Layered Decoding and Layered Secrecy (DBC-NLD-LS).}\label{fig:dbcnldls}
\end{figure}

In this model (see Figure~\ref{fig:dbcnldls}), the transmitter wishes to send $M$ messages $\{ W_{m} \}_{m = 1}^M$ to the $K$ legitimate receivers. The non-layered decoding structure requires the legitimate receiver $k \in [1,K]$ to reliably decode all $M$ messages, and the layered secrecy structure requires the eavesdropper $m \in [1,M]$ to be kept ignorant about messages $\{ W_{i} \}_{{i} = m}^M$. Consider a $(\ceil{2^{nR_1}},\dots, \ceil{2^{nR_M}},n)$ code for the DBC-NLD-LS, where $W_{m} \in [\ceil{2^{nR_m}}]$ for any $m \in [1,M]$. The reliability condition to be satisfied by this code is measured in terms of the average probability of error at each legitimate receiver and is given by:
\begin{align}
\lim_{n \rightarrow \infty} \mathbb{P} \left[ (\hat{W}_1, \dots, \hat{W}_M) \neq ({W}_1, \dots, {W}_M) \right] = 0, \quad \text{for any legitimate receiver } k \in [1,K]. \label{eq:reliabilitycondition_2}
\end{align}
On the other hand, the {strong} secrecy condition to be satisfied by the code is measured in terms of the information leakage at each eavesdropper and is given by:
\begin{align}
\lim_{n \rightarrow \infty} I (W_m, W_{m+1}, \dots,W_M ; Z_m^n) = 0 ,\quad \text{for the eavesdropper } m \in [1, M]. \label{eq:secrecycondition_2}
\end{align}
A tuple of rates $(R_1,\dots, R_M) \in \mathbb{R}_{+}^M$ is achievable for the DBC-NLD-LS if there exists a sequence of $(\ceil{2^{nR_1}},\dots, \ceil{2^{nR_M}},n)$ codes satisfying Equations~\eqref{eq:reliabilitycondition_2}~and~\eqref{eq:secrecycondition_2}. 

\begin{Proposition}[Adapted from \cite{7264977,6687232}]
\label{prop:SCR_2}
The achievable region of the DBC-NLD-LS is the union of all $M$-tuples of rates $(R_1,\dots, R_M) \in \mathbb{R}_{+}^M$ satisfying the following inequalities,
\begin{align*}
\sum_{{i} = m}^M R_{i} \leq I(X;Y_1) - I(X; Z_{m}), \qquad m = 1,\dots,M,
\end{align*}
where the union is taken over all distributions $p_X$.
\end{Proposition} 
The proof for the case of only one legitimate receiver in the context of the fading wiretap channel is provided in \cite{6687232}, where the information-theoretic achievable scheme is based on embedded coding, stochastic encoding and rate sharing. Due to the degradedness condition of Equation~\eqref{eq:degch}, by applying the data processing inequality and Fano's inequality, an achievable scheme ensuring the reliability condition in Equation~\eqref{eq:reliabilitycondition_2} for the legitimate Receiver 1 will satisfy it for any legitimate receiver $k \in [2,K]$.

\begin{Corollary}
\label{coro:SCR_2}
The achievable subregion of the DBC-NLD-LS without considering rate sharing is a $K$-orthotope defined by the closure of all $K$-tuples of rates $(R_1,\dots, R_M) \in \mathbb{R}_{+}^M$ satisfying:
\begin{align*}
R_m & \leq I(X; Z_{m+1}) - I(X; Z_{m} ), \qquad m =1,\dots,M-1, \\
R_M & \leq I(X; Y_1) - I(X; Z_{M} ).
\end{align*}
\end{Corollary}

\subsection{Degraded Broadcast Channel with Layered Decoding and Non-Layered Secrecy}\label{sec:SM_dbcldnls}
\begin{figure}[H]
 \resizebox{0.95\linewidth}{!}{%
\centering
\begin{tikzpicture}
\node (E) at (-4.7,0) [shape=rectangle,thick,rounded corners=8pt,inner sep=0pt,minimum width=2cm, minimum height=0.8cm, draw] {\small Encoder};
\node (C) at (-1.2,0) [shape=rectangle,thick,rounded corners=8pt,inner sep=0pt,minimum width=3cm, minimum height=4.8cm,align=center,draw] {DBC \\ {\small $p_{Y_K \dots Y_1 Z_M \dots Z_1|X}$} \\ {\small satisfying} \\ {\small Equation~\eqref{eq:degch}} };
\node (E1) at (2.75,-2.1) [shape=rectangle,thick,rounded corners=4pt,minimum width=3cm, inner xsep=6pt, inner ysep=2pt, draw] {\footnotesize Eavesdropper 1};
\node (EM) at (2.75,-0.8) [shape=rectangle,thick,rounded corners=4pt,minimum width=3cm, inner xsep=6pt, inner ysep=2pt, draw] {\footnotesize Eavesdropper $M$};
\node (RK) at (2.75,2.1) [shape=rectangle,thick,rounded corners=4pt,minimum width=3cm, inner xsep=6pt, inner ysep=2pt, draw] {\footnotesize Leg. Receiver $K$};
\node (R2) at (2.75,0.8) [shape=rectangle,thick,rounded corners=4pt,minimum width=3cm, inner xsep=6pt, inner ysep=2pt, draw] {\footnotesize Leg. Receiver 2};
\node (R1) at (2.75,0) [shape=rectangle,thick,rounded corners=4pt,minimum width=3cm, inner xsep=6pt, inner ysep=2pt, draw] {\footnotesize Leg. Receiver 1};
\node (Wt) at (2.7,1.52) [] {$\vdots$};
\node (Wt) at (2.7,-1.36) [] {$\vdots$};
\node (E1t) at (6.3,-2.1) [] {\footnotesize $\Ccancel[red]{W_1}, \Ccancel[red]{W_2},\dots,\Ccancel[red]{W_K}$};
\node (EMt) at (6.3,-0.8) [] {\footnotesize $\Ccancel[red]{W_1}, \Ccancel[red]{W_2},\dots,\Ccancel[red]{W_K}$};
\node (RKt) at (6.3,2.1) [] {\footnotesize $W_1, W_2,\dots,W_K$};
\node (R2t) at (6.3,0.8) [] {\footnotesize $W_1, W_2$\color{white}{$,\dots,W_K$}};
\node (R1t) at (6.3,0) [] {\footnotesize $W_1$\color{white}{$,W_2,\dots,W_K$}};
\draw [arrow] (-7.9,0) -- (E) node [midway, above] (TextNode) {\footnotesize $W_1,\dots,W_K$};
\draw [arrow] (E) -- (C) node [midway, above] (TextNode) {\footnotesize $X^n$};
\draw [arrow] (0.3,2.1) -- (RK) node [midway, above] (TextNode) {\footnotesize $Y_K^n$};
\draw [arrow] (0.3,0.8) -- (R2) node [midway, above] (TextNode) {\footnotesize $Y_2^n$};
\draw [arrow] (0.3,0) -- (R1) node [midway, above] (TextNode) {\footnotesize $Y_1^n$};
\draw [arrow] (0.3,-0.8) -- (EM) node [midway, above] (TextNode) {\footnotesize $Z_M^n$};
\draw [arrow] (0.3,-2.1) -- (E1) node [midway, above] (TextNode) {\footnotesize $Z_1^n$};
\draw [arrow] (RK) -- (RKt) node [midway, above] (TextNode) {};
\draw [arrow] (R2) -- (R2t) node [midway, above] (TextNode) {};
\draw [arrow] (R1) -- (R1t) node [midway, above] (TextNode) {};
\draw [arrow] (EM) -- (EMt) node [midway, above] (TextNode) {};
\draw [arrow] (E1) -- (E1t) node [midway, above] (TextNode) {};
\end{tikzpicture}
}
\caption{DBC with Layered Decoding and Non-Layered Secrecy (DBC-LD-NLS).}\label{fig:dbcldnls}
\end{figure}
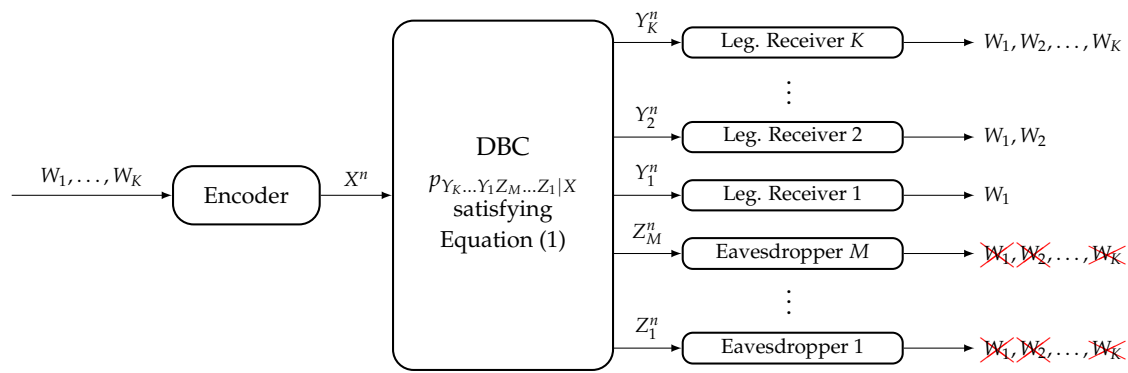

In this model (see Figure~\ref{fig:dbcldnls}), the transmitter wishes to send $K$ messages $\{W_{\ell} \}_{{\ell}=1}^{K}$ to the $K$ legitimate receivers. The layered decoding structure requires the legitimate receiver $k \in [1,K]$ to reliably decode the messages $\{W_{{\ell}} \}_{{{\ell}}=1}^{k}$, and the non-layered secrecy structure requires the eavesdropper $m \in [1,M]$ to be kept ignorant of all $K$ messages. Consider a $(\ceil{2^{nR_1}},\dots, \ceil{2^{nR_K}},n)$ code for the DBC-LD-NLS, where $W_{\ell} \in [\ceil{2^{nR_{\ell}}}]$ for any $\ell \in [1,K]$. The reliability condition to be satisfied by this code is: 
\begin{align}
\lim_{n \rightarrow \infty} \mathbb{P} \left[ (\hat{W}_1, \dots, \hat{W}_{k-1}, \hat{W}_k) \neq ({W}_1, \dots, {W}_{k-1}, {W}_k) \right] = 0, \quad \text{for the legitimate receiver } k \in [1, K], \label{eq:reliabilitycondition_1}
\end{align}
and the {strong} secrecy condition is given by:
\begin{align}\label{eq:secrecycondition_1}
\lim_{n \rightarrow \infty} I(W_1,\dots,W_K ; Z_m^n) = 0, \quad \text{for any eavesdropper } m \in [1, M].
\end{align}
A tuple of rates $(R_1,\dots, R_K) \in \mathbb{R}_{+}^K$ is achievable for the DBC-LD-NLS if there exists a sequence of $(\ceil{2^{nR_1}},\dots, \ceil{2^{nR_K}},n)$ codes such that they satisfy Equations~\eqref{eq:reliabilitycondition_1}~and~\eqref{eq:secrecycondition_1}.

\begin{Proposition}[ Adapted from \cite{7264977,Ekrem1661312}]
\label{prop:SCR_1}
The achievable region of the DBC-LD-NLS is the union of all $K$-tuples of rates $(R_1,\dots, R_K) \in \mathbb{R}_{+}^K$ satisfying the following inequalities,
\begin{align*}
\sum_{{\ell} = 1}^k R_{\ell} \leq \sum_{{\ell}=1}^k I(V_{\ell}; Y_{\ell} | V_{{\ell}-1}) - I(V_k, Z_M), \qquad k = 1,\dots,K,
\end{align*}
where $V_0 \triangleq \varnothing$ and $V_K \triangleq X$, and the union is taken over all distributions $p_{V_1\dots V_K}$ such that $V_1 - V_2 - \dots - V_{K} $ forms a Markov chain.
\end{Proposition} 
The proof for the case of only one eavesdropper is provided in \cite{Ekrem1661312}, where the information-theoretic achievable scheme is based on superposition coding, stochastic encoding and rate sharing.~Due to the degradedness condition of Equation~\eqref{eq:degch}, note that any achievable scheme ensuring the strong secrecy condition in Equation~\eqref{eq:secrecycondition_1} for the eavesdropper $M$ will also satisfy it for any eavesdropper $m \in [1,M-1]$.

\begin{Corollary}
\label{coro:SCR_1}
The achievable subregion of the DBC-LD-NLS without considering rate sharing is a $K$-orthotope defined by the closure of all $K$-tuples of rates $(R_1,\dots, R_K) \in \mathbb{R}_{+}^K$ satisfying:
\begin{align*}
R_{\ell} & \leq I(V_{\ell}; Y_{\ell} | V_{{\ell}-1}) - I(V_{\ell}; Z_M | V_{{\ell}-1}), \qquad {\ell} =1,\dots,K.
\end{align*}
\end{Corollary}

\section{Review of Polar Codes}\label{sec:rev}
Let $(\mathcal{X} \times \mathcal{Y}, p_{XY})$ be a Discrete Memoryless Source (DMS), where ${X} \in \{0,1\}$ (see Endnote~\cite{Note1}---which refers to References \cite{5513555,csasouglu2009polarization}) and $Y \in \mathcal{Y}$.~The polar transform over the $n$-sequence $X^n$, $n$ being any power of two, is defined as \mbox{$U^n \triangleq X^n G_n$}, where $G_n \triangleq \begin{brsm}1 & 1\\1 & 0\end{brsm}^{\otimes n}$ is the source polarization matrix \cite{arikan2010source}.~Since $G_n = G_n^{-1}$, \mbox{then~$X^n = U^n G_n$}.


The polarization theorem for source coding with side information \cite{arikan2010source} ({Theorem} 1) states that the polar transform extracts the randomness of $X^n$ in the sense that, as $n \rightarrow \infty$, the set of indices $j \in [n]$ can be divided practically into two disjoint sets, namely $\mathcal{H}_{X|Y}^{(n)}$ and $\mathcal{L}_{X|Y}^{(n)}$, such that $U(j)$ for $j \in \mathcal{H}_{X|Y}^{(n)}$ is practically independent of $(U^{1:j-1},Y^n)$ and uniformly distributed, i.e., $H ({U(j) | U^{1:j-1}, Y^n} ) \rightarrow 1$, and~$U(j)$ for $j \in \mathcal{L}_{X|Y}^{(n)}$ is almost determined by $(U^{1:j-1}, Y^n)$, i.e., $H ( U(j) | U^{1:j-1}, Y^n ) \rightarrow 0$. Formally, let: 
\begin{align}
\mathcal{H}_{X|Y}^{(n)} & \triangleq \big\{ j \in [n]: H \big( U(j) \big| U^{1:j-1},Y^n \big) \geq 1-\delta_n \big\}, \nonumber \\
\mathcal{L}_{X|Y}^{(n)} & \triangleq \big\{ j \in [n]: H \big( U(j) \big| U^{1:j-1},Y^n \big) \leq \delta_n \big\}, \nonumber
\end{align}
\textls[-15]{where $\delta_n \triangleq 2^{-n^{\beta}}$ for some $\beta \in (0, \frac{1}{2})$. Then, by \cite{arikan2010source} ({Theorem} 1), we have $\lim_{n \rightarrow \infty} \frac{1}{n} | \mathcal{H}_{X|Y}^{(n)} | = H(X|Y)$} and $\lim_{n \rightarrow \infty} \frac{1}{n} | \mathcal{L}_{X|Y}^{(n)} | = 1 - H(X|Y)$, which imply that $\lim_{n \rightarrow \infty} \frac{1}{n} |( \mathcal{H}_{X|Y}^{(n)} )^{\text{{C}}} \cap ( \mathcal{L}_{X|Y}^{(n)} )^{\text{{C}}}| = 0$, i.e., the number of elements that {have not been polarized} is asymptotically negligible in terms of rate.~Furthermore, \cite{arikan2010source}~({Theorem} 2) states that given $U[(\mathcal{L}_{X|Y}^{(n)} )^{\text{C}}]$ and $Y^n$, $U[\mathcal{L}_{X|Y}^{(n)}]$ can be reconstructed using SC decoding with error probability in $O(n \delta_n)$.~Alternatively, the previous sets can be defined based on the Bhattacharyya parameters $\{ Z ( U(j) \big| U^{1:j-1},Y^n) \}_{j=1}^n$ because both parameters {polarize} simultaneously \cite{arikan2010source} (Proposition~2).~It~is worth mentioning that both the entropy terms and the Bhattacharyya parameters required to define these sets can be obtained deterministically from $p_{XY}$ and the algebraic properties of $G_n$ \cite{tal2013construct,vangala2015comparative,6601656}.

Similarly to $\mathcal{H}_{X|Y}^{(n)}$ and $\mathcal{L}_{X|Y}^{(n)}$, the sets $\mathcal{H}_{X}^{(n)}$ and $\mathcal{L}_{X}^{(n)}$ can be defined by considering that observations $Y^n$ are absent.~A discrete memoryless channel $(\mathcal{X}, p_{Y|X}, \mathcal{Y})$ with some arbitrary $p_X$ can be seen as a~DMS $(\mathcal{X} \times \mathcal{Y}, p_{X}p_{Y|X})$.~In channel polar coding, first, we define $\mathcal{H}_{X|Y}^{(n)}$, $\mathcal{L}_{X|Y}^{(n)}$, $\mathcal{H}_{X}^{(n)}$ and $\mathcal{L}_{X}^{(n)}$ from the target distribution $p_{X}p_{Y|X}$ ({polar construction}).~Then, based on the previous sets, the encoder somehow constructs $\tilde{U}^n$ and applies the inverse polar transform $\tilde{X}^n = \tilde{U}^n G_n$, with~distribution $\tilde{q}_{X^n}$ (since the polar-based encoder will construct random variables that must approach the target distribution of the DMS, throughout this paper, we use a {tilde} above the random variables to emphasize this purpose). Afterwards, the transmitter sends $\tilde{X}^n$ over the channel, which induces $\tilde{Y}^n \sim \tilde{q}_{Y^n}$. \mbox{If $\mathbb{V} (\tilde{q}_{X^nY^n}, p_{X^nY^n}) \rightarrow 0$}, then the receiver can reliably reconstruct $\tilde{U}[\mathcal{L}_{X|Y}^{(n)}]$ from $\tilde{Y}^n$ and $\tilde{U}[(\mathcal{L}_{X|Y}^{(n)} )^{\text{C}}]$ by using SC decoding~\cite{korada2010polar}.

To conclude this part, the following lemma provides a useful property of polar codes for the DBC.

\begin{Lemma}[ Subset property, adapted from {\cite{6975233} ({Lemma~4})}]\label{lemma:subsetproperty}
Let $(X, Y_2, Y_1)$ be random variables such that $X - Y_{2} - Y_1 $ forms a Markov chain. Then, the following property holds for the polar transform $U^n = X^n G_n$,
\begin{gather}
H \big( U(j) \big| U^{1:j-1} \big) \geq H \big( U(j) \big| U^{1:j-1}, Y_1^n \big) \geq H \big( U(j) \big| U^{1:j-1}, Y_2^n \big) \quad \forall j \in [n], \quad \text{which implies} \nonumber \\
\mathcal{L}_{X}^{(n)} \subseteq \mathcal{L}_{X|Y_1}^{(n)} \subseteq \mathcal{L}_{X|Y_2}^{(n)}, \quad \text{and} \quad \mathcal{H}_{X| Y_{2}}^{(n)} \subseteq \mathcal{H}_{X|Y_{1}}^{(n)} \subseteq \mathcal{H}_{X}^{(n)}. \nonumber 
\end{gather}
\end{Lemma}

\begin{Remark}
The subset property also holds if the sets are defined based on the Bhattacharyya parameters because, under the previous Markov chain condition, $Z \big( U(j) \big| U^{1:j-1} \big) \geq Z \big( U(j) \big| U^{1:j-1}, Y_1^n \big) \geq Z \big( U(j) \big| U^{1:j-1}, Y_2^n \big)$.
\end{Remark}

\begin{Remark} \label{re:subp}
According to \cite{6975233} (Lemma~4), the subset property also holds if the channels are stochastically degraded. Therefore, since the construction of the polar codes proposed in the following sections is based basically on Lemma~\ref{lemma:subsetproperty}, the polar coding schemes are suitable for physically- and stochastically-degraded channels.
\end{Remark}

\section{Polar Coding Scheme For the DBC-NLD-LS}\label{sec:PCS_dbcnldls}
The polar coding scheme provided in this section is designed to achieve the supremum of the achievable rates given in Corollary~\ref{coro:SCR_2} (secrecy-capacity without rate sharing).~Thus, consider the DMS $\big( \mathcal{X} \times \mathcal{Y}_K \times \cdots \times \mathcal{Y}_1 \times \mathcal{Z}_M \times \cdots \times \mathcal{Z}_1, p_{X Y_K\dots Y_1 Z_M \dots Z_1} \big)$ that represents the input and output random variables involved in the achievable subregion of Corollary~\ref{coro:SCR_2}, where $\mathcal{X} = \{0,1\}$. Let~$(X^n,Y_K^n,\dots,Y_1^n,Z_M^n,\dots,Z_1^n)$ be an i.i.d. $n$-sequence of this source. We define the polar transform $U^n \triangleq X^n G_n$, whose distribution is $p_{U^n} (u^n ) = p_{X^n} (u^n G_n)$ (due to the invertibility of $G_n$), and we write:
\begin{align}
\label{eq:PCS_dbcnldls_jointU}
p_{U^n} (u^n ) \triangleq \prod_{j = 1}^n p_{U(j) | U^{1:j-1}} ( u(j) \big| u^{1:j-1} ) .
\end{align} 

\subsection{Polar Code Construction}\label{sec:PCS_dbcnldls_pcc}
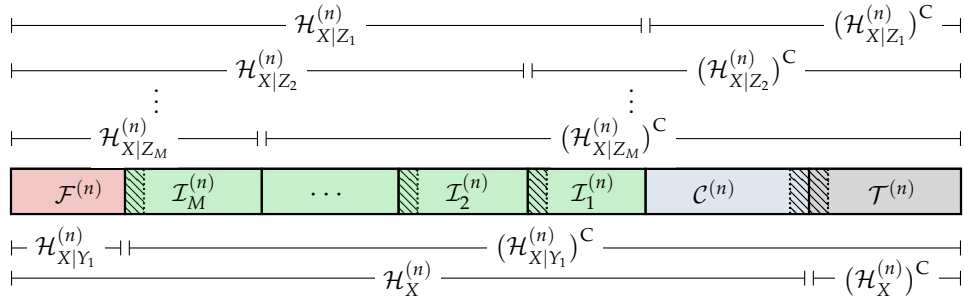
\begin{figure}[H]
\centering
\begin{tikzpicture}
\node at (0,-0.6) [shape=rectangle,thick,inner sep=0pt,minimum width=12.5cm, minimum height=0.6cm, draw] {};
\draw[|-|] (-6.25cm,-1.75cm) -- node[fill=white,inner sep=1mm,midway] {\small $\mathcal{H}_{X}^{(n)}$} (4.2cm,-1.75cm);
\draw[|-|] (4.3cm,-1.75cm) -- node[fill=white,inner sep=1mm,midway] {\small $\big(\mathcal{H}_{X}^{(n)}\big)^{\text{C}}$} (6.25cm,-1.75cm);
\draw[|-|] (-6.25cm,-1.35cm) -- node[fill=white,inner sep=1mm,midway] {\small $\mathcal{H}_{X|Y_1}^{(n)}$} (-4.8cm,-1.35cm);
\draw[|-|] (-4.7cm,-1.35cm) -- node[fill=white,inner sep=1mm,midway] {\small $\big(\mathcal{H}_{X| Y_1}^{(n)}\big)^{\text{C}}$} (6.25cm,-1.35cm);
\draw[thick,densely dotted] (-4.5cm,-0.3cm) -- (-4.5cm,-0.9cm); 
\node at (-5.5,-0.6) [shape=rectangle,thick,inner sep=0pt,minimum width=1.5cm, minimum height=0.6cm, draw,fill={rgb:red,20;green,2;blue,2},fill opacity=0.24] {};
\fill[pattern=north west lines] (-4.5,-0.3) rectangle (-4.75,-0.9);
\draw[|-|] (-6.25cm,0.1cm) -- node[fill=white,inner sep=1mm,midway] {\small $\mathcal{H}_{X|Z_M}^{(n)}$} (-3cm,0.1cm);
\draw[|-|] (-2.9cm,0.1cm) -- node[fill=white,inner sep=1mm,midway] {\small $\big(\mathcal{H}_{X|Z_M}^{(n)}\big)^{\text{C}}$} (6.25cm,0.1cm);
\node at (-3.85,-0.6) [shape=rectangle,thick,inner sep=0pt,minimum width=1.8cm, minimum height=0.6cm, draw,fill={rgb:red,2;green,20;blue,4},fill opacity=0.24] {};
\node at (-2.05,-0.6) [shape=rectangle,thick,inner sep=0pt,minimum width=1.8cm, minimum height=0.6cm, draw,fill={rgb:red,2;green,20;blue,4},fill opacity=0.24] {};
\draw[thick,densely dotted] (-0.9cm,-0.3cm) -- (-0.9cm,-0.9cm);
\fill[pattern=north west lines] (-1.15,-0.3) rectangle (-0.9,-0.9);
\node at (-0.3,-0.6) [shape=rectangle,thick,inner sep=0pt,minimum width=1.7cm, minimum height=0.6cm, draw,fill={rgb:red,2;green,20;blue,4},fill opacity=0.24] {};
\node[text width=0.2cm, anchor=west] at (-4.5cm,0.7) {\small $\vdots$};
\node[text width=0.2cm, anchor=west] at (1.75cm,0.7) {\small $\vdots$};
\draw[|-|] (-6.25cm,1cm) -- node[fill=white,inner sep=1mm,midway] {\small $\mathcal{H}_{X|Z_2}^{(n)}$} (0.5cm,1cm);
\draw[|-|] (0.6cm,1cm) -- node[fill=white,inner sep=1mm,midway] {\small $\big(\mathcal{H}_{X|Z_2}^{(n)}\big)^{\text{C}}$} (6.25cm,1cm);
\node at (1.325,-0.6) [shape=rectangle,thick,inner sep=0pt,minimum width=1.55cm, minimum height=0.6cm, draw,fill={rgb:red,2;green,20;blue,4},fill opacity=0.24] {};
\fill[pattern=north west lines] (0.55,-0.3) rectangle (0.8,-0.9);
\draw[thick,densely dotted] (0.8cm,-0.3cm) -- (0.8cm,-0.9cm);
\draw[|-|] (-6.25cm,1.6cm) -- node[fill=white,inner sep=1mm,midway] {\small $\mathcal{H}_{X|Z_1}^{(n)}$} (2.05cm,1.6cm);
\draw[|-|] (2.15cm,1.6cm) -- node[fill=white,inner sep=1mm,near end] {\small $\big(\mathcal{H}_{X|Z_1}^{(n)}\big)^{\text{C}}$} (6.25cm,1.6cm);
\node at (3.175,-0.6) [shape=rectangle,thick,inner sep=0pt,minimum width=2.15cm, minimum height=0.6cm, draw,fill={rgb:red,5;green,10;blue,20},fill opacity=0.14] {};
\draw[thick,densely dotted] (4cm,-0.3cm) -- (4cm,-0.9cm);
\fill[pattern=north west lines] (4,-0.3) rectangle (4.25,-0.9);
\node at (5.25,-0.6) [shape=rectangle,thick,inner sep=0pt,minimum width=2cm, minimum height=0.6cm, draw,fill={rgb:red,2;green,2;blue,2},fill opacity=0.24] {};
\draw[thick,densely dotted] (4.5cm,-0.3cm) -- (4.5cm,-0.9cm);
\fill[pattern=north west lines] (4.25,-0.3) rectangle (4.5,-0.9);
\node[text width=1cm, anchor=west] at (-5.8cm,-0.6) {\small $\mathcal{F}^{(n)}$};
\node[text width=1cm, anchor=west] at (-4.25cm,-0.6) {\small $\mathcal{I}_M^{(n)}$};
\node[text width=1cm, anchor=west] at (-2.45cm,-0.65) {$\cdots$};
\node[text width=1cm, anchor=west] at (-0.65cm,-0.6) {\small $\mathcal{I}_2^{(n)}$};
\node[text width=1cm, anchor=west] at (1cm,-0.6) {\small $\mathcal{I}_1^{(n)}$};
\node[text width=1cm, anchor=west] at (2.6cm,-0.6) {\small $\mathcal{C}^{(n)}$};
\node[text width=1cm, anchor=west] at (4.9cm,-0.6) {\small $\mathcal{T}^{(n)}$};
\end{tikzpicture}
\caption{Polar code construction for DBC-NLD-LS. The hatched area represents those indices $j \in ( \mathcal{H}_{X|Y_1}^{(n)} )^{\text{C}} \cap ( \mathcal{L}_{X|Y_1}^{(n)} )^{\text{C}}$, which can belong to the sets $\mathcal{I}_m^{(n)}$ ($m \in [1,M]$), $\mathcal{C}^{(n)}$, $\mathcal{F}^{(n)}$ or $\mathcal{T}^{(n)}$.}\label{fig:pc_dbcnldls}
\end{figure}

Let $\delta_n \triangleq 2^{-n^{\beta}}$, where $\beta \in (0, \frac{1}{2})$. Based on $p_{X Y_K \dots Y_1 Z_M \dots Z_1}$, we define:
\begin{align}
\mathcal{H}_{X}^{(n)} & \triangleq \big\{ j \in [n]: H \big( U(j) \big| U^{1:j-1} \big) \geq 1-\delta_n \big\}, \label{eq:Aset1} \\
\mathcal{L}_{X}^{(n)} & \triangleq \big\{ j \in [n]: H \big( U(j) \big| U^{1:j-1} ) \leq \delta_n \big\}, \label{eq:Aset2} \\
\mathcal{L}_{X | Y_{k}}^{(n)} & \triangleq \big\{ j \in [n]: H \big( U(j) \big| U^{1:j-1}, Y_{k}^n \big) \leq \delta_n \big\}, \quad k = 1,\dots,K, \label{eq:Aset3} \\
\mathcal{H}_{X | Y_{k}}^{(n)} & \triangleq \big\{ j \in [n]: H \big( U(j) \big| U^{1:j-1}, Y_{k}^n \big) \geq 1-\delta_n \big\}, \quad k = 1,\dots,K, \label{eq:Aset4} \\
\mathcal{H}_{X| Z_{m}}^{(n)} & \triangleq \big\{ j \in [n]: H \big( U(j) \big| U^{1:j-1}, Z_m^n \big) \geq 1-\delta_n \big\}, \quad m = 1,\dots,M. \label{eq:Aset5}
\end{align}
Then, based on the previous sets, we define the following partition of the universal set $[n]$,
\begin{align}
\label{eq:setIM} \mathcal{I}_M^{(n)} & \triangleq \mathcal{H}_{X|Z_{M}}^{(n)} \cap \big( \mathcal{H}_{X|Y_1}^{(n)} \big)^{\text{C}} , \\
\label{eq:setIm} \mathcal{I}_m^{(n)} & \triangleq \mathcal{H}_{X|Z_{m}}^{(n)} \cap \big( \mathcal{H}_{X|Z_{m+1}}^{(n)} \big)^{\text{C}}, \quad m = 1,\dots, M-1, \\
\label{eq:setF1} \mathcal{F}^{(n)} & \triangleq \mathcal{H}_{X|Y_1}^{(n)} , \\
\label{eq:setC1} \mathcal{C}^{(n)} & \triangleq \mathcal{H}_{X}^{(n)} \cap \big( \mathcal{H}_{X|Z_{1}}^{(n)} \big)^{\text{C}}, \\
\label{eq:setT1} \mathcal{T}^{(n)} & \triangleq \big( \mathcal{H}_{X}^{(n)} \big)^{\text{C}},
\end{align}
which is graphically represented in Figure~\ref{fig:pc_dbcnldls}.~Roughly speaking, in order to ensure reliability and strong secrecy, the distribution of $\tilde{U}^n$ after the encoding process must be close in terms of statistical distance to the distribution given in Equation~\eqref{eq:PCS_dbcnldls_jointU} corresponding to the original DMS. Hence, the~elements $U(j)$ such that $j \in \mathcal{H}_{X}^{(n)}$ will be suitable for storing uniformly-distributed random sequences. On the other hand, $U[\mathcal{T}^{(n)}]$ will not, and the elements $U(j)$ such that $j \in \mathcal{T}^{(n)}$ will be constructed somehow from $U^{1:j-1}$ and the distribution $p_{U(j) | U^{1:j-1}}$.~The set $\mathcal{I}_m^{(n)}$ ($m \in [1,M]$) belongs to $\mathcal{H}_{X|Z_m}^{(n)}$, and by~Lemma~\ref{lemma:subsetproperty}, we have $\mathcal{H}_{X|Z_{m}}^{(n)} \subseteq \mathcal{H}_{X|Z_{m^{\prime}}}^{(n)}$ for any $m^{\prime} < m$.~Thus, $U [ \mathcal{I}_m^{(n)}]$ will be suitable for storing information to be secured from Eavesdroppers 1--$m$. Since $\mathcal{C}^{(n)} \subseteq (\mathcal{H}_{X|Z_m}^{(n)})^{\text{C}}$ for any $m \in [1,M]$, the sequence $U \big[ \mathcal{C}^{(n)} \big]$ cannot contain information to be secured from any eavesdropper, and it will be used to store the {local randomness} \cite{6034749} required to confuse the eavesdroppers (the local randomness in polar codes plays the same role as the stochastic encoding used in \cite{6772207,1055892}). According to \cite{arikan2010source} ({Theorem~2}), the~legitimate Receiver 1 will be able to reliably infer $U [ \mathcal{L}_{X|Y_{1}}^{(n)} ]$ given $Y_{1}^n$ and $U [ (\mathcal{L}_{X|Y_{1}}^{(n)})^{\text{C}} ]$. Hence, if the polar coding scheme somehow make the entries $U(j)$ such that $j$ belongs to $\mathcal{F}^{(n)}$ and $( \mathcal{H}_{X|Y_{1}}^{(n)} )^{\text{C}} \cap ( \mathcal{L}_{X|Y_{1}}^{(n)} )^{\text{C}}$ (hatched areas in Figure~\ref{fig:pc_dbcnldls}) available to the legitimate Receiver 1, this receiver will be able to reliably infer the entire sequence $U^n$. In this sense, $U [ \mathcal{F}^{(n)} ]$ will be used to store the uniformly-distributed random sequence provided by a source of common randomness that will be available to all parties. Since $\mathcal{F}^{(n)} \subseteq \mathcal{H}_{X|Z_m}^{(n)}$ for any $m \in [1,M]$, the knowledge of $U[ \mathcal{F}^{(n)}]$ of the eavesdroppers will not compromise the strong secrecy condition. On the other hand, $U [ ( \mathcal{H}_{X|Y_{1}}^{(n)} )^{\text{C}} \cap ( \mathcal{L}_{X|Y_{1}}^{(n)} )^{\text{C}} ]$ will contain secret information or elements that cannot be known directly by all the eavesdroppers. Therefore, the transmitter somehow will secretly send it to the legitimate receivers. Nevertheless, as will be seen, this additional transmission will incur an asymptotically negligible rate penalty. Finally, by Lemma~\ref{lemma:subsetproperty}, we have $(\mathcal{L}_{X|Y_{1}}^{(n)} )^{\text{C}} \supseteq ( \mathcal{L}_{X|Y_{k}}^{(n)} )^{\text{C}}$ for any $k > 1$. Hence, given $U [ (\mathcal{L}_{X|Y_{1}}^{(n)})^{\text{C}} ]$, all the legitimate receivers will be able to reliably infer the entire sequence $U^n$ from their own channel observations. 

\begin{Remark}\label{remark:construction}
The goal of the polar code construction is to obtain the entropy terms $\{ H( U(j) \big| U^{1:j-1}) \}_{j=1}^n$, $\{ H( U(j) \big| U^{1:j-1}, Y_1^n) \}_{j=1}^n$ and $\{ H( U(j) \big| U^{1:j-1}, Z_m^n) \}_{j=1}^n$ for all $m \in [1,M]$ required to define the sets in Equations~\eqref{eq:Aset1}--\eqref{eq:Aset5} and, consequently, to obtain the partition of $[n]$ given in Equations~\eqref{eq:setIM}--\eqref{eq:setT1}. In Section~\ref{sec:results}, we discuss further how to construct polar codes under both reliability and secrecy constraints.
\end{Remark}

\subsection{Polar Encoding}\label{sec:PCS_dbcnldls_encoder}
The polarization-based encoder aims to construct the sequence $\tilde{U}^n$ and, consequently, $\tilde{X}^n = \tilde{U}^n G_n$. Let $W_m$ for all $m \in [1,M]$ and $C$ be uniformly-distributed random vectors of size $|\mathcal{I}_{m}^{(n)}|$ and $|\mathcal{C}^{(n)}|$, respectively, where $C$ represents the local randomness required to confuse the eavesdroppers, and recall that $W_m$ represents the message $m$ that is intended for all legitimate receivers. Let $F$ be a given uniformly-distributed random $|\mathcal{F}^{(n)}|$-sequence, which represents the source of common randomness that is available to all parties.~The encoder constructs the sequence $\tilde{u}^n$ as follows.~Consider the realizations $w_m$ for all $ m \in [1,M]$, $c$ and $f$, whose elements have been indexed by the set of indices $\mathcal{I}_{m}^{(n)}$, $\mathcal{C}^{(n)}$ and $\mathcal{F}^{(n)}$, respectively. The encoder draws $\tilde{u}^n$ from the distribution:
\begin{align}
\tilde{q}_{U(j) | U^{1:j-1}} ( \tilde{u}(j) | \tilde{u}^{1:j-1} ) \triangleq \left\{ 
\begin{array}{ll}
\mathds{1} \big\{ \tilde{u}(j) = w_m(j) \big\} & \text{if } j \in \mathcal{I}_{m}^{(n)}, \, \, \, m = 1,\dots,M, \\
\mathds{1} \big\{ \tilde{u}(j) = c(j) \big\} & \text{if } j \in \mathcal{C}^{(n)}, \\
\mathds{1} \big\{ \tilde{u}(j) = f(j) \big\} & \text{if } j \in \mathcal{F}^{(n)}, \\
p_{U(j) | U^{1:j-1}} \! \left( \tilde{u}(j) | \tilde{u}^{1:j-1} \right) & \text{if } j \in \big( \mathcal{H}_{X}^{(n)} \big)^{\text{C}} \cap \big( \mathcal{L}_{X}^{(n)} \big)^{\text{C}}, \\
 \mathds{1} \big\{ \tilde{u}(j) = \xi^{(j)}(\tilde{u}^{1:j-1}) \big\} & \text{if } j \in \mathcal{L}_{X}^{(n)},
\end{array}
\right. \label{eq:distrenc1}
\end{align}
where: 
\begin{align} 
\xi^{(j)} ( \tilde{u}^{1:j-1} ) \triangleq \argmax_{u \in \mathcal{X}} p_{U(j) | U^{1:j-1} } ( u | \tilde{u}^{1:j-1} ), \label{eq:argm2a}
\end{align}
$p_{U(j) | U^{1:j-1}}$ being the distribution induced by the original DMS. Note that $\mathcal{T}^{(n)} = ( ( \mathcal{H}_{X}^{(n)} )^{\text{C}} \cap ( \mathcal{L}_{X}^{(n)} )^{\text{C}} ) \cup \mathcal{L}_{X}^{(n)}$, and according to Equation~\eqref{eq:distrenc1}, $\tilde{U} [ \mathcal{L}_{X}^{(n)} ]$ is constructed deterministically by adapting the SC encoding algorithm in \cite{7447169}, while $\tilde{U} [ ( \mathcal{H}_{X}^{(n)} )^{\text{C}} \cap ( \mathcal{L}_{X}^{(n)} )^{\text{C}} ]$ is constructed randomly. {By \cite{arikan2010source} ({Theorem~1}), we have that the amount of randomness for SC encoding will be asymptotically negligible in terms of rate}.~Then, the encoder computes $\tilde{X}^n = \tilde{U}^n G_n$ and transmits it over the DBC, inducing $(\tilde{Y}_K, \dots, \tilde{Y}_1, \tilde{Z}_M, \dots, \tilde{Z}_1)$.

Finally, besides the sequence $\tilde{X}^n$, the encoder outputs the following additional secret sequence,
\begin{align}
\label{eq:secretmessage2}
\Phi \triangleq \tilde{U} \Big[ \big( \mathcal{H}_{X|Y_1}^{(n)} \big)^{\text{C}} \cap \big( \mathcal{L}_{X|Y_1}^{(n)} \big)^{\text{C}} \Big].
\end{align}
This sequence $\Phi$ must be additionally transmitted to all legitimate receivers keeping it masked from the eavesdroppers. To do so, the transmitter can perform a modulo-two addition between $\Phi$ and a~uniformly-distributed secret key that is privately shared with the legitimate receivers and somehow additionally send it to them. Nevertheless, by \cite{arikan2010source} ({Theorem~1}), we know that this additional transmission is asymptotically negligible in terms of rate.

\begin{Remark}\label{remarkA}
The additional secret sequence $\Phi$ can be divided into two parts: $\tilde{U} [ \mathcal{H}_{X}^{(n)} \cap ( \mathcal{H}_{X|Y_1}^{(n)} )^{\text{C}} \cap ( \mathcal{L}_{X|Y_1}^{(n)} )^{\text{C}} ]$, which will be uniformly distributed according to Equation~\eqref{eq:distrenc1}, and the remaining part that will not. The transmitter could make the uniformly-distributed part available to the legitimate receivers by using a chaining structure as the one presented in \cite{6620400}. However, such a scheme requires the transmission to take place over several blocks of size $n$. Moreover, it requires having a large memory capacity on either the transmitter or the legitimate receivers, which can make the polar coding scheme unpractical in communication systems. 
\end{Remark}



\subsection{Polar Decoding}\label{sec:PCS_dbcnldls_decoder}
Before the decoding process, consider that the realization of the source of common randomness $F$ is available to all parties and the sequence $\Phi$ has been successfully received by the legitimate receivers.

The legitimate receiver $k \in [1,K]$ forms an estimate $\hat{U}^n$ of the sequence $\tilde{U}^n$ as follows. Given that $\Phi$ and $F$ are available, notice that it knows $\tilde{U}[( \mathcal{L}_{X|Y_1}^{(n)} )^{\text{C}} ]$. Moreover, by Lemma~\ref{lemma:subsetproperty}, \mbox{$( \mathcal{L}_{X|Y_1}^{(n)})^{\text{C}} \supseteq ( \mathcal{L}_{X|Y_k}^{(n)})^{\text{C}}$} for any $k > 1$.~Thus, the $k$-th legitimate receiver performs SC decoding for source coding with side information \cite{arikan2010source} to construct $\tilde{U}^n$ from $\tilde{U}[( \mathcal{L}_{X|Y_1}^{(n)} )^{\text{C}} ]$ and its channel output observations $\tilde{Y}_k$. In~Section~\ref{sec:m2_rel}, we show formally that the reliability condition in Equation~\eqref{eq:reliabilitycondition_2} is satisfied at each legitimate receiver $k \in [1,K]$.

\subsection{Information Leakage}\label{sec:PCS_dbcnldls_sec}
Besides the observations $\tilde{Z}_m^n$, the eavesdropper $m \in [1,M]$ has access to the common randomness $F = \tilde{U}[ \mathcal{F}^{(n)} ]$. Thus, the information about the messages $\{W_i\}_{i=m}^M$ leaked to this eavesdropper is: 
\begin{align}
I(W_m, \dots, W_M; F, \tilde{Z}_{m}^n) = I \big( \tilde{U}\big[ \cup_{i=m}^M \mathcal{I}_{i}^{(n)} \big] ; \tilde{U}\big[ \mathcal{F}^{(n)} \big], \tilde{Z}_m^n \big) . \label{eq:leakageM2}
\end{align}
In Section~\ref{sec:m2_sec}, we prove that $(W_m, W_{m+1}, \dots, W_M)$ is asymptotically statistically independent of~$(F,\tilde{Z}_m^n)$.

\subsection{Performance of the Polar Coding Scheme}\label{sec:performance2}
The analysis of the polar coding scheme described previously leads to the following theorem.

\begin{Theorem}\label{th:theorem_dbcnldls}
Consider an arbitrary DBC $\big( \mathcal{X}, p_{Y_K\dots Y_1 Z_M \dots Z_1|X}, \mathcal{Y}_K \times \cdots \times \mathcal{Y}_1 \times \mathcal{Z}_M \times \cdots \times \mathcal{Z}_1 \big)$ such that $\mathcal{X} \in \{0,1\}$ and $p_{Y_K\dots Y_1 Z_M \dots Z_1 | X}$ satisfies the Markov chain condition $X-Y_K- \dots - Y_1 - Z_M - \dots - Z_1$. The polar coding scheme described in Sections~\ref{sec:PCS_dbcnldls_pcc}--\ref{sec:PCS_dbcnldls_sec} achieves any rate tuple of the region defined in Corollary~\ref{coro:SCR_2}, satisfying the reliability and strong secrecy conditions given in Equations~\eqref{eq:reliabilitycondition_2}~and~\eqref{eq:secrecycondition_2}, respectively. 
\end{Theorem}

\begin{Corollary}\label{coro:entireregion_dbcnldls}
Since $\tilde{U}[ \mathcal{I}_m^{(n)} ]$ for some $m \in [1,M]$ can contain information to be secured from Eavesdroppers 1--$m$, the polar coding scheme described in Sections~\ref{sec:PCS_dbcnldls_pcc}--\ref{sec:PCS_dbcnldls_sec} can achieve the entire region considering rate sharing of Proposition~\ref{prop:SCR_2} by storing part of any message $W_{m^{\prime}}$ such that $m^{\prime} < m$ into $\tilde{U}[ \mathcal{I}_m^{(n)} ]$ instead of part of $W_m$.
\end{Corollary}

\begin{Corollary}\label{coro:F_dbcnldls}
If we consider a communication scenario requiring transmissions over several blocks of size $n$, the same realization of the source of common randomness $F$ that is known by all parties could be used at each block, and the reliability and the strong secrecy conditions would still be ensured.
\end{Corollary}

{The proof of Theorem~\ref{th:theorem_dbcnldls} follows in four steps with similar reasoning as in \cite{7426806} and is provided in Sections~\ref{sec:m2_ar}--\ref{sec:m2_sec}. The proof of Corollary~\ref{coro:entireregion_dbcnldls} is immediate, and the proof of Corollary~\ref{coro:F_dbcnldls} is provided in Section~\ref{sec:m2_F}.}

\subsubsection{Transmission Rates}\label{sec:m2_ar}
In this step, we prove that the polar coding scheme approaches the corner point of the subregion defined in Corollary~\ref{coro:SCR_2}. For any $m \in [1,M-1]$, the rate $R_m$ corresponding to the message $W_m$ satisfies:
\begin{align*}
\lim_{n \rightarrow \infty} R_{m} = \lim_{n \rightarrow \infty} \frac{1}{n} \big| \mathcal{I}_{m}^{(n)} \big| & \stackrel{(a)}{=}\lim_{n \rightarrow \infty} \frac{1}{n} \Big| \mathcal{H}_{X|Z_{m}}^{(n)} \cap \big( \mathcal{H}_{X|Z_{m+1}}^{(n)} \big)^{\text{C}} \Big| \\
& \stackrel{(b)}{=} \lim_{n \rightarrow \infty} \frac{1}{n} \Big( \big| \mathcal{H}_{X|Z_{m}}^{(n)} \big| - \big| \mathcal{H}_{X|Z_{m+1}}^{(n)} \big| \Big) \\
& \stackrel{(c)}{=} I(X;Z_{m+1}) - I(X;Z_m),
\end{align*}
\textls[-15]{where $(a)$ follows from the definition of the set $\mathcal{I}_{m}^{(n)}$ in Equation~\eqref{eq:setIm}, $(b)$ holds because, by Lemma~\ref{lemma:subsetproperty}, $\mathcal{H}_{X|Z_{m}}^{(n)} \supseteq \mathcal{H}_{X|Z_{m+1}}^{(n)}$, and $(c)$ follows from \cite{arikan2010source} (Theorem 1).~Similarly, according to Equation~\eqref{eq:setIM}, we~obtain:}
\begin{align*}
\lim_{n \rightarrow \infty} R_{M} & = \lim_{n \rightarrow \infty} \frac{1}{n} \big| \mathcal{I}_{M}^{(n)} \big| = \lim_{n \rightarrow \infty} \frac{1}{n} \Big| \mathcal{H}_{X|Z_{M}}^{(n)} \cap \big( \mathcal{H}_{X|Y_1}^{(n)} \big)^{\text{C}} \Big| = I(X;Y_1) - I(X;Z_M).
\end{align*}

\subsubsection{Distribution of the DMS after the Polar Encoding}
Let $\tilde{q}_{U^n}$ be the distribution of $\tilde{U}^n$ after the encoding in Section~\ref{sec:PCS_dbcnldls_encoder}.~The following lemma shows that $\tilde{q}_{U^n}$ and the distribution $p_{U^n}$ in Equation~\eqref{eq:PCS_dbcnldls_jointU} of the original DMS are nearly statistically indistinguishable for sufficiently large $n$ and, consequently, so are the overall distributions $\tilde{q}_{X Y_K \dots Y_1 Z_M \dots Z_1}$ and $p_{X Y_K \dots Y_1 Z_M \dots Z_1}$.

\begin{Lemma}\label{lemma:distDMS_2}
Let $\delta_n = 2^{-n^{\beta}}$ for some $\beta \in (0, \frac{1}{2})$. Then, \vspace{-6pt}
\begin{align*}
\mathbb{V} (\tilde{q}_{U^n}, p_{U^n}) & \leq \delta_{\text{{nld-ls}}}^{(n)}, \\
\mathbb{V} (\tilde{q}_{X^n Y_K^n \dots Y_1^n Z_M^n \dots Z_1^n}, p_{X Y_K \dots Y_1 Z_M \dots Z_1}) = \mathbb{V} (\tilde{q}_{U^n}, p_{U^n}) & \leq \delta_{\text{{nld-ls}}}^{(n)},
\end{align*}
where $\delta_{\text{{nld-ls}}}^{(n)} \triangleq n \sqrt{ 4 \sqrt{n \delta_n \ln 2} ( 2n - \log (2 \sqrt{n \delta_n \ln 2} ) ) + \delta_n} + \sqrt{2 n \delta_n \ln 2}$.
\end{Lemma}
\begin{proof}
See Appendix~\ref{app:distributionDMS}, setting $L = 1$. 
\end{proof}

\begin{Remark}\label{remark:distortion}
The first term of $\delta_{\text{{nld-ls}}}^{(n)}$ bounds the impact on the total variation distance of using the deterministic SC encoding in Equation~\eqref{eq:argm2a} for the entries $\tilde{U} [ \mathcal{L}_{X}^{(n)} ]$, while the second term bounds the impact of storing uniformly-distributed random sequences (messages, local randomness and common randomness) into the entries~$\tilde{U} [ \mathcal{H}_{X}^{(n)} ]$. 
\end{Remark}

As will be seen in the following subsections, an encoding process satisfying Lemma~\ref{lemma:distDMS_2} is crucial for the reliability and the secrecy performance of the polar code.

\subsubsection{Reliability Performance}\label{sec:m2_rel}
Consider the probability of incorrectly decoding all messages $\{ W_{m} \}_{m=1}^M$ at the legitimate receiver $k \in [1,K]$. Let $\tilde{q}_{X^n Y_k^n}$ and $p_{X^n Y_k^n}$ be the marginal distributions of $\tilde{q}_{X^n Y_K^n \dots Y_1^n Z_M^n \dots Z_1^n}$ and $p_{X^n Y_K^n \dots Y_1^n Z_M^n \dots Z_1^n}$, respectively. Consider an optimal coupling \cite{levin2009markov} (Proposition~4.7) between $\tilde{q}_{X^n Y_k^n}$ and $p_{X^n Y_k^n}$ such that: 
\begin{align}
\mathbb{P} \big[ \mathcal{E}_{X^n Y_{k}^n} \big] = \mathbb{V} (\tilde{q}_{X^n Y_k^n}, p_{X^n Y_k^n}), \nonumber
\end{align}
where $\mathcal{E}_{X^n Y_{k}^n} \triangleq \{ (\tilde{X}^n, \tilde{Y}_{k}^n ) \neq (X^n, Y_{k}^n ) \}$ or, equivalently, $\mathcal{E}_{X^n Y_{k}^n} \triangleq \{ (\tilde{U}^n, \tilde{Y}_{k}^n ) \neq (U^n, Y_{k}^n ) \}$ because of the invertibility of $G_n$. Thus, for the legitimate receiver $k \in [1,K]$, we obtain:
\begin{align}
\mathbb{P} \Big[ (\hat{W}_1, \dots \hat{W}_M) \neq (W_1, \dots, W_M) \Big] & \leq \mathbb{P} \big[ \hat{U}^n \neq \tilde{U}^n \big] \nonumber \\
& = \mathbb{P} \big[ \hat{U}^n \neq \tilde{U}^n \big| \mathcal{E}^{\text{C}}_{X^n Y_{k}^n} \big] \mathbb{P} \big[ \mathcal{E}^{\text{C}}_{X^n Y_{k}^n} \big] + \mathbb{P} \big[ \hat{U}^n \neq \tilde{U}^n \big| {\mathcal{E}}_{X^n Y_{k}^n} \big] \mathbb{P} \big[ {\mathcal{E}}_{X^n Y_{k}^n} \big] \nonumber \\
& \leq \mathbb{P} \big[ \hat{U}^n \neq \tilde{U}^n \big| \mathcal{E}^{\text{C}}_{X^n Y_{k}^n} \big] + \mathbb{P} \big[ {\mathcal{E}}_{X^n Y_{k}^n} \big] \nonumber \\
& \stackrel{(a)}{\leq} \sum_{j \in \mathcal{L}_{X|Y_1}^{(n)}} Z \big({U}(j) \big| U^{1:j-1}, {Y}_k^n \big) + \mathbb{P} \big[ {\mathcal{E}}_{X^n Y_{k}^n} \big] \nonumber \\
& \stackrel{(b)}{\leq} n \sqrt{\delta_n} + \mathbb{P} \big[ {\mathcal{E}}_{X^n Y_{k}^n} \big] \nonumber \\
& \stackrel{(c)}{\leq} n \sqrt{\delta_n} + \delta_{\text{{nld-ls}}}^{(n)}, \label{eq:reliability2}
\end{align}
\textls[-15]{where $(a)$ holds by \cite{arikan2010source} (Theorem 2) because $\tilde{U} [ ( \mathcal{L}_{X | Y_1}^{(n)} )^{\text{C}} ]$ is available to all receivers, $(b)$~holds by Lemma~\ref{lemma:subsetproperty}, that is, $Z ({U}(j) | U^{1:j-1}, {Y}_k^n ) \leq Z ({U}(j) | U^{1:j-1}, {Y}_1^n )$ for any $k > 1$, and by the definition of $\mathcal{L}_{X|Y_1}^{(n)}$ in Equation~\eqref{eq:Aset3} and \cite{arikan2010source} (Proposition 2), that is $Z ({U}(j) | U^{1:j-1}, {Y}_1^n ) \leq (H ({U}(j) | U^{1:j-1}, {Y}_1^n) )^{1/2}$,} and $(c)$ holds by the optimal coupling and Lemma~\ref{lemma:distDMS_2} because $\mathbb{V} (\tilde{q}_{X^n Y_k^n}, p_{X^n Y_k^n}) \leq \mathbb{V} (\tilde{q}_{X^n Y_K^n \dots Y_1^n Z_M^n \dots Z_1^n}, p_{X^n Y_K^n \dots Y_1^n Z_M^n \dots Z_1^n})$. Therefore, the polar coding scheme satisfies the reliability condition given in Equation~\eqref{eq:reliabilitycondition_2}.

\subsubsection{Secrecy Performance}\label{sec:m2_sec}
\textls[-15]{Consider the information leakage at the eavesdropper $m \in [1,M]$ given in Equation~\eqref{eq:leakageM2}. We~obtain:}
\begin{align}
I(W_m, \dots, W_M; F, \tilde{Z}_{m}^n) & = H \big(\tilde{U}\big[ \cup_{i=m}^M \mathcal{I}_{i}^{(n)} \big] \big) + H \big(\tilde{U}\big[ \mathcal{F}^{(n)} \big] \big| \tilde{Z}_m^n \big) - H \big(\tilde{U}\big[ \big( \cup_{i=m}^M \mathcal{I}_{i}^{(n)} \big) \cup \mathcal{F}^{(n)} \big] \big| \tilde{Z}_m^n \big) \nonumber \\
& \leq \sum_{i=m}^M \big| \mathcal{I}_{i}^{(n)} \big| + \big| \mathcal{F}^{(n)} \big| - H \big(\tilde{U}\big[ \big( \cup_{i=m}^M \mathcal{I}_{i}^{(n)} \big) \cup \mathcal{F}^{(n)} \big] \big| \tilde{Z}_m^n \big). \label{eq:leakage2}
\end{align}
\textls[-25]{Now, we provide a lower-bound for the conditional entropy term of Equation~\eqref{eq:leakage2}. First, for large enough $n$,}
\begin{align}
& \Big| H \big(\tilde{U}\big[ \big( \cup_{i=m}^M \mathcal{I}_{i}^{(n)} \big) \cup \mathcal{F}^{(n)} \big] \big| \tilde{Z}_m^n \big) - H \big({U}\big[ \big( \cup_{i=m}^M \mathcal{I}_{i}^{(n)} \big) \cup \mathcal{F}^{(n)} \big] \big| {Z}_m^n \big) \Big| \nonumber \\
& \quad \stackrel{(a)}{\leq} \big| H \big( \tilde{Z}_m^n \big) - H \big( {Z}_m^n \big) \big| + \Big| H \big(\tilde{U}\big[ \big( \cup_{i=m}^M \mathcal{I}_{i}^{(n)} \big) \cup \mathcal{F}^{(n)} \big] , \tilde{Z}_m^n \big) - H \big({U}\big[ \big( \cup_{i=m}^M \mathcal{I}_{i}^{(n)} \big) \cup \mathcal{F}^{(n)} \big] , {Z}_m^n \big) \Big| \nonumber \\
& \quad \stackrel{(b)}{\leq} \mathbb{V} (\tilde{q}_{Z_m^n}, p_{Z_m^n}) \log \frac{2^n}{\mathbb{V} (\tilde{q}_{Z_m^n}, p_{Z_m^n})} \nonumber \\
& \qquad + \mathbb{V} (\tilde{q}_{U[( \cup_{i=m}^M \mathcal{I}_{i}^{(n)} ) \cup \mathcal{F}^{(n)} ] Z_m^n}, p_{U[( \cup_{i=m}^M \mathcal{I}_{i}^{(n)} ) \cup \mathcal{F}^{(n)} ] Z_m^n}) \log \frac{2^{( n + | ( \cup_{i=m}^M \mathcal{I}_{i}^{(n)} ) \cup \mathcal{F}^{(n)} | )}}{\mathbb{V} (\tilde{q}_{U[( \cup_{i=m}^M \mathcal{I}_{i}^{(n)} ) \cup \mathcal{F}^{(n)} ] Z_m^n}, p_{U[( \cup_{i=m}^M \mathcal{I}_{i}^{(n)} ) \cup \mathcal{F}^{(n)} ] Z_m^n})} \nonumber \\
& \quad \stackrel{(c)}{\leq} 3 n \delta_{\text{nld-ls}}^{(n)} - 2 \delta_{\text{nld-ls}}^{(n)} \log \delta_{\text{nld-ls}}^{(n)} , \label{eq:entropy2bound}
\end{align}

\noindent where $(a)$ holds by the chain rule of entropy and the triangle inequality, $(b)$ holds by \cite{csiszar2011information} (Lemma~2.9) and $(c)$ holds because the function $x \mapsto x \log x$ is decreasing for $x > 0$ small enough and by Lemma~\ref{lemma:distDMS_2} because $\mathbb{V} (\tilde{q}_{Z_m^n}, p_{Z_m^n}) \leq \mathbb{V} (\tilde{q}_{X^n Y_K^n \dots Y_1^n Z_M^n \dots Z_1^n}, p_{X^n Y_K^n \dots Y_1^n Z_M^n \dots Z_1^n})$, as well as by the invertibility of $G_n$, $\mathbb{V} (\tilde{q}_{U[( \cup_{i=m}^M \mathcal{I}_{i}^{(n)} ) \cup \mathcal{F}^{(n)} ] Z_m^n}, p_{U[( \cup_{i=m}^M \mathcal{I}_{i}^{(n)} ) \cup \mathcal{F}^{(n)} ] Z_m^n}) \leq \mathbb{V} (\tilde{q}_{X^n Y_K^n \dots Y_1^n Z_M^n \dots Z_1^n}, p_{X^n Y_K^n \dots Y_1^n Z_M^n \dots Z_1^n})$. Hence, we~have:
\begin{align}
H \big(\tilde{U}\big[ \big( \cup_{i=m}^M \mathcal{I}_{i}^{(n)} \big) \cup \mathcal{F}^{(n)} \big] \big| \tilde{Z}_m^n \big) & \geq H \big({U}\big[ \big( \cup_{i=m}^M \mathcal{I}_{i}^{(n)} \big) \cup \mathcal{F}^{(n)} \big] \big| {Z}_m^n \big) - ( 3 n \delta_{\text{nld-ls}}^{(n)} - 2 \delta_{\text{nld-ls}}^{(n)} \log \delta_{\text{nld-ls}}^{(n)} ) \nonumber \\
& \stackrel{(a)}{\geq} \sum_{j \in \big( \cup_{i = m}^M \mathcal{I}_{i}^{(n)} \big) \cup \mathcal{F}^{(n)}} \! \! \! \! \! \! \! H \big({U}(j) \big| U^{1:j-1}, {Z}_m^n \big) - ( 3 n \delta_{\text{nld-ls}}^{(n)} - 2 \delta_{\text{nld-ls}}^{(n)} \log \delta_{\text{nld-ls}}^{(n)} ) \nonumber \\
& \stackrel{(b)}{\geq} \left( \sum_{i = m}^M \big| \mathcal{I}_{i}^{(n)} \big| + \big| \mathcal{F}^{(n)} \big| \right) (1 - \delta_n) - ( 3 n \delta_{\text{nld-ls}}^{(n)} - 2 \delta_{\text{nld-ls}}^{(n)} \log \delta_{\text{nld-ls}}^{(n)} ), \label{eq:secrecy2b}
\end{align}
where $(a)$ holds because conditioning does not increase the entropy and $(b)$ holds because, according to Equations~\eqref{eq:setIM}--\eqref{eq:setF1} and Lemma~\ref{lemma:subsetproperty}, $( \cup_{i=m}^M \mathcal{I}_i^{(n)} ) \cup \mathcal{F}^{(n)} \subseteq \mathcal{H}_{X | Z_m}^{(n)}$, as well as by the definition of $\mathcal{H}_{X | Z_m}^{(n)}$ in Equation~\eqref{eq:Aset5}.

Finally, by substituting Equation~\eqref{eq:secrecy2b} into Equation~\eqref{eq:leakage2}, for $n$ sufficiently large, we obtain:
\begin{align}
I (W_m, \dots, W_M; F, \tilde{Z}_{m}^n) \leq n \delta_n + 3 n \delta_{\text{nld-ls}}^{(n)} - 2 \delta_{\text{nld-ls}}^{(n)} \log \delta_{\text{nld-ls}}^{(n)} , \label{eq:secrecy2c}
\end{align}
Hence, the polar code satisfies the strong secrecy condition in Equation~\eqref{eq:secrecycondition_2}, and the proof of Theorem~\ref{th:theorem_dbcnldls} is concluded.

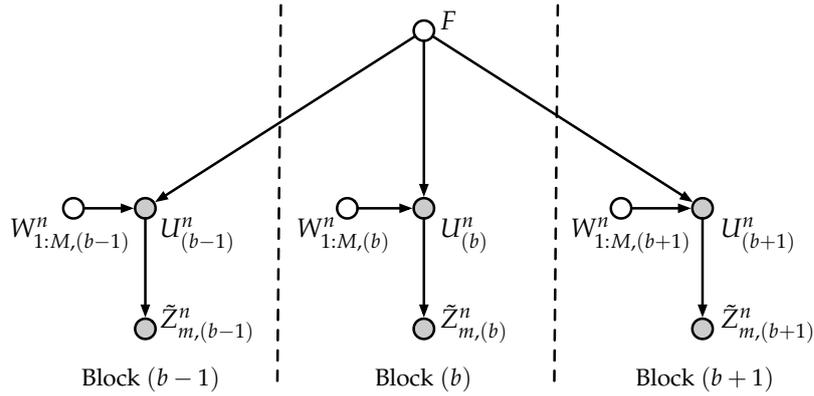
\begin{figure}[H]
\centering
\begin{overpic}[width=0.55\textwidth]{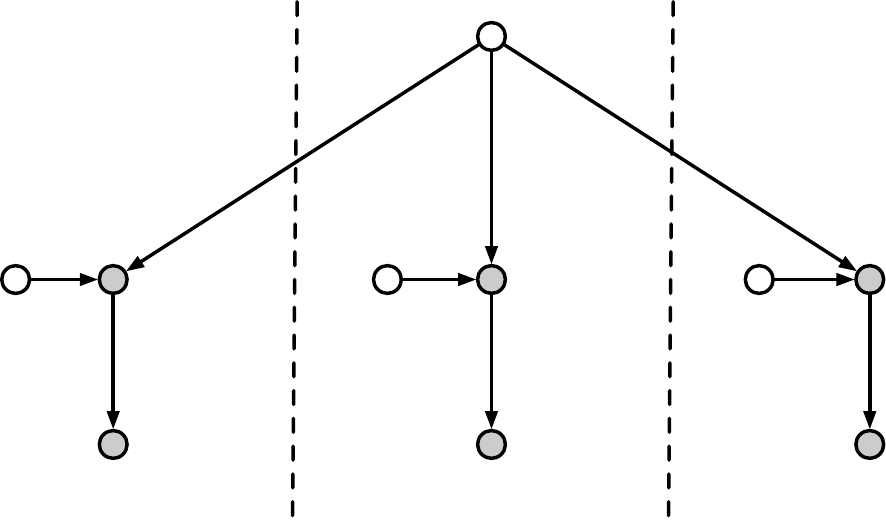}
\put (3,1) {\small Block $(b-1)$}
\put (48,1) {\small Block $(b)$}
\put (88,1) {\small Block $(b+1)$}
\put (58,56) {$F$}
\put (15,24) {$U_{(b-1)}^n$}
\put (58,24) {$U_{(b)}^n$}
\put (101,24) {$U_{(b+1)}^n$}
\put (15,10) {$\tilde{Z}_{m,(b-1)}^n$}
\put (58,10) {$\tilde{Z}_{m,(b)}^n$}
\put (101,10) {$\tilde{Z}_{m,(b+1)}^n$}
\put (-8,24) {$W_{1:M,(b-1)}^n$}
\put (36,24) {$W_{1:M,(b)}^n$}
\put (78,24) {$W_{1:M,(b+1)}^n$}
\end{overpic}
\caption{Bayesian graph plotting the dependencies between the random variables of different blocks that are involved in the secrecy analysis when we consider a transmission over several blocks of size $n$.}\label{fig:depen}
\end{figure}

{
\subsubsection{Reuse of the Source of Common Randomness}\label{sec:m2_F}
Consider that the transmission takes place over $B$ blocks of size $n$. We use the subscript $b \in [1,B]$ between parentheses to denote random variables associated with the block $b$. From Lemma~\ref{lemma:distDMS_2}, we have $\mathbb{V} (\tilde{q}_{U_{(b)}^n}, p_{U^n}) \leq \delta_{\text{{nld-ls}}}^{(n)}$ for any $b \in [1,B]$ because we use the same encoding of Equation~\eqref{eq:distrenc1} at each block. Hence, by the union bound, the polar code satisfies the reliability condition given in Equation~\eqref{eq:reliabilitycondition_2} because:
\begin{align*}
\mathbb{P} \Big[ \cup_{b =1}^B \big\{ \hat{U}_{(b)}^n \neq \tilde{U}_{(b)}^n \big\} \Big] & \leq \sum_{b =1}^B \mathbb{P} \big[ \hat{U}_{(b)}^n \neq \tilde{U}_{(b)}^n \big] \leq B ( n \sqrt{\delta_n} + \delta_{\text{{nld-ls}}}^{(n)} ),
\end{align*}
where the last inequality follows from the fact that, since $F$ and $\Phi_{(b)}$ are perfectly known, \mbox{$\mathbb{P} \big[ \hat{U}_{(b)}^n \neq \tilde{U}_{(b)}^n \big]$} only depends on the decoding at block $b$ and, consequently, can be bounded as in Equation~\eqref{eq:reliability2}.

With a slight abuse of notation, let $W_{m:M,(b_1:b_2)}$, where $1 \leq b_1 \leq b_2 \leq B$, denote the messages $\{ (W_{m,(b)}, \dots, W_{M,(b)}) \}_{b=b_1}^{b_2}$. It remains to show that $W_{m:M,(1:B)}$ is asymptotically statistically independent of $(F, \tilde{Z}_{m,(1:B)}^n)$. Since $F$ is reused at each block, we have to consider the dependencies between the random variables of different blocks that are involved in the secrecy analysis. According to these dependencies, which are represented in the Bayesian graph of Figure~\ref{fig:depen}, we obtain:
\begin{align*}
 I( W_{m:M,(1:B)} ; \tilde{Z}_{m,(1:B)}^n, F) & \stackrel{(a)}{=} I( W_{m:M,(1:B)} ; \tilde{Z}_{m,(1:B)}^n | F) \\
& = \sum_{b=0}^{B-1} I( W_{m:M,(1:B)} ; \tilde{Z}_{m,(b+1)}^n | F, \tilde{Z}_{m,(1:b)}^n) \\
& \stackrel{(b)}{\leq} B \big( n \delta_n + 3 n \delta_{\text{nld-ls}}^{(n)} - 2 \delta_{\text{nld-ls}}^{(n)} \log \delta_{\text{nld-ls}}^{(n)}\big) ,
\end{align*}
where $(a)$ follows from the independence between $W_{m:M,(1:B)}$ and $F$, and $(b)$ holds because:
\begin{align*}
& I( W_{m:M,(1:B)} ; \tilde{Z}_{m,(b+1)}^n | F, \tilde{Z}_{m,(1:b)}^n) \\
& \quad = I( W_{m:M,(1:b+1)} ; \tilde{Z}_{m,(b+1)}^n | F, \tilde{Z}_{m,(1:b)}^n) + I( W_{m:M,(b+2:B)} ; \tilde{Z}_{m,(b+1)}^n | F, \tilde{Z}_{m,(1:b)}^n, W_{m:M,(1:b+1)}) \\
& \quad \leq I( W_{m:M,(1:b+1)}, F, \tilde{Z}_{m,(1:b)}^n ; \tilde{Z}_{m,(b+1)}^n ) + I( W_{m:M,(b+2:B)} ; \tilde{Z}_{m,(1:b+1)}^n , F, W_{m:M,(1:b+1)}) \\
& \quad \stackrel{(a)}{=} I( W_{m:M,(1:b+1)}, F, \tilde{Z}_{m,(1:b)}^n ; \tilde{Z}_{m,(b+1)}^n ) \\
& \quad \leq I( W_{m:M,(b+1)}, F ; \tilde{Z}_{m,(b+1)}^n ) + I( W_{m:M,(1:b)}, \tilde{Z}_{m,(1:b)}^n ; \tilde{Z}_{m,(b+1)}^n | W_{m:M,(b+1)}, F ) \\ 
& \quad \stackrel{(b)}{\leq} \big( n \delta_n + 3 n \delta_{\text{nld-ls}}^{(n)} - 2 \delta_{\text{nld-ls}}^{(n)} \log \delta_{\text{nld-ls}}^{(n)}\big) + I( W_{m:M,(1:b)}, \tilde{Z}_{m,(1:b)}^n ; W_{m:M,(b+1)}, \tilde{Z}_{m,(b+1)}^n | F ) \\
& \quad \stackrel{(c)}{=} n \delta_n + 3 n \delta_{\text{nld-ls}}^{(n)} - 2 \delta_{\text{nld-ls}}^{(n)} \log \delta_{\text{nld-ls}}^{(n)} ,
\end{align*}
where $(a)$ holds because the messages at blocks $b+2$--$B$ are independent of $F$ and all the random variables of the previous blocks, $(b)$ follows from Equation~\eqref{eq:secrecy2c} and $(c)$ holds by applying {d-separation}~\cite{pearl2009causality} over the graph of Figure~\ref{fig:depen} because $(W_{m:M,(1:b)}, \tilde{Z}_{m,(1:b)}^n) \leftarrow F \rightarrow ( W_{m:M,(b+1)}, \tilde{Z}_{m,(b+1)}^n)$ forms a {common cause} and, consequently, $(W_{m:M,(1:b)}, \tilde{Z}_{m,(1:b)}^n)$ and $( W_{m:M,(b+1)}, \tilde{Z}_{m,(b+1)}^n)$ are independent given $F$.
}

\vspace{-12pt}

\section{Polar Coding Scheme for the DBC-LD-NLS}\label{sec:PCS_dbcldnls}
The polar coding scheme provided in this section is designed to achieve the supremum of the achievable rates given in Corollary~\ref{coro:SCR_1} (secrecy-capacity without rate sharing). In this model, there are $K$ input random variables $\{V_{\ell}\}_{\ell=1}^K$ (where $V_K \triangleq X$), each one corresponding to a different superposition layer.~Consider the DMS $\big( \mathcal{V}_1 \times \cdots \times \mathcal{V}_K \times \mathcal{Y}_K \times \cdots \times \mathcal{Y}_1 \times \mathcal{Z}_M \times \cdots \times \mathcal{Z}_1, p_{V_1 \dots V_K Y_K\dots Y_1 Z_M \dots Z_1} \big)$ that represents the input and output random variables involved in the achievable subregion of Corollary~\ref{coro:SCR_1}, where $\mathcal{V}_{\ell} = \{0,1\}$ for any $\ell \in [1,K]$. Let $(V_1^n, \dots, V_K^n,Y_K^n,\dots,Y_1^n,Z_M^n,\dots,Z_1^n)$ be an i.i.d. $n$-sequence of this source. Then, we define the $K$ polar transforms $U^n_{\ell} \triangleq V_{\ell}^n G_n$, where $\ell \in [1,K]$. \mbox{Since $V_1 - V_2 - \cdots - V_K$} and, consequently, $U_1 - U_2 - \cdots - U_K$ (by the invertibility of $G_n$) form a~Markov chain, the~joint distribution of $(U_1^n, \dots, U_K^n)$ satisfies''
\begin{align}
 p_{U_1^n \dots U_K^n} (u_1^n,\dots,u_K^n) \triangleq \prod_{\ell =1}^K \prod_{j=1}^n p_{U_{\ell}(j) | U_{\ell}^{1:j-1} V_{{\ell}-1}^n} \big( u_{\ell}(j) \big| u_{\ell}^{1:j-1}, u_{{\ell}-1}^n G_n \big). \label{eq:PCS_dbcldnls_jointU}
\end{align}

\subsection{Polar Code Construction}\label{sec:PCS_dbcldnls_pcc}
\begin{figure}[H]
\centering
\begin{tikzpicture}
\node at (-4.7,-0.6) [shape=rectangle,thick,inner sep=0pt,minimum width=3.1cm, minimum height=0.6cm, draw,fill={rgb:red,20;green,2;blue,2},fill opacity=0.24] {};
\node at (-1.425,-0.6) [shape=rectangle,thick,inner sep=0pt,minimum width=3.45cm, minimum height=0.6cm, draw,fill={rgb:red,2;green,20;blue,4},fill opacity=0.24] {};
\node at (1.9,-0.6) [shape=rectangle,thick,inner sep=0pt,minimum width=3.2cm, minimum height=0.6cm, draw,fill={rgb:red,5;green,10;blue,20},fill opacity=0.14] {};
\node at (4.875,-0.6) [shape=rectangle,thick,inner sep=0pt,minimum width=2.75cm, minimum height=0.6cm, draw,fill={rgb:red,2;green,2;blue,2},fill opacity=0.24] {};
\node at (0,-0.6) [shape=rectangle,thick,inner sep=0pt,minimum width=12.5cm, minimum height=0.6cm, draw] {};
 
\fill[pattern=north west lines] (-3.15,-0.3) rectangle (-2.9,-0.9);
\fill[pattern=north west lines] (3.25,-0.3) rectangle (3.5,-0.9);
\fill[pattern=north west lines] (3.5,-0.3) rectangle (3.75,-0.9);

\draw[thick] (0.3cm,-0.3cm) -- (0.3cm,-0.9cm);
\draw[thick] (3.5cm,-0.3cm) -- (3.5cm,-0.9cm) ;
\draw[thick,densely dotted] (-2.9cm,-0.3cm) -- (-2.9cm,-0.9cm);
\draw[thick,densely dotted] (3.75cm,-0.3cm) -- (3.75cm,-0.9cm);
\draw[thick,densely dotted] (3.25cm,-0.3cm) -- (3.25cm,-0.9cm);
\draw[|-|] (-6.25cm,0.75cm) -- node[fill=white,inner sep=1mm,midway] {\small $\mathcal{H}_{V_{\ell}|V_{{\ell}-1}Z_M}^{(n)}$} (0.25cm,0.75cm);
\draw[|-|] (0.35cm,0.75cm) -- node[fill=white,inner sep=1mm,midway] {\small $\big(\mathcal{H}_{V_{\ell}|V_{{\ell}-1}Z_M}^{(n)}\big)^{\text{C}}$} (6.25cm,0.75cm);
\draw[|-|] (-6.25cm,0.1cm) -- node[fill=white,inner sep=1mm,midway] {\small $\mathcal{H}_{V_{\ell}|V_{{\ell}-1}Y_{\ell}}^{(n)}$} (-3.2cm,0.1cm);
\draw[|-|] (-3.1cm,0.1cm) -- node[fill=white,inner sep=1mm,midway] {\small $\big(\mathcal{H}_{V_{\ell}|V_{{\ell}-1}Y_{\ell}}^{(n)}\big)^{\text{C}}$} (6.25cm,0.1cm);
\draw[|-|] (-6.25cm,-1.35cm) -- node[fill=white,inner sep=1mm,midway] {\small $\mathcal{H}_{V_{\ell}|V_{{\ell}-1}}^{(n)}$} (3.45cm,-1.35cm);
\draw[|-|] (3.55cm,-1.35cm) -- node[fill=white,inner sep=1mm,midway] {\small $\big(\mathcal{H}_{V_{\ell}|V_{{\ell}-1}}^{(n)}\big)^{\text{C}}$} (6.25cm,-1.35cm);
\node[text width=1cm, anchor=west] at (-5cm,-0.6) {\small $\mathcal{F}_{\ell}^{(n)}$};
\node[text width=1cm, anchor=west] at (-1.6cm,-0.6) {\small $\mathcal{I}_{\ell}^{(n)}$};
\node[text width=1cm, anchor=west] at (1.6cm,-0.6) {\small $\mathcal{C}_{\ell}^{(n)}$};
\node[text width=1cm, anchor=west] at (4.4cm,-0.6) {\small $\mathcal{T}_{\ell}^{(n)}$};
\end{tikzpicture}
\caption{Polar code construction for the DBC-LD-NLS at the ${\ell}$-th layer. The hatched area represents those indices $j \in ( \mathcal{H}_{V_{\ell}|V_{{\ell}-1}Y_{\ell}}^{(n)} )^{\text{C}} \cap ( \mathcal{L}_{V_{\ell}|V_{{\ell}-1}Y_{\ell}}^{(n)} )^{\text{C}}$, which can belong to the sets $\mathcal{I}_{\ell}^{(n)}$, $\mathcal{C}_{\ell}^{(n)}$ or $\mathcal{T}_{\ell}^{(n)}$.}\label{fig:pc_dbcldnls}
\end{figure}
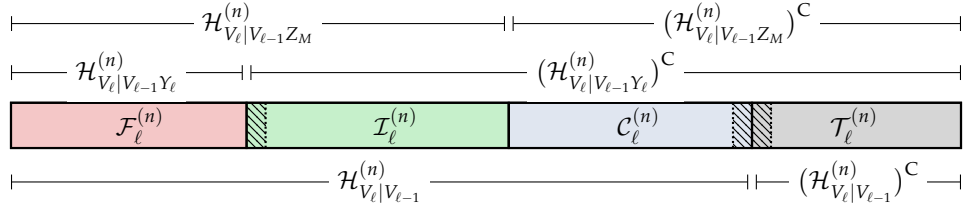

Based on $p_{V_1 \dots V_KY_K \dots Y_1 Z_M \dots Z_1}$, the construction is carried out similarly at each superposition layer. Consider the polar construction at layer $\ell \in [1,K]$.
Let $\delta_n \triangleq 2^{-n^{\beta}}$, where $\beta \in (0, \frac{1}{2})$.~For the polar transform $U_{{\ell}}^n = V_{{\ell}}^n G_n$ associated with the ${\ell}$-th layer, we define the sets:
\begin{align}
\mathcal{H}_{V_{\ell}|V_{{\ell}-1}}^{(n)} & \triangleq \big\{ j \in [n]: H \big( U_{\ell}(j) \big| U_{\ell}^{1:j-1}, V_{\ell-1}^n \big) \geq 1-\delta_n \big\}, \label{eq:pol1b1} \\
\mathcal{L}_{V_{\ell}|V_{{\ell}-1}}^{(n)} & \triangleq \big\{ j \in [n]: H \big( U_{\ell}(j) \big| U_{\ell}^{1:j-1}, V_{\ell-1}^n \big) \leq \delta_n \big\}, \label{eq:pol1b2} \\
\mathcal{L}_{V_{\ell} | V_{{\ell}-1}Y_{k}}^{(n)} & \triangleq \big\{ j \in [n]: H \big( U_{\ell}(j) \big| U_{\ell}^{1:j-1}, V_{\ell-1}^n, Y_{k}^n \big) \leq \delta_n \big\}, \quad k = \ell ,\dots, K, \label{eq:pol1a} \\
\mathcal{H}_{V_{\ell} | V_{{\ell}-1}Y_{k}}^{(n)} & \triangleq \big\{ j \in [n]: H \big( U_{\ell}(j) \big| U_{\ell}^{1:j-1}, V_{\ell-1}^n, Y_{k}^n \big) \geq 1- \delta_n \big\}, \quad k = \ell ,\dots, K, \label{eq:pol1a2} \\
\mathcal{H}_{V_{\ell}|V_{{\ell}-1}Z_{m}}^{(n)} & \triangleq \big\{ j \in [n]: H \big( U_{\ell}(j) \big| U_{\ell}^{1:j-1}, V_{\ell-1}^n, Z_m^n \big) \geq 1-\delta_n \big\}, \quad m = 1,\dots, M, \label{eq:pol1c}
\end{align}
where we recall that $V_0 = \varnothing$ when $\ell = 1$ and $V_K \triangleq X$ when $\ell = K$. At each layer $\ell \in [1,K]$, based on these previous sets, we define the following partition of the universal set $[n]$,
\begin{align}
\label{eq:e1_setI} \mathcal{I}_{\ell}^{(n)} & \triangleq \mathcal{H}_{V_{\ell}|V_{{\ell}-1}Z_{M}}^{(n)} \cap \big( \mathcal{H}_{V_{\ell}|V_{{\ell}-1}Y_{\ell}}^{(n)} \big)^{\text{C}}, \\
\label{eq:e1_setF} \mathcal{F}_{\ell}^{(n)} & \triangleq \mathcal{H}_{V_{\ell}|V_{{\ell}-1}Y_{\ell}}^{(n)}, \\
\label{eq:e1_setC} \mathcal{C}_{\ell}^{(n)} & \triangleq \mathcal{H}_{V_{\ell} | V_{{\ell}-1}}^{(n)} \cap \big( \mathcal{H}_{V_{\ell} | V_{{\ell}-1}Z_M}^{(n)} \big)^{\text{C}} , \\
\label{eq:e1_setT} \mathcal{T}_{\ell}^{(n)} & \triangleq \big( \mathcal{H}_{V_{\ell} | V_{{\ell}-1}}^{(n)} \big)^{\text{C}} ,
\end{align}
which is graphically represented in Figure~\ref{fig:pc_dbcldnls}.~The way we define this partition at the $\ell$-th layer follows similar reasoning as the one to define the partition in Section~\ref{sec:PCS_dbcnldls_pcc} for the DBC-NLD-LS. In this sense, $U_{\ell} [ \mathcal{H}_{V_{\ell}|V_{\ell-1}}^{(n)} ]$ will be suitable for storing uniformly-distributed random sequences. Otherwise, $U_{\ell} [ \mathcal{T}_{\ell}^{(n)} ]$ will not and $U_{\ell}(j)$ such that $j \in \mathcal{T}_{\ell}^{(n)}$ will be constructed somehow from $({U}_{\ell}^{1:j-1},V_{\ell-1})$ and the distribution $p_{U_{\ell}(j) | U_{\ell}^{1:j-1} V_{{\ell}-1}^n}$. Now, $U_{\ell} [ \mathcal{I}_{\ell}^{(n)} ]$ will be suitable for storing information to be secured from all eavesdroppers because $\mathcal{I}_{\ell}^{(n)}$ belongs to $\mathcal{H}_{V_{\ell}|V_{{\ell}-1}Z_{M}}^{(n)}$, and by Lemma~\ref{lemma:subsetproperty}, $\mathcal{H}_{V_{\ell}|V_{{\ell}-1}Z_{M}}^{(n)} \subseteq \mathcal{H}_{V_{\ell}|V_{{\ell}-1}Z_{m^{\prime}}}^{(n)}$ for any $m^{\prime} \in [1,M-1]$. Since $\mathcal{C}_{\ell}^{(n)} \subseteq ( \mathcal{H}_{V_{\ell}|V_{{\ell}-1}Z_{M}}^{(n)} )^{\text{C}}$, $U [ \mathcal{C}_{\ell}^{(n)} ]$ will be used to store the local randomness required to confuse all eavesdroppers about the secret information carried on this layer. According to~\cite{arikan2010source} (Theorem 2), the legitimate receiver $k \in [1,K]$ will be able to reliably infer $U_{\ell}[ \mathcal{L}_{V_{\ell}|V_{\ell-1} Y_{k}}^{(n)} ]$ given $Y_{k}^n$ and $U_{\ell}[ ( \mathcal{L}_{V_{\ell}|V_{\ell-1}Y_k}^{(n)} )^{\text{C}} ]$. By Lemma~\ref{lemma:subsetproperty}, we have $( \mathcal{L}_{V_{\ell}|V_{\ell-1}Y_{\ell}}^{(n)} )^{\text{C}} \supseteq ( \mathcal{L}_{V_{\ell}|V_{\ell-1}Y_k}^{(n)} \big)^{\text{C}}$ for any $\ell < k$. Therefore, given $U_{\ell} [ ( \mathcal{L}_{V_{\ell}|V_{\ell-1}Y_{\ell}}^{(n)} )^{\text{C}} ]$, the legitimate receivers $\ell$--$K$ will be able to reliably reconstruct $U_{\ell}^n$ from its own channel observations. In this sense, $U_{\ell} [ \mathcal{F}_{\ell}^{(n)} ]$ will be used to store the random sequence provided by the source of common randomness. Since $\mathcal{F}_{\ell}^{(n)} \subseteq \mathcal{H}_{V_{\ell}|V_{{\ell}-1}Z_{M}}^{(n)}$, the strong secrecy condition will not be compromised. On the other hand, $U[ ( \mathcal{H}_{V_{\ell}|V_{\ell-1}Y_{\ell} }^{(n)} )^{\text{C}} \cap ( \mathcal{L}_{V_{\ell}|V_{\ell-1}Y_{\ell}}^{(n)} )^{\text{C}}]$ (hatched areas in Figure~\ref{fig:pc_dbcldnls}) will contain secret information or elements that cannot be known directly by the eavesdroppers. Therefore, the transmitter somehow will make those elements available to the legitimate receivers $\ell$--$K$ keeping them masked from all eavesdroppers by incurring an asymptotically-negligible rate penalty. 

As mentioned in Remark~\ref{remark:construction}, the goal of the polar construction is to obtain the entropy terms associated with the sets in Equations~\eqref{eq:pol1b1}--\eqref{eq:pol1c} and then define the partition of $[n]$ given in Equations~\eqref{eq:e1_setI}--\eqref{eq:e1_setT}.

\subsection{Polar Encoding}\label{sec:PCS_dbcldnls_encoder}
The superposition-based polar encoder will consist of $K$ encoding blocks operating sequentially at each superposition layer, the block at layer $\ell \in [1,K]$ being responsible for the construction of $\tilde{U}_{\ell}^n$. In~order to construct $\tilde{U}_{\ell}^n$ for some $\ell \in [2,K]$, the encoder block needs $\tilde{V}_{\ell-1}^n = \tilde{U}_{\ell-1}^n G_n$, which have been constructed previously by the encoding block operating at the $(\ell-1)$-th layer.

Consider the encoding procedure at layer $\ell \in [1,K]$. Let $W_{\ell}$ and $C_{\ell}$ be uniformly-distributed random vectors of size $|\mathcal{I}_{{\ell}}^{(n)}|$ and $|\mathcal{C}_{{\ell}}^{(n)}|$, respectively, where $W_{\ell}$ represents the message intended for receivers ${\ell}$--$K$ and $C_{\ell}$ the local randomness required at the ${\ell}$-th layer to confuse all eavesdroppers about this message. Let $F_{\ell}$ be a given uniformly-distributed random $|\mathcal{F}_{{\ell}}^{(n)}|$-sequence, which represents the source of common randomness that is available to all parties. The $\ell$-th encoding block constructs the sequence $\tilde{u}_{\ell}^n$ as follows. Given the realizations $w_{\ell}$, $c_{\ell}$ and $f_{\ell}$, whose elements have been indexed by the set of indices $\mathcal{I}_{{\ell}}^{(n)}$, $\mathcal{C}_{{\ell}}^{(n)}$ and $\mathcal{F}_{{\ell}}^{(n)}$, respectively, and given $\tilde{v}^n_{{\ell}-1} = \tilde{u}_{{\ell}-1}^n G_n$ provided by the previous encoding block (recall that $\tilde{v}^n_{0} \triangleq \varnothing$ at the first layer), the $\ell$-th encoding block draws $\tilde{u}_{\ell}^n$ from:
\begin{align}
& \tilde{q}_{U_{\ell}(j) | U_{\ell}^{1:j-1}V_{{\ell}-1}^n} \! \big( \tilde{u}_{\ell}(j) | \tilde{u}_{\ell}^{1:j-1}, \tilde{v}_{{\ell}-1}^n \big) \! \nonumber \\
& \triangleq \! \left\{ 
\begin{array}{ll}
\mathds{1} \big\{ \tilde{u}_{\ell}(j) = w_{\ell}(j) \big\} & \text{if } j \in \mathcal{I}_{\ell}^{(n)}, \\
\mathds{1} \big\{ \tilde{u}_{\ell}(j) = c_{\ell}(j) \big\} & \text{if } j \in \mathcal{C}_{\ell}^{(n)}, \\
\mathds{1} \big\{ \tilde{u}_{\ell}(j) = f_{\ell}(j) \big\} & \text{if } j \in \mathcal{F}_{\ell}^{(n)}, \\
\! p_{U_{\ell}(j) | U_{\ell}^{1:j-1}V_{{\ell}-1}^n} \! \big( \tilde{u}_{\ell}(j) | \tilde{u}_{\ell}^{1:j-1}, \tilde{v}_{{\ell}-1}^n \big)& \text{if } j \in \big( \mathcal{H}_{V_{\ell}|V_{\ell-1}}^{(n)} \big)^{\text{C}} \cap \big( \mathcal{L}_{V_{\ell}|V_{\ell-1}}^{(n)} \big)^{\text{C}}, \\
\! \mathds{1} \big\{ \tilde{u}_{\ell}(j) = \xi_{\ell}^{(j)}(\tilde{u}_{\ell}^{1:j-1}, \tilde{v}_{{\ell}-1}^n) \big\} & \text{if } j \in \mathcal{L}_{V_{\ell} | V_{{\ell}-1}}^{(n)}, 
\end{array}
\right. \label{eq:distrenc2}
\end{align}
where:
\begin{align}
\xi_{\ell}^{(j)} \big( \tilde{u}_{\ell}^{1:j-1}, \tilde{v}_{{\ell}-1}^n \big) \triangleq \argmax_{u \in \mathcal{V}_{\ell}} p_{U_{\ell}(j) | U_{\ell}^{1:j-1} V_{{\ell}-1}^n} \big( u \big| \tilde{u}_{\ell}^{1:j-1}, \tilde{v}_{{\ell}-1}^n \big), \label{eq:argmax1}
\end{align}
$p_{U_{\ell}(j) | U_{\ell}^{1:j-1} V_{\ell-1}^n}$ being the distribution induced by the original DMS. Notice that $\mathcal{T}_{\ell}^{(n)} = ( ( \mathcal{H}_{V_{\ell}|V_{\ell-1}}^{(n)} )^{\text{C}} \cap ( \mathcal{L}_{V_{\ell}|V_{\ell-1}}^{(n)} )^{\text{C}} ) \cup \mathcal{L}_{V_{\ell}|V_{\ell-1}}^{(n)}$, and similarly to the previous model, $\tilde{U} [ \mathcal{L}_{V_{\ell}|V_{\ell-1}}^{(n)} ]$ is constructed in a~deterministic way by adapting the SC encoding algorithm in \cite{7447169}; and $\tilde{U} [ ( \mathcal{H}_{V_{\ell}|V_{\ell-1}}^{(n)} )^{\text{C}} \cap ( \mathcal{L}_{V_{\ell}|V_{\ell-1}}^{(n)} )^{\text{C}} ]$ is constructed randomly. {By \cite{arikan2010source} (Theorem 1), the rate of the amount of randomness for SC encoding will be asymptotically negligible}. After constructing $\tilde{U}_{\ell}^n$, the ${\ell}$-th encoding block computes the sequence $\tilde{V}_{\ell}^n = \tilde{U}_{\ell}^n G_n$ and delivers it to the next encoding block.~If ${\ell} = K$, then $\tilde{V}_K^n \triangleq \tilde{X}^n$, and the encoder transmits it over the DBC, which induces the channel outputs $(\tilde{Y}_K^n, \dots, \tilde{Y}_1^n, \tilde{Z}_M^n, \dots, \tilde{Z}_1^n)$.

Finally, besides the sequence $\tilde{X}^n$, the encoder outputs the following additional secret sequences,
\begin{align}
\label{eq:secretmessage1}
\Phi_{\ell} \triangleq \tilde{U}_{\ell} \Big[ \big( \mathcal{H}_{V_{\ell}|V_{\ell-1}Y_{\ell}}^{(n)} \big)^{\text{C}} \cap \big( \mathcal{L}_{V_{\ell}|V_{\ell-1}Y_{\ell}}^{(n)} \big)^{\text{C}} \Big], \quad \ell = 1, \dots, K, 
\end{align}
{The sequence $\Phi_{\ell}$ corresponding to the layer $\ell \in [1,K]$ must be additionally transmitted to the legitimate receivers $\ell$--$K$ keeping it masked from the eavesdroppers. To do so, the transmitter can perform a~modulo-two addition between $\{\Phi_{\ell}\}_{\ell=1}^K$ and a uniformly-distributed secret key privately shared with the legitimate receivers and somehow additionally send it to them. If $K \ll n$, by \cite{arikan2010source} (Theorem 1), we have that the overall rate required to transmit these additional secret sequences is asymptotically negligible, i.e., $\lim_{n \rightarrow \infty} \sum_{\ell=1}^K \frac{ | \Phi_{\ell} |}{n} = 0$. As for the previous model, the uniformly-distributed part of any $\Phi_{\ell}$ could be made available to the corresponding legitimate receivers by using a chaining structure as in \cite{6620400}. However, this approach will present the same disadvantages as those mentioned in Remark~\ref{remarkA}.}

\subsection{Polar Decoding}\label{sec:PCS_dbcldnls_decoder}
Consider that the realizations of $\{F_{\ell}\}_{{\ell}=1}^K$ are available to all parties, and the sequences $\{\Phi_{\ell}\}_{{\ell}=1}^K$ have been successfully received by the corresponding legitimate receivers before the decoding process. 

Consider the decoding at the legitimate receiver $k \in [1,K]$. This receiver forms the estimates $\{ \hat{U}_{\ell}^n \}_{\ell=1}^k$ of the sequences $\{ \tilde{U}_{\ell}^n \}_{\ell=1}^k$ in a successive manner from $\hat{U}_{1}^n$-$ \hat{U}_{k}^n$, and the procedure to estimate $\tilde{U}_{\ell}^n$ for some $\ell \in [1,k]$ is as follows. First, given that $\Phi_{\ell}$ and $F_{\ell}$ are available, the receiver knows $\tilde{U}_{\ell}[( \mathcal{L}_{V_{\ell}|V_{{\ell}-1}Y_{\ell}}^{(n)} )^{\text{C}} ]$. Moreover, by Lemma~\ref{lemma:subsetproperty}, $( \mathcal{L}_{V_{\ell}|V_{{\ell}-1}Y_k}^{(n)}\big)^{\text{C}} \subseteq \big( \mathcal{L}_{V_{\ell}|V_{{\ell}-1}Y_{\ell}}^{(n)}\big)^{\text{C}}$ for any $\ell < k$. Thus, given $\tilde{U}_{\ell}[( \mathcal{L}_{V_{\ell}|V_{{\ell}-1}Y_{\ell}}^{(n)} )^{\text{C}} ]$, the $k$-th legitimate receiver performs SC decoding for source coding with side information \cite{arikan2010source} to construct $\hat{U}_{\ell} [ \mathcal{L}_{V_{\ell}|V_{{\ell}-1}Y_{\ell}}^{(n)} ]$ from $\tilde{Y}_k^n$, and from $\hat{V}_{{\ell}-1}^n = \hat{U}_{{\ell}-1}^n G_n$ estimated previously. In Section~\ref{sec:m1_rel}, we show formally that the polar coding scheme satisfies the reliability condition in Equation~\eqref{eq:reliabilitycondition_1}.

\subsection{Information Leakage}\label{sec:PCS_dbcldnls_sec}
Besides the observations $\tilde{Z}_m^n$, the eavesdropper $m \in [1,M]$ has access to the common randomness $\{ F_{\ell} \}_{{\ell}=1}^K$. Therefore, the information about all messages leaked to the $m$-th eavesdropper is: 
\begin{align}
I(W_1, \dots, W_K; F_1, \dots , F_K, \tilde{Z}_{m}^n) = I \big( \tilde{U}_1\big[ \mathcal{I}_{1}^{(n)} \big], \dots, \tilde{U}_K\big[ \mathcal{I}_{K}^{(n)} \big] ; \tilde{U}_1\big[ \mathcal{F}_{1}^{(n)} \big], \dots, \tilde{U}_K\big[ \mathcal{F}_{K}^{(n)} \big], \tilde{Z}_m^n \big). \label{eq:leakage1}
\end{align}
In Section~\ref{sec:m1_sec}, we prove that $(W_1, \dots, W_K)$ is asymptotically statistically independent of $(F_1,\dots,F_K, \tilde{Z}_m^n)$.

\subsection{Performance of the Polar Coding Scheme}\label{sec:performance1}
The analysis of the polar coding scheme leads to the following theorem. 

\begin{Theorem}\label{th:theorem_dbcldnls}
Consider an arbitrary DBC $\big( \mathcal{X}, p_{Y_K\dots Y_1 Z_M \dots Z_1|X}, \mathcal{Y}_K \times \cdots \times \mathcal{Y}_1 \times \mathcal{Z}_M \times \cdots \times \mathcal{Z}_1 \big)$ such that $\mathcal{X} \in \{0,1\}$ and $p_{Y_K\dots Y_1 Z_M \dots Z_1|X}$ satisfies the Markov chain condition $X-Y_K- \dots - Y_1 - Z_M - \dots - Z_1$. The polar coding scheme described in Sections~\ref{sec:PCS_dbcldnls_pcc}--\ref{sec:PCS_dbcldnls_sec} achieves any rate tuple of the achievable region defined in Corollary~\ref{coro:SCR_1}, satisfying the reliability and strong secrecy conditions in Equations~\eqref{eq:reliabilitycondition_1}~and~\eqref{eq:secrecycondition_1}, respectively.
\end{Theorem}

\begin{Corollary}\label{coro:entireregion_dbcldnls}
Since $\tilde{U}_{\ell}[ \mathcal{I}_{\ell}^{(n)} ]$ for some $\ell \in [1,K]$ can contain any information to be reliably decoded by the legitimate receivers $\ell$--$K$, the coding scheme in Sections~\ref{sec:PCS_dbcldnls_pcc}--\ref{sec:PCS_dbcldnls_sec} can achieve the entire region considering the rate sharing of Proposition~\ref{prop:SCR_1} by storing part of any message $W_{\ell^{\prime}}$ such that $\ell^{\prime} > \ell$ into $\tilde{U}_{\ell}[ \mathcal{I}_{\ell}^{(n)} ]$ instead of part of $W_{\ell}$.
\end{Corollary}

\begin{Corollary}\label{coro:F_dbcldnls}
If we consider a communication scenario requiring transmissions over several blocks of size $n$, the same realization of the source of common randomness $(F_1, \dots, F_K)$ that is known by all parties could be used at each block, and the reliability and the strong secrecy conditions would still be ensured.
\end{Corollary}

{As in Theorem~\ref{th:theorem_dbcnldls}, the proof of Theorem~\ref{th:theorem_dbcldnls} follows in four steps and is provided in Sections~\ref{sec:m2_ar}--\ref{sec:m2_sec}. The proof of Corollary~\ref{coro:entireregion_dbcldnls} is immediate. The proof of Corollary~\ref{coro:F_dbcldnls} is omitted because it follows similar reasoning as in Corollary~\ref{coro:F_dbcnldls}. Despite that in this model, we have different superposition layers, the dependencies between the random variables at different blocks have the same structure of those graphically represented in Figure~\ref{fig:depen}.}

\subsubsection{Transmission Rates}\label{sec:m1_ar}
We prove that the polar coding scheme approaches the corner point of the subregion defined in Corollary~\ref{coro:SCR_1}. For any $\ell \in [1,K]$, the transmission rate $R_{\ell}$ corresponding to the message $W_{\ell}$ satisfies:
\begin{align}
\lim_{n \rightarrow \infty} R_{{\ell}} & = \lim_{n \rightarrow \infty} \frac{1}{n} \big| \mathcal{I}_{{\ell}}^{(n)} \big| \stackrel{(a)}{=} \lim_{n \rightarrow \infty} \frac{1}{n} \Big| \mathcal{H}_{V_{\ell}|V_{{\ell}-1}Z_{M}}^{(n)} \cap \big( \mathcal{H}_{V_{\ell}|V_{{\ell}-1}Y_{\ell}}^{(n)} \big)^{\text{C}} \Big| \nonumber \\
& \stackrel{(b)}{=} \lim_{n \rightarrow \infty} \frac{1}{n} \big| \mathcal{H}_{V_{\ell}|V_{{\ell}-1}Z_M}^{(n)} \big| - \lim_{n \rightarrow \infty} \frac{1}{n} \big| \mathcal{H}_{V_{\ell}|V_{{\ell}-1}Y_{{\ell}}}^{(n)} \big| \nonumber \\
& \stackrel{(c)}{=} I(V_{\ell}; Y_{\ell} | V_{{\ell}-1}) - I(V_{\ell}; Z_M | V_{{\ell}-1}), \label{eq:messagerates_1}
\end{align}
where $(a)$ follows from the definition of the set $\mathcal{I}_{{\ell}}^{(n)}$ in Equation~\eqref{eq:e1_setI}, $(b)$ holds because, by Lemma~\ref{lemma:subsetproperty}, $\mathcal{H}_{V_{\ell}|V_{{\ell}-1}Z_{M}}^{(n)} \supseteq\mathcal{H}_{V_{\ell}|V_{{\ell}-1}Y_{\ell}}^{(n)}$, and $(c)$ holds by \cite{arikan2010source} (Theorem 1).

\subsubsection{Distribution of the DMS after the Polar Encoding}
Let $\tilde{q}_{U_1^n \dots U_K^n}$ be the distribution of $(\tilde{U}_1^n, \dots, \tilde{U}_K^n)$ after the encoding in Section~\ref{sec:PCS_dbcldnls_encoder}. The following lemma shows that $\tilde{q}_{U_1^n \dots U_K^n}$ and $p_{U_1^n \dots U_K^n}$ of the DMS are nearly statistically indistinguishable for sufficiently large $n$ and, consequently, so are the overall distributions $\tilde{q}_{V_1^n \dots V_K^n Y_K^n \dots Y_1^n Z_M^n \dots Z_1^n}$ and $p_{V_1^n \dots V_K^n Y_K^n \dots Y_1^n Z_M^n \dots Z_1^n}$.

\begin{Lemma}\label{lemma:distDMS_1}
Let $\delta_n = 2^{-n^{\beta}}$ for some $\beta \in (0, \frac{1}{2})$. Then, \vspace{-6pt}
\begin{align*}
\mathbb{V} (\tilde{q}_{U_1^n \dots U_K^n}, p_{U_1^n \dots U_K^n}) & \leq \delta_{\text{{ld-nls}}}^{(n)}, \\
\mathbb{V} (\tilde{q}_{V_1^n \dots V_K^n Y_K^n \dots Y_1^n Z_M^n \dots Z_1^n}, p_{V_1^n \dots V_K^n Y_K^n \dots Y_1^n Z_M^n \dots Z_1^n}) = \mathbb{V} (\tilde{q}_{U_1^n \dots U_K^n}, p_{U_1^n \dots U_K^n}) & \leq \delta_{\text{{ld-nls}}}^{(n)},
\end{align*}
where $\delta_{\text{{ld-nls}}}^{(n)}\triangleq K n \sqrt{ 4 \sqrt{n \delta_n \ln 2} \big( 2n - \log \big(2 \sqrt{n \delta_n \ln 2} \big) \big) + \delta_n} + \sqrt{K 2 n \delta_n \ln 2}$.
\end{Lemma}
\begin{proof}
See Appendix~\ref{app:distributionDMS} setting $L = K$. 
\end{proof}

\begin{Remark}\label{remark:TV1}
The first term of $\delta_{\text{{ld-nls}}}^{(n)}$ bounds the impact on the total variation distance of using the deterministic SC encoding in Equation~\eqref{eq:argmax1} for $\tilde{U}_{\ell} \big[ \mathcal{L}_{V_{\ell}|V_{\ell-1}}^{(n)} \big]$ at each layer $\ell \in [1,K]$. The second term bounds the impact of storing uniformly-distributed random sequences that are independent of $\tilde{V}_{\ell-1}^n$ into $\tilde{U}_{\ell} \big[ \mathcal{H}_{V_{\ell}|V_{\ell-1}}^{(n)} \big]$. 
\end{Remark}

\subsubsection{Reliability Performance}\label{sec:m1_rel}
Consider the probability of incorrectly decoding $\{ W_{\ell} \}_{\ell=1}^k$ at the legitimate receiver $k \in [1,K]$. Let~$\tilde{q}_{V_{\ell}^nY_k^n}$ and $p_{V_{\ell}^n Y_k^n}$ for any $\ell \leq k$ be marginals of $\tilde{q}_{V_1^n \dots V_K^n Y_K^n \dots Y_1^n Z_M^n \dots Z_1^n}$ and $p_{V_1^n \dots V_K^n Y_K^n \dots Y_1^n Z_M^n \dots Z_1^n}$, respectively. Consider an optimal coupling \cite{levin2009markov} (Proposition 4.7) between $\tilde{q}_{V_{\ell}^nY_k^n}$ and $p_{V_{\ell}^nY_k^n}$ such~that: 
\begin{align}
\mathbb{P} \big[ \mathcal{E}_{V_{\ell}^nY_k^n} \big] = \mathbb{V} (\tilde{q}_{V_{\ell}^nY_k^n}, p_{V_{\ell}^nY_k^n}), \nonumber
\end{align}
where $\mathcal{E}_{V_{\ell}^nY_k^n} \triangleq \{ ( \tilde{V}_{\ell}^n, \tilde{Y}_k^n ) \neq ( V_{\ell}^n, Y_k^n ) \}$ or, equivalently, $\mathcal{E}_{V_{\ell}^nY_k^n} \triangleq \{ ( \tilde{U}_{\ell}^n, \tilde{Y}_k^n ) \neq ( U_{\ell}^n, Y_k^n ) \}$ due to the invertibility of $G_n$. Furthermore, for all $\ell \in [1,k]$, we define the error events $\mathcal{E}_{\hat{V}_{\ell}^n} \triangleq \{ \hat{V}_{\ell}^n \neq \tilde{V}_{\ell}^n \}$ or, equivalently, $\mathcal{E}_{\hat{V}_{\ell}^n} \triangleq \{ \hat{U}_{\ell}^n \neq \tilde{U}_{\ell}^n \}$; and we define $\mathcal{E}_{\hat{V}_{0}^n} \triangleq \emptyset$. Hence, for any $\ell \in [1,k]$, the average probability of incorrectly decoding the message $W_{\ell}$ at the $k$-th receiver can be upper-bounded as:
\vspace{-18pt}
\begin{align}
 \mathbb{P} [ \hat{W}_{\ell} \neq W_{\ell} ] & \leq \mathbb{P} \big[ \hat{U}_{\ell}^n \neq \tilde{U}_{\ell}^n \big] \nonumber \\
 & = \mathbb{P} \big[ \hat{U}_{\ell}^n \neq \tilde{U}_{\ell}^n \big| \mathcal{E}^{\text{C}}_{V_{\ell}^n Y_k^n} \cap \mathcal{E}^{\text{C}}_{\hat{V}_{\ell-1}^n} \big] \mathbb{P} \big[ \mathcal{E}^{\text{C}}_{V_{\ell}^n Y_k^n} \cap \mathcal{E}^{\text{C}}_{\hat{V}_{\ell-1}^n} \big] \nonumber \\
& \quad + \mathbb{P} \big[ \hat{U}_{\ell}^n \neq \tilde{U}_{\ell}^n \big| \mathcal{E}_{V_{\ell}^n Y_k^n} \cup \mathcal{E}_{\hat{V}_{\ell-1}^n} \big] \mathbb{P} \big[ \mathcal{E}_{V_{\ell}^n Y_k^n} \cup \mathcal{E}_{\hat{V}_{\ell-1}^n} \big] \nonumber \\
& \leq \mathbb{P} \big[ \hat{U}_{\ell}^n \neq \tilde{U}_{\ell}^n \big| \mathcal{E}^{\text{C}}_{V_{\ell}^n Y_k^n} \cap \mathcal{E}^{\text{C}}_{\hat{V}_{\ell-1}^n} \big] + \mathbb{P} \big[ \mathcal{E}_{V_{\ell}^n Y_k^n} \big] + \mathbb{P} \big[ \mathcal{E}_{\hat{V}_{\ell-1}^n} \big] \nonumber \\
& \stackrel{(a)}{\leq} \sum_{j \in \mathcal{L}_{V_{\ell}|V_{\ell-1}Y_{\ell}}^{(n)}} Z \big({U}_{\ell}(j) \big| U_{\ell}^{1:j-1}, V_{\ell-1}^n, {Y}_k^n \big) + \mathbb{P} \big[ \mathcal{E}_{V_{\ell}^n Y_k^n} \big] + \mathbb{P} \big[ \mathcal{E}_{\hat{V}_{\ell-1}^n} \big] \nonumber \\ 
& \stackrel{(b)}{\leq} n \sqrt{\delta_n} + \mathbb{P} \big[ \mathcal{E}_{V_{\ell}^n Y_k^n} \big] + \mathbb{P} \big[ \mathcal{E}_{\hat{V}_{\ell-1}^n} \big] \nonumber \\ 
& \stackrel{(c)}{\leq} n \sqrt{\delta_n} + \delta_{\text{{ld-nls}}}^{(n)} + \mathbb{P} \big[ \mathcal{E}_{\hat{V}_{\ell-1}^n} \big] \label{eq:reliability1}
\end{align}
where $(a)$ holds by \cite{arikan2010source} (Theorem 2) because $\tilde{U}_{\ell} [ ( \mathcal{L}_{V_{\ell} | V_{\ell-1} Y_{\ell}}^{(n)} )^{\text{C}} ]$ for any $\ell \leq k$ is available to the $k$-th receiver, $(b)$ holds by Lemma~\ref{lemma:subsetproperty}, by the definition of the set $\mathcal{L}_{V_{\ell} | V_{\ell-1} Y_{1}}^{(n)}$ in Equation~\eqref{eq:pol1a} and by applying~\cite{arikan2010source} (Proposition 2) and $(c)$ holds by the optimal coupling and Lemma~\ref{lemma:distDMS_1} because $\mathbb{V} (\tilde{q}_{V_{\ell}^n Y_k^n}, p_{V_{\ell}^nY_k^n}) \leq \mathbb{V} (\tilde{q}_{V_1^n \dots V_K^n Y_K^n \dots Y_1^n Z_M^n \dots Z_1^n}, p_{V_1^n \dots V_K^n Y_K^n \dots Y_1^n Z_M^n \dots Z_1^n})$. Thus, by induction, we obtain:\vspace{-4pt}
\begin{align}
\mathbb{P} \big[ (\hat{W}_1, \dots \hat{W}_k) \neq (W_1, \dots, W_k) \big] & \leq \sum_{\ell=1}^k \mathbb{P} [ \hat{U}_{\ell} \neq \tilde{U}_{\ell} ] \leq \frac{k (k+1)}{2} \big( n \sqrt{\delta_n} + \delta_{\text{{ld-nls}}}^{(n)} \big). \label{eq:reliability1b}
\end{align} 
Consequently, if $K \ll n$, the polar coding scheme satisfies the reliability condition in Equation~\eqref{eq:reliabilitycondition_1}.

\subsubsection{Secrecy Performance}\label{sec:m1_sec}
{Consider the leakage at the eavesdropper $m \in [1,M]$ given in Equation~\eqref{eq:leakage1}. As in Equation~\eqref{eq:leakage2}, we~obtain:}
\begin{align}
I(W_1, \dots, W_K; F_1, \dots, F_K, \tilde{Z}_{m}^n) \! \leq \! \sum_{\ell=1}^K \! \big| \mathcal{I}_{\ell}^{(n)} \! \cup \! \mathcal{F}_{\ell}^{(n)} \big| \! - \! H \big(\tilde{U}_{1} \big[ \mathcal{I}_{1}^{(n)} \! \cup \! \mathcal{F}_{1}^{(n)} \big], \dots, \tilde{U}_{K} \big[ \mathcal{I}_{K}^{(n)} \! \cup \! \mathcal{F}_{K}^{(n)} \big] \big| \tilde{Z}_m^n \big). \label{eq:leakage1b}
\end{align}
Following similar reasoning as in Equation~\eqref{eq:entropy2bound}, for $n$ large enough, we have:
\begin{align}
& \Big| H \big(\tilde{U}_{1} \big[ \mathcal{I}_{1}^{(n)} \! \cup \! \mathcal{F}_{1}^{(n)} \big], \dots, \tilde{U}_{K} \big[ \mathcal{I}_{K}^{(n)} \! \cup \! \mathcal{F}_{K}^{(n)} \big] \big| \tilde{Z}_m^n \big) - H \big({U}_{1} \big[ \mathcal{I}_{1}^{(n)} \! \cup \! \mathcal{F}_{1}^{(n)} \big], \dots, {U}_{K} \big[ \mathcal{I}_{K}^{(n)} \! \cup \! \mathcal{F}_{K}^{(n)} \big] \big| {Z}_m^n \big) \Big| \nonumber \\
& \quad \stackrel{(a)}{\leq} \mathbb{V} (\tilde{q}_{Z_m^n}, p_{Z_m^n}) \log \frac{2^n}{\mathbb{V} (\tilde{q}_{Z_m^n}, p_{Z_m^n})} + \mathbb{V}^{\dagger} \log \frac{2^{( n + \sum_{\ell=1}^K | \mathcal{I}_{\ell}^{(n)} \cup \mathcal{F}_{\ell}^{(n)} | )}}{ \mathbb{V}^{\dagger} } \nonumber \\
& \quad \stackrel{(b)}{\leq} (K+2) n \delta_{\text{ld-nls}}^{(n)} - 2 \delta_{\text{ld-nls}}^{(n)} \log \delta_{\text{ld-nls}}^{(n)} , \label{eq:entropy1bound}
\end{align}
where $(a)$ holds by defining $\mathbb{V}^{\dagger} \triangleq \mathbb{V}(\tilde{q}_{{U}_{1} [ \mathcal{I}_{1}^{(n)} \cup \mathcal{F}_{1}^{(n)} ], \dots, {U}_{K} [ \mathcal{I}_{K}^{(n)} \cup \mathcal{F}_{K}^{(n)} ] {Z}_m^n}, {p}_{{U}_{1} [ \mathcal{I}_{1}^{(n)} \cup \mathcal{F}_{1}^{(n)} ], \dots, {U}_{K} [ \mathcal{I}_{K}^{(n)} \cup \mathcal{F}_{K}^{(n)} ] {Z}_m^n} )$ and \cite{csiszar2011information} (Lemma~2.9) and $(b)$ follows from Lemma~\ref{lemma:distDMS_2} by using similar reasoning as in Equation~\eqref{eq:entropy2bound} and~because the function $x \mapsto x \log x$ is decreasing for $x > 0$ small enough. Hence, we obtain:

\vspace{-18pt}
\begin{align}
& H \big( \tilde{U}_1 \big[ \mathcal{I}_{1}^{(n)} \cup \mathcal{F}_{1}^{(n)} \big], \dots, \tilde{U}_K \big[ \mathcal{I}_{K}^{(n)} \cup \mathcal{F}_{K}^{(n)} \big] \big| \tilde{Z}_m^n \big) \nonumber \\
 & \quad \geq H \big( {U}_1 \big[ \mathcal{I}_{1}^{(n)} \cup \mathcal{F}_{1}^{(n)} \big], \dots, {U}_K \big[ \mathcal{I}_{K}^{(n)} \cup \mathcal{F}_{K}^{(n)} \big] \big| {Z}_m^n \big) - ((K+2) n \delta_{\text{ld-nls}}^{(n)} - 2 \delta_{\text{ld-nls}}^{(n)} \log \delta_{\text{ld-nls}}^{(n)}) \nonumber \\
 & \quad \stackrel{(a)}{\geq} \sum_{{\ell}=1}^K \sum_{j \in \mathcal{I}_{{\ell}}^{(n)} \cup \mathcal{F}_{{\ell}}^{(n)}} H \big( {U}_{\ell}(j) \big| U_{\ell}^{1:j-1} , V_{{\ell}-1}^n, Z_m^n \big) - ((K+2) n \delta_{\text{ld-nls}}^{(n)} - 2 \delta_{\text{ld-nls}}^{(n)} \log \delta_{\text{ld-nls}}^{(n)}) \nonumber \\
 & \quad \stackrel{(b)}{\geq} \sum_{\ell=1}^K  \big| \mathcal{I}_{\ell}^{(n)} \cup \mathcal{F}_{\ell}^{(n)} \big| \left( 1 - 2 \delta_n \right) - ((K+2) n \delta_{\text{ld-nls}}^{(n)} - 2 \delta_{\text{ld-nls}}^{(n)} \log \delta_{\text{ld-nls}}^{(n)}) , \label{eq:leakage1c}
\end{align} 
where $(a)$ holds because conditioning does not increase the entropy and because $U_1^n - \dots - U_{K-1}^n - U_K^n$ forms a Markov chain and the invertibility of $G_n$ and $(b)$ holds because, according to Equations \eqref{eq:e1_setI} and \eqref{eq:e1_setF}, $\mathcal{I}_{\ell}^{(n)} \cup \mathcal{F}_{\ell}^{(n)} \subseteq \mathcal{H}_{V_{\ell} | V_{\ell-1} Z_M}^{(n)}$ for all $\ell \in [1,K]$, because by Lemma~\ref{lemma:subsetproperty}, we have $\mathcal{H}_{V_{\ell} | V_{\ell-1} Z_M}^{(n)} \subseteq \mathcal{H}_{V_{\ell} | V_{\ell-1} Z_m}^{(n)}$ for any $m \in [1,M-1]$, and by the definition of the set $\mathcal{H}_{V_{\ell} | V_{\ell-1} Z_m}^{(n)}$ given in Equation~\eqref{eq:pol1c}.

Finally, by substituting Equation~\eqref{eq:leakage1c} into Equation~\eqref{eq:leakage1b}, we obtain:
\begin{align}
I(W_1, \dots, W_K; F_1, \dots, F_K, \tilde{Z}_{m}^n) \leq n \delta_n + (K+2) n \delta_{\text{ld-nls}}^{(n)} - 2 \delta_{\text{ld-nls}}^{(n)} \log \delta_{\text{ld-nls}}^{(n)},
\end{align}
Hence, if $K \ll n$, the polar code satisfies the secrecy condition in Equation~\eqref{eq:secrecycondition_1}, and the proof is~concluded.
 
\section{Polar Construction and Performance Evaluation}\label{sec:results}
In this section, we discuss further how to construct the polar codes for the DBC-NLD-LS and DBC-LD-NLS proposed in Sections~\ref{sec:PCS_dbcnldls}~and~\ref{sec:PCS_dbcldnls}, respectively. Moreover, we evaluate the reliability and the secrecy performance of both polar coding schemes according to different parameters involved in the polar code construction. Although the construction of polar codes has been covered in a large number of references (see, for instance, \cite{tal2013construct,vangala2015comparative,6601656}), they only focus on polar codes under reliability constraints. 

For the DBC-NLD-LS, we consider the Binary Erasure Broadcast Channel (BE-BC), where each individual channel of the DBC is a Binary Erasure Channel (BEC). For this model, we propose a construction of the polar code that is based on the Bhattacharyya parameters instead of the corresponding entropy terms.~The reason is that, for the BE-BC, the Bhattacharyya parameters associated with the sets in Equations~\eqref{eq:Aset1}--\eqref{eq:Aset5} can be computed exactly \cite{arikan2009channel} (Proposition 5).~Then, we~evaluate the reliability and the secrecy performance of the code, and we focus on how different parameters involved in the proposed polar code construction impact its performance.

On the other hand, for the DBC-LD-NLS, we consider the Binary Symmetric Broadcast Channel (BS-BC), where each individual channel is a Binary Symmetric Channel (BSC). From \cite{arikan2009channel} (Proposition~5), we know that the method to compute the exact values of the Bhattacharyya parameters for a BEC provides an upper-bound on the Bhattacharyya parameters of the BSC. Although this method can be useful to construct polar codes under reliability constraints \cite{tal2013construct,vangala2015comparative,6601656}, it fails when the code must guarantee some secrecy condition based on the information leakage. Indeed, in order to upper-bound the information leakage in Equation~\eqref{eq:leakage1}, according to Equation~\eqref{eq:leakage1c}, notice that we need a lower-bound on the entropy terms (or Bhattacharyya parameters). Hence, for this model, we focus more on proposing a new polar code construction that is based directly on the entropy terms associated with the sets in Equations~\eqref{eq:pol1b1}--\eqref{eq:pol1c}.

Throughout this section, as in \cite{arikan2009channel}, we say that a channel or a conditional distribution $p_{Y|X}(y|x)$ with $x \in \mathcal{X} \triangleq \{0,1\}$ and $y \in \mathcal{Y} \triangleq \{ 0, \dots, |\mathcal{Y}|-1 \}$ is symmetric if the columns of the probability transition matrix 
$\mathbf{P}_{Y|X} \triangleq \Big[ 
\begin{footnotesize}
\begin{array}{ccc}
p_{Y|X}(0|0) & \cdots & p_{Y|X}( |\mathcal{Y}|-1 | 0 ) \\
p_{Y|X}(0|1) & \cdots & p_{Y|X}( |\mathcal{Y}|-1 | 1 )
\end{array} 
\end{footnotesize} \Big]$
can be grouped into sub-matrices such that for each sub-matrix, each row is a permutation of each other row and each column is a permutation of each other column. Therefore, the individual channels of both BE-BC and the BS-BC are symmetric.
 
Due to the symmetry of BE-BC, we will see that the distribution induced by the encoding described in Section~\ref{sec:PCS_dbcnldls_encoder} for the DBC-NLD-LS will approach exactly the optimum distribution of the original DMS used in the polar code construction. Consequently, the performance of the polar code will depend only on the parameters involved in the construction. On the other hand, despite the symmetry of the BS-BC, due to its superposition-based structure, the encoding described in Section~\ref{sec:PCS_dbcldnls_encoder} for the DBC-NLD-LS only approaches the target distribution asymptotically. Hence, this encoding will impact the reliability and secrecy performance of the polar code when we consider a finite blocklength. 
 
\subsection{DBC-NLD-LS}
For this model, we consider BE-BC with two legitimate receivers ($K=2$) and two eavesdroppers ($M=2$).~Therefore, each individual channel is a BEC with $\mathcal{X} \triangleq \{ 0,1\}$ and $\mathcal{Y}_k = \mathcal{Z}_m \triangleq \{ 0 ,1, E\}$, $E$ being the {erasure symbol} and $k,m \in \{1,2\}$. The individual channels are defined simply by their {erasure probability}, which is denoted by $\epsilon_{Y_k}$ for the corresponding legitimate receiver $k$ ($\mathbb{P}[Y_k = E] = \epsilon_{Y_k}$) and $\epsilon_{Z_m}$ for the eavesdropper $m$ ($\mathbb{P}[Z_m = E] = \epsilon_{Z_m}$).~Due to the degradedness condition of the broadcast channel given in Equation~\eqref{eq:degch}, we have $\epsilon_{Y_2} < \epsilon_{Y_1} < \epsilon_{Z_2} < \epsilon_{Z_1}$. By properly applying \cite{bloch2011physical} (Proposition 3.2), it is easy to shown that the secrecy-capacity achieving distribution $p_X^{\star}$ for this model is the uniform, i.e., $p_X^{\star}(x) = \frac{1}{2}$ $\forall x \in \{0,1\}$. For the simulations, we consider a BE-BC such that $\epsilon_{Y_2} = 0.01$, $\epsilon_{Y_1} = 0.04$, $\epsilon_{Z_2} = 0.2$ and $\epsilon_{Z_1} = 0.35$. According to Corollary~\ref{coro:SCR_2} and since $p_X^{\star}(x)$ is uniform, we obtain that the capacity without considering rate sharing is $R_1^{\star} = 0.15$ and $R_2^{\star} = 0.16$.

\subsubsection{Practical Polar Code Construction}\label{sec:pccm1}
Given the blocklength $n$ and the distribution $p^{\star}_{XY_2Y_1Z_2Z_1} = p_X^{\star} p_{Y_2Y_1Z_2Z_1|X}$, the goal of the polar code construction is to obtain the partition of the universal set $[n]$ defined in Equations~\eqref{eq:setIM}--\eqref{eq:setT1} and graphically represented in Figure~\ref{fig:pc_dbcnldls}.~Hence, we need to define first the required sets of Equations~\eqref{eq:Aset1}--\eqref{eq:Aset5}, which means having to compute the entropy terms $\{H(U(j)|U^{1:j-1}) \}_{j=1}^n$, $\{H(U(j)|U^{1:j-1}, Y_1^n) \}_{j=1}^n$ and $\{H(U(j)|U^{1:j-1}, Z_m^n) \}_{j=1}^n$ $\forall m \in \{1,2\}$ associated with the polar transform $U^n = X^n G_n$. Alternatively, as mentioned in Section~\ref{sec:rev}, we can define the sets in Equations~\eqref{eq:Aset1}--\eqref{eq:Aset5} from the corresponding Bhattacharyya parameters. Indeed, since each individual channel is a BEC, by \cite{arikan2009channel} (Proposition 5), we can compute with very low complexity the exact values of $\{Z(U(j)|U^{1:j-1}) \}_{j=1}^n$, $\{Z(U(j)|U^{1:j-1}, Y_1^n) \}_{j=1}^n$ and $\{Z(U(j)|U^{1:j-1}, Z_m^n) \}_{j=1}^n$ $\forall m \in \{1,2\}$.~To do so, we use the recursive algorithm \cite{vangala2015comparative} (PCC-0) adapted to the BEC, which, for instance, will obtain $\{Z(U(j)|U^{1:j-1}, Y_1^n) \}_{j=1}^n$ from the initial value $Z(X | Y_1) = \epsilon_{Y_1}$ (the entire code in MATLAB used for this section is provided as Supplementary Material---see Endnote \cite{Note2}). Regarding $\{Z(U(j)|U^{1:j-1}) \}_{j=1}^n$, since $p_X^{\star}$ is uniform, it is clear that $Z(U(j)|U^{1:j-1} ) = H(U(j)|U^{1:j-1} ) = 1$ for all $j \in [n]$, which means $\mathcal{H}_{X}^{(n)} = [n]$. Consequently, the set $\mathcal{T}^{(n)} = \emptyset$, and according to Equation~\eqref{eq:distrenc1}, neither random, nor deterministic SC encoding will be needed.

In order to compare the performance of the polar coding scheme according to different parameters and to provide more flexibility in the design, instead of using only $\delta_n$ to define the sets in Equations~\eqref{eq:Aset1}--\eqref{eq:Aset5}, we introduce the pair $(\delta_n^{(\text{r})}, \delta_n^{(\text{s})})$, where $\delta_n^{(\text{r})} \triangleq 2^{-n^{\beta^{(\text{r})}}}$ and $\delta_n^{(\text{s})} \triangleq 2^{-n^{\beta^{(\text{s})}}}$ for some $\beta^{(\text{r})},\beta^{(\text{s})} \in (0,\frac{1}{2})$. Let $R^{\prime}_{1} \in [0,R_{1}^{\star}]$ and $R^{\prime}_{2} \in [0,R_{2}^{\star}]$ denote the target rates that the polar coding scheme must approach. We obtain the partition defined in Equations~\eqref{eq:setIM}--\eqref{eq:setT1} as follows. First, we define $\big( \mathcal{H}_{X|Y_1}^{(n)} \big)^{\text{C}} \triangleq \big\{ j \in [n]: H\left( U(j) \left| U^{1:j-1}, Y_1^n \right. \right) \leq 1 - \delta_n^{(\text{s})} \big\}$, where one can notice that we have used $\delta_n^{(\text{s})}$. Then, we choose $\mathcal{I}_2^{(n)}$ by taking the $\ceil{n R_2^{\prime}}$ indices $j \in \big( \mathcal{H}_{X|Y_1}^{(n)} \big)^{\text{C}}$ that correspond to the highest Bhattacharyya parameters $\{Z(U(j)|U^{1:j-1}, Z_2^n) \}_{j=1}^n$ for Eavesdropper 2. Second, we choose $\mathcal{I}_1^{(n)}$ by taking the $\ceil{n R_1^{\prime}}$ indices $j \in \big( \mathcal{H}_{X|Y_1}^{(n)} \big)^{\text{C}}\setminus \mathcal{I}_2^{(n)}$ that correspond to the highest Bhattacharyya parameters $\{Z(U(j)|U^{1:j-1}, Z_1^n) \}_{j=1}^n$ for Eavesdropper 1. Finally, we obtain $\mathcal{C}^{(n)} = \big( \mathcal{H}_{X|Y_1}^{(n)} \big)^{\text{C}} \setminus \big( \mathcal{I}_1^{(n)} \cup \mathcal{I}_2^{(n)} \big)$ and $\mathcal{F}^{(n)} = \mathcal{H}_{X|Y_1}^{(n)}$.~Furthermore, in order to evaluate the reliability performance of the code, we~define $\mathcal{L}_{X|Y_1}^{(n)} \triangleq \big\{ j \in [n]: H \left( U(j) \left| U^{1:j-1}, Y_1^n \right. \right) \leq\delta_n^{(\text{r})} \big\}$, where one can notice that we have used $\delta_n^{(\text{r})}$. Since the additional secret sequence $\Phi$ corresponds to those entries belonging to $
\big( \mathcal{H}_{X|Y_1}^{(n)} \big)^{\text{C}} \cap \big( \mathcal{L}_{X|Y_1}^{(n)} \big)^{\text{C}}$, its length will depend on $(\delta_n^{(\text{r})}, \delta_n^{(\text{s})})$. According to the polar code construction proposed in this section, notice~that $\delta_n^{(\text{s})}$ must be small enough to guarantee that $\big| \big( \mathcal{H}_{X|Y_1}^{(n)} \big)^{\text{C}} \big| \geq R^{\prime}_{1} + R^{\prime}_{2}$.

\subsubsection{Performance Evaluation}\label{sec:pe1}
First, notice that the encoding of Section~\ref{sec:PCS_dbcnldls_encoder} will induce a distribution $\tilde{q}_{X^nY_2^nY_1^nZ_2^nZ_1^n} = p^{\star}_{X^nY_2^nY_1^nZ_2^nZ_1^n}$ because $\mathcal{T}^{(n)} = \emptyset$ (we do not use SC encoding), and the encoder will store uniformly-distributed sequences into the entries $U(j)$ that satisfy $H(U(j)|U^{1:j-1} ) = 1$ for all $j \in \mathcal{H}_{X}^{(n)} = [n]$.~Hence, $\mathbb{V}(\tilde{q}_{X^nY_2^nY_1^nZ_2^nZ_1^n},{p}^{\star}_{X^nY_2^nY_1^nZ_2^nZ_1^n}) = 0$, and the performance will only depend on the code construction.

To evaluate the reliability performance, we obtain an upper-bound $P_{\text{b}}^{\text{ub}(1)}$ on the average bit error probability at the legitimate Receiver 1. Since $\mathbb{V}(\tilde{q}_{X^nY_2^nY_1^nZ_2^nZ_1^n},{p}^{\star}_{X^nY_2^nY_1^nZ_2^nZ_1^n}) = 0$, from Equation~\eqref{eq:reliability2}, we~have:
\begin{align}
P_{\text{b}}^{\text{ub} (1)} \triangleq \frac{1}{\big| \mathcal{L}_{X|Y_1}^{(n)} \big|} \sum_{j \in \mathcal{L}_{X|Y_1}^{(n)}} Z \big({U}(j) \big| U^{1:j-1}, {Y}_1^n \big). \label{eq:r1_Pb} 
\end{align}
Due to the degradedness condition of the BE-BC and, consequently, by Lemma~\ref{lemma:subsetproperty}, the average bit error probability at the legitimate Receiver 2 will be always less than the one at the legitimate Receiver 1. Since the legitimate receivers must estimate the entries belonging to $\mathcal{L}_{X|Y_1}^{(n)}$ regardless of $\big( \mathcal{H}_{X|Y_1}^{(n)} \big)^{\text{C}}$ and the target rates $(R^{\prime}_{1},R^{\prime}_{2})$, the reliability performance only depends on the pair $(n,\delta_n^{(\text{r})})$. 

In order to evaluate the secrecy performance, we compute an upper-bound on the information leakage $I(W_1,W_2; F,\tilde{Z}_1^n)$ and an upper-bound on the information leakage $I(W_2; F, \tilde{Z}_2^n)$. Since~$\mathbb{V}(\tilde{q}_{X^nY_2^nY_1^nZ_2^nZ_1^n},{p}^{\star}_{X^nY_2^nY_1^nZ_2^nZ_1^n}) = 0$, from Equations~\eqref{eq:leakage2} and \eqref{eq:secrecy2b}, we obtain:
\begin{align}
I^{\text{ub}}(W_1, W_2; F, \tilde{Z}_{1}^n) & \triangleq \sum_{i=1}^2 \big| \mathcal{I}_{i}^{(n)} \big| + \big| \mathcal{F}^{(n)} \big| - \sum_{j \in \mathcal{I}_{1}^{(n)} \cup \mathcal{I}_{2}^{(n)} \cup \mathcal{F}^{(n)}} Z \big({U}(j) \big| U^{1:j-1}, {Z}_1^n \big)^2, \label{eq:r1_I1} \\
I^{\text{ub}}(W_2; F, \tilde{Z}_{2}^n) & \triangleq \big| \mathcal{I}_{2}^{(n)} \big| + \big| \mathcal{F}^{(n)} \big| - \sum_{j \in \mathcal{I}_{2}^{(n)} \cup \mathcal{F}^{(n)}} Z \big({U}(j) \big| U^{1:j-1}, {Z}_2^n \big)^2, \label{eq:r1_I2} 
\end{align}
where we have used \cite{arikan2010source} (Proposition 2) to express the information leakage in terms of the Bhattacharyya parameters because $H(U(j)|U^{1:j-1}, Z^n_m) \geq Z (U(j)|U^{1:j-1}, Z^n_m)^2$. According to the proposed polar code construction, the secrecy performance will depend on $(n,\delta_n^{(\text{s})})$ and the rates $(R^{\prime}_{1},R^{\prime}_{2})$, but not on $\delta_n^{(\text{r})}$. 

Additionally, we evaluate the rate of the additional sequence $\Phi$ simply by computing:
\begin{align}
\frac{1}{n} |\Phi | = \frac{1}{n} \Big| \big( \mathcal{H}_{X|Y_1}^{(n)} \big)^{\text{C}} \cap \big( \mathcal{L}_{X|Y_1}^{(n)} \big)^{\text{C}} \Big|, \label{eq:r1_sm}
\end{align}
which will depend on the triple $(n,\delta_n^{(\text{r})},\delta_n^{(\text{s})})$, but not on $(R^{\prime}_{1},R^{\prime}_{2})$.

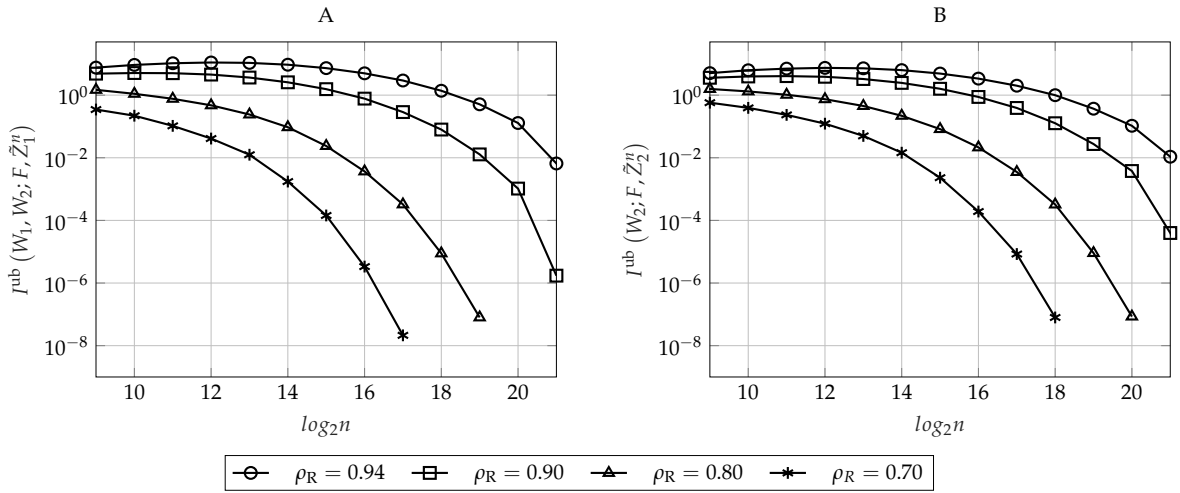
\begin{figure}[H]
 \centering
 \resizebox{\linewidth}{!}{%
 \begin{tikzpicture}

\begin{groupplot}[
group style={
group name=myplot,
group size= 2 by 1,
horizontal sep=1in
},
]
        
\nextgroupplot[
		title={A},
        width=3in,
	    height=2.2in,
		scale only axis,
		xmin=9,
		xmax=21,
		xlabel style={font=\color{white!15!black}},
		xlabel={$log_2 n$},
		ymode=log,
		ymin=1e-9,
		ymax=50,
		yminorticks=true,
		ylabel style={font=\color{white!15!black}},
		ylabel={$I^{\text{ub}}\left(W_1,W_2;F,\tilde{Z}_1^n \right)$},
		axis background/.style={fill=white},
		xmajorgrids,
		ymajorgrids,
		yminorgrids,
		legend style = { column sep = 10pt, legend columns = -1, legend to name = grouplegend,}]
        ]
       \addplot [color=black, line width=1.0pt, mark size=2.8pt, mark=o, mark options={solid, black}]
  	    table[row sep=crcr]{%
8	6.04775396978924\\
9	7.55972561297711\\
10	9.221273435048\\
11	10.429306005593\\
12	11.0182912645696\\
13	10.7149765516304\\
14	9.36189219610133\\
15	7.28127302483572\\
16	4.97015802495107\\
17	2.93015105304312\\
18	1.37999819302895\\
19	0.508857361836999\\
20	0.128712934178475\\
21	0.00657046140016367\\
22	0.000133484834805131\\
};
\addlegendentry{$\rho_{\text{R}} = 0.94$}

\addplot [color=black, line width=1.0pt, mark size=2.8pt, mark=square, mark options={solid, black}]
  table[row sep=crcr]{%
8	4.18779468626122\\
9	4.85192089875346\\
10	5.08016886864714\\
11	5.02229759581275\\
12	4.5329453815241\\
13	3.64311870708929\\
14	2.5672284052639\\
15	1.55269461149933\\
16	0.774004767550378\\
17	0.287723318000099\\
18	0.0796298641434987\\
19	0.0128162700137182\\
20	0.00103763656661613\\
21	1.72104531278213e-06\\
22	0\\
};
\addlegendentry{$\rho_{\text{R}} = 0.90$}

\addplot [color=black, line width=1.0pt, mark size=2.8pt, mark=triangle, mark options={solid, black}]
  table[row sep=crcr]{%
8	1.79178852434898\\
9	1.4759880035708\\
10	1.10273182344325\\
11	0.764501617835524\\
12	0.473151069748937\\
13	0.24092524614079\\
14	0.0945748565310851\\
15	0.0237950803292733\\
16	0.00366758106720226\\
17	0.000324493900188827\\
18	8.85548924998147e-06\\
19	7.98150722403079e-08\\
20	0\\
21	0\\
22	0\\
};
\addlegendentry{$\rho_{\text{R}} = 0.80$}

\addplot [color=black, line width=1.0pt, mark size=2.8pt, mark=asterisk, mark options={solid, black}]
  table[row sep=crcr]{%
8	0.58048518870163\\
9	0.346725233212017\\
10	0.220405816647872\\
11	0.10444922175883\\
12	0.0411136501926819\\
13	0.0125144686560361\\
14	0.0017397253961235\\
15	0.000143791955565575\\
16	3.34910428136936e-06\\
17	2.1362939151004e-08\\
18	0\\
19	0\\
20	0\\
21	0\\
22	0\\
};
\addlegendentry{$\rho_{R} = 0.70$}


\nextgroupplot[
		title ={B},
        width=3in,
		height=2.2in,
		scale only axis,
		xmin=9,
		xmax=21,
		xlabel style={font=\color{white!15!black}},
		xlabel={$log_2 n$},
		ymode=log,
		ymin=1e-9,
		ymax=50,
		yminorticks=true,
		ylabel style={font=\color{white!15!black}},
		ylabel={$I^{\text{ub}}\left(W_2;F,\tilde{Z}_2^n \right)$},
		axis background/.style={fill=white},
		xmajorgrids,
		ymajorgrids,
		yminorgrids,
		]
\addplot [color=black, line width=1.0pt, mark size=2.8pt, mark=o, mark options={solid, black}]
  table[row sep=crcr]{%
8	4.0145415820006\\
9	5.0378615800388\\
10	6.2247161637519\\
11	7.01784041342707\\
12	7.42210918004041\\
13	7.19429795855058\\
14	6.28961065850231\\
15	4.88472350920151\\
16	3.37023217162432\\
17	2.00086716867245\\
18	0.995949047819158\\
19	0.368832017258683\\
20	0.105014568363549\\
21	0.0108022921485826\\
22	0.00112230330705643\\
};

\addplot [color=black, line width=1pt, mark size=2.8pt, mark=square, mark options={solid, black}]
  table[row sep=crcr]{%
8	3.11352419388914\\
9	3.58315359454625\\
10	3.94751088304562\\
11	4.04910268430251\\
12	3.85373211432444\\
13	3.27025205256572\\
14	2.45847936844123\\
15	1.59561201085776\\
16	0.874839162333137\\
17	0.389389820201905\\
18	0.126756153688984\\
19	0.0274444158523693\\
20	0.00377875784324715\\
21	3.96170653402805e-05\\
22	9.94186848402023e-08\\
};

\addplot [color=black, line width=1pt, mark size=2.8pt, mark=triangle, mark options={solid, black}]
  table[row sep=crcr]{%
8	1.73310941504521\\
9	1.57468894115524\\
10	1.31340741348453\\
11	1.02770409539761\\
12	0.740765668264089\\
13	0.450645352393678\\
14	0.219009330602216\\
15	0.0823728803058657\\
16	0.0214544004556956\\
17	0.00351220840875612\\
18	0.000321661775524262\\
19	9.06568238860927e-06\\
20	8.50719516165555e-08\\
21	0\\
22	0\\
};

\addplot [color=black, line width=1pt, mark size=2.8pt, mark=asterisk, mark options={solid, black}]
  table[row sep=crcr]{%
8	0.79973386594947\\
9	0.573882347034875\\
10	0.391906053382779\\
11	0.234080253005644\\
12	0.122681747561273\\
13	0.0498896956179635\\
14	0.0145182102199783\\
15	0.00231313003744162\\
16	0.000192157224682887\\
17	8.43489588078228e-06\\
18	7.95813321019523e-08\\
19	0\\
20	0\\
21	0\\
22	0\\
};

\end{groupplot}

\node at ($(8,-1.6)$) {\ref{grouplegend}};

\end{tikzpicture}
 }
 \caption{Secrecy 
 performance of the polar coding scheme for DBC-NLD-LS over BE-BC as a~function of the blocklength $n$ and the normalized target rate $\rho_{\text{R}}$ when we set $\beta^{(\text{r})}=0.16$ and $\beta^{(\text{s})} = 0.30$. {(\textbf{A})} Upper-bound on the information about $(W_1,W_2)$ leaked to Eavesdropper 1 defined as in Equation~\eqref{eq:r1_I1}. {(\textbf{B})} Upper-bound on the information about $W_2$ leaked to Eavesdropper 2 defined as in Equation~\eqref{eq:r1_I2}.}\label{fig:sec1}
 \end{figure}

Let $\rho_{\text{R}}$ be the normalized target rate in which the polar coding scheme operates, that is \mbox{$\rho_{\text{R}} \triangleq \frac{R_1^{\prime}}{R_1^{\star}} = \frac{R_2^{\prime}}{R_2^{\star}}$}. In Figure~\ref{fig:sec1}A,B, we evaluate the upper-bounds on the information leakage defined in Equations~\eqref{eq:r1_I1}~and~\eqref{eq:r1_I2}, respectively, as a function of the blocklength $n$ for different values of $\rho_{\text{R}}$. To~do so, we set $\beta^{(\text{r})}=0.16$ and $\beta^{(\text{s})} = 0.30$, which defines a particular pair $(\delta_n^{(\text{r})},\delta_n^{(\text{s})})$ for each value of $n$ (recall that $\delta_n^{(\text{r})}$ does not impact on the secrecy performance of the polar code). {As we proved in Section~\ref{sec:m2_sec}, for large enough $n$, the secrecy performance improves as $n$ increases. Moreover, to achieve a particular secrecy performance level, the polar code will require a larger blocklength $n$ as the rates approach the capacity. This happens because, given $(n,\delta_n^{(\text{s})})$ and, consequently, $\big( \mathcal{H}_{X|Y_1}^{(n)} \big)^{\text{C}}$, the parameter $\rho_{\text{R}}$ only determines the amount of indices that will belong to $\mathcal{I}_1^{(n)} \cup \mathcal{I}_2^{(n)} \subseteq \big( \mathcal{H}_{X|Y_1}^{(n)} \big)^{\text{C}}$. Since, by construction, we take those indices corresponding to the highest Bhattacharyya parameters associated with the eavesdroppers, taking more elements always increases the corresponding leakage. For rates approaching the capacity and small values of $n$, notice that we obtain a secrecy performance that is getting worse as $n$ increases (for instance, for $\rho_{\text{R}} = 0.94$, we obtain that the information leakage is increasing from $n=2^9$ to $n=2^{12}$). This behavior is mainly explained because the elements of $U^n$ have not been polarized enough for small values of $n$. Consequently, for a given value of $\beta^{(\text{s})}$, not all the Bhattacharyya parameters associated with the eavesdroppers corresponding to the sets $\mathcal{I}_1^{(n)}$ and $\mathcal{I}_2^{(n)}$ are sufficiently close to one. Since, for a given $\rho_{\text{R}}$, the cardinality of $\mathcal{I}_1^{(n)}$ and $\mathcal{I}_2^{(n)}$ increases with $n$, then~the information leakage can increase with $n$ when $n$ is not large enough. Moreover, since operating at lower rates means taking a fewer number of indices in $\mathcal{I}_1^{(n)}$ and $\mathcal{I}_2^{(n)}$, but taking those that are closest to one, this behavior appears only for large values of $\rho_{\text{R}}$}.

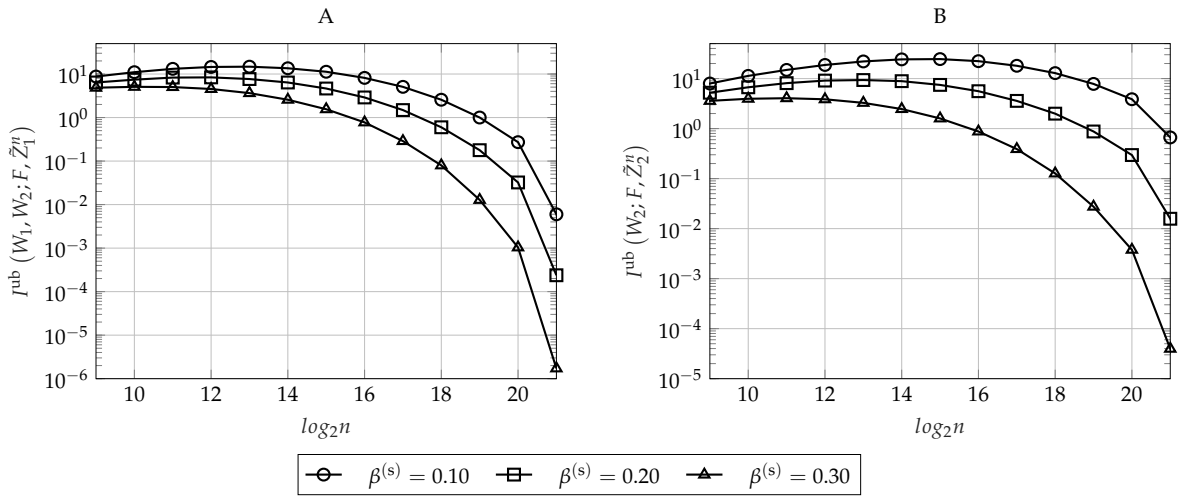
\begin{figure}[H]
 \centering
 \resizebox{\linewidth}{!}{%
 \begin{tikzpicture}

\begin{groupplot}[
group style={
group name=myplot,
group size= 2 by 1,
horizontal sep=1in
},
]
        
\nextgroupplot[
		title={A},
        width=3in,
	    height=2.2in,
		scale only axis,
		xmin=9,
		xmax=21,
		xlabel style={font=\color{white!15!black}},
		xlabel={$log_2 n$},
		ymode=log,
		ymin=1e-6,
		ymax=50,
		yminorticks=true,
		ylabel style={font=\color{white!15!black}},
		ylabel={$I^{\text{ub}}\left(W_1,W_2;F,\tilde{Z}_1^n \right)$},
		axis background/.style={fill=white},
		xmajorgrids,
		ymajorgrids,
		legend style = { column sep = 10pt, legend columns = -1, legend to name = grouplegend,}]
        ]
       \addplot [color=black, line width=1.0pt, mark size=2.8pt, mark=o, mark options={solid, black}]
  	    table[row sep=crcr]{%
8	6.04775396978924\\
9	8.75784877572083\\
10	11.0211342200312\\
11	13.1611960213132\\
12	14.5441531791075\\
13	14.7519901657315\\
14	13.5687041561273\\
15	11.2799381463927\\
16	8.15819628307727\\
17	5.07350585651766\\
18	2.5662387131335\\
19	1.00129346869508\\
20	0.271977703987795\\
21	0.00598276131010304\\
22	4.24415338784456e-06\\
};
\addlegendentry{$\beta^{(\text{s})} = 0.10$}

\addplot [color=black, line width=1.0pt, mark size=2.8pt, mark=square, mark options={solid, black}]
  table[row sep=crcr]{%
8	4.61464936717604\\
9	6.39394301934742\\
10	7.42942874655058\\
11	8.28461643507298\\
12	8.37804011696034\\
13	7.67423780036372\\
14	6.39044498428993\\
15	4.64657085622439\\
16	2.89743823109266\\
17	1.48559579315879\\
18	0.597664533853822\\
19	0.178024887618812\\
20	0.0322704287580564\\
21	0.000237875967286527\\
22	0\\
};
\addlegendentry{$\beta^{(\text{s})} = 0.20$}

\addplot [color=black, line width=1.0pt, mark size=2.8pt, mark=triangle, mark options={solid, black}]
  table[row sep=crcr]{%
8	4.18779468626122\\
9	4.85192089875346\\
10	5.08016886864714\\
11	5.02229759581275\\
12	4.5329453815241\\
13	3.64311870708929\\
14	2.5672284052639\\
15	1.55269461149933\\
16	0.774004767550378\\
17	0.287723318000099\\
18	0.0796298641434987\\
19	0.0128162700137182\\
20	0.00103763656661613\\
21	1.72104531278213e-06\\
22	0\\
};
\addlegendentry{$\beta^{(\text{s})} = 0.30$}


\nextgroupplot[
		title ={B},
        width=3in,
		height=2.2in,
		scale only axis,
		xmin=9,
		xmax=21,
		xlabel style={font=\color{white!15!black}},
		xlabel={$log_2 n$},
		ymode=log,
		ymin=1e-5,
		ymax=50,
		ylabel style={font=\color{white!15!black}},
		ylabel={$I^{\text{ub}}\left(W_2;F,\tilde{Z}_2^n \right)$},
		axis background/.style={fill=white},
		xmajorgrids,
		ymajorgrids,
		]
\addplot [color=black, line width=1.0pt, mark size=2.8pt, mark=o, mark options={solid, black}]
  table[row sep=crcr]{%
8	5.1732207836185\\
9	7.9641227586841\\
10	11.3130761666392\\
11	14.8952176868488\\
12	18.7140535189818\\
13	22.0043635701974\\
14	24.2067597174112\\
15	24.5259243646532\\
16	22.2841625844982\\
17	18.0194295671779\\
18	12.8570249882185\\
19	7.81815618950786\\
20	3.86480596469337\\
21	0.665623837917034\\
22	0.0248445247998461\\
};

\addplot [color=black, line width=1pt, mark size=2.8pt, mark=square, mark options={solid, black}]
  table[row sep=crcr]{%
8	3.50403256346543\\
9	5.21178238249703\\
10	6.69960264021146\\
11	8.10335679596505\\
12	9.112244795029\\
13	9.34958956960653\\
14	8.87779872378207\\
15	7.47129037643758\\
16	5.5917519082298\\
17	3.57891390093701\\
18	1.98940386732538\\
19	0.875644418050069\\
20	0.295829607284395\\
21	0.0157524254270053\\
22	9.31185204535723e-05\\
};

\addplot [color=black, line width=1pt, mark size=2.8pt, mark=triangle, mark options={solid, black}]
  table[row sep=crcr]{%
8	3.11352419388914\\
9	3.58315359454625\\
10	3.94751088304562\\
11	4.04910268430251\\
12	3.85373211432444\\
13	3.27025205256572\\
14	2.45847936844123\\
15	1.59561201085776\\
16	0.874839162333137\\
17	0.389389820201905\\
18	0.126756153688984\\
19	0.0274444158523693\\
20	0.00377875784324715\\
21	3.96170653402805e-05\\
22	9.94186848402023e-08\\
};

\end{groupplot}

\node at ($(8,-1.6)$) {\ref{grouplegend}};

\end{tikzpicture}
 }
 \caption{Secrecy 
performance of the polar coding scheme for DBC-NLD-LS over BE-BC as a function of $n$ and $\beta^{(\text{s})}$, which defines $\delta_n^{(\text{s})}$ for each $n$, when we set $\beta^{(\text{r})}=0.16$ and $\rho_{\text{R}} = 0.90$. {(\textbf{A})} Upper-bound on the information about $(W_1,W_2)$ leaked to Eavesdropper 1 defined as in Equation~\eqref{eq:r1_I1}. {(\textbf{B})} Upper-bound on the information about $W_2$ leaked to Eavesdropper 2 defined as in Equation~\eqref{eq:r1_I2}.}\label{fig:sec2}
 \end{figure}

The impact of $\delta_n^{(\text{s})}$ on the secrecy performance is graphically represented in Figure~\ref{fig:sec2}A,B, where the former plots the upper-bound defined in Equation~\eqref{eq:r1_I1} and the latter the upper-bound in Equation~\eqref{eq:r1_I2} as a function of the blocklength $n$ for different values of $\beta^{(\text{s})}$. Now, we set $\beta^{(\text{r})}=0.16$ and $\rho_{\text{R}}=0.90$. As can be seen in Figure~\ref{fig:sec2}, the secrecy performance improves as the value of $\beta^{(\text{s})}$ increases (or equivalently, as $\delta_n^{(\text{s})}$ decreases). This behavior is as expected because notice that $\delta_n^{(\text{s})}$ defines the value of the highest Bhattacharyya parameter $Z\big( U(j) \big| U^{1:j-1}, Y_1^n \big)$ that will belong to $\big( \mathcal{H}_{X|Y_1}^{(n)} \big)^{\text{C}}$, that is the set containing the possible candidates for $\mathcal{I}_1^{(n)} \cup \mathcal{I}_2^{(n)}$. Since the polar construction chooses the indices that will belong to $\mathcal{I}_1^{(n)}$ and $\mathcal{I}_2^{(n)}$ by taking the ones corresponding to the highest Bhattacharyya parameters associated with the eavesdroppers and since, by Lemma~\ref{lemma:subsetproperty}, $Z\big( U(j) \big| U^{1:j-1}, Z_1^n \big) \geq Z\big( U(j) \big| U^{1:j-1}, Z_2^n \big) \geq Z\big( U(j) \big| U^{1:j-1}, Y_1^n \big)$ for any $j \in [n]$, the sums in Equations~\eqref{eq:r1_I1}~and~\eqref{eq:r1_I2} over the indices $j \in \mathcal{I}_1^{(n)} \cup \mathcal{I}_2^{(n)}$ will be larger as $\beta^{(\text{s})}$ increases (as $\delta_n^{(\text{s})}$ decreases), while their cardinality remains the same for a given $\rho_{\text{R}}$. Furthermore, notice that $\delta_n^{(\text{s})}$ also defines $\mathcal{F}^{(n)} = \mathcal{H}_{X|Y_1}^{(n)} = \{ j \in [n] : Z\big( U(j) \big| U^{1:j-1}, Y_1^n \big) > 1-\delta_n^{(\text{s})} \}$. Thus, the larger is the value of $\beta^{(\text{s})}$ (the lower is $\delta_n^{(\text{s})}$), the smaller is the cardinality of $\mathcal{F}^{(n)}$ and the higher are the Bhattacharyya parameters associated with the eavesdroppers that belong to this set.

Figure~\ref{fig:rel} plots the upper-bound on the average bit error probability at the legitimate Receiver 1 defined in Equation~\eqref{eq:r1_Pb} as a function of the blocklength $n$ for different values of $\beta^{(\text{r})}$ (which defines a~particular $\delta_n^{(\text{r})}$ for each $n$). For this figure, we set $\beta^{(\text{s})} = 0.30$ and $\rho_{\text{R}} = 0.90$. As can be seen in Figure~\ref{fig:rel}, the higher is the value of $\beta^{(\text{r})}$ (the smaller is the value of $\delta_n^{(\text{r})}$), the better is the reliability performance of the polar code. This is because $\delta_n^{(\text{r})}$ defines the higher Bhattacharyya parameter associated with the legitimate Receiver 1 whose corresponding index will belong to the set $\mathcal{L}_{X|Y_1}^{(n)}$ (recall that this set contains the indices of those entries that the legitimate receivers have to estimate). Hence, it is clear that the upper-bound in Equation~\eqref{eq:r1_Pb} is decreasing as $\delta_n^{(\text{r})}$ decreases (as $\beta^{(\text{r})}$ increases). Moreover, as~we have proven in Section~\ref{sec:m2_rel}, we~can see that the reliability performance is always improving as $n$~increases.

Finally, how the values of the pair $(\beta^{(\text{r})},\beta^{(\text{s})})$, or equivalently, the values of $(\delta_n^{(\text{r})},\delta_n^{(\text{s})})$, impact~the rate of the additional secret sequence $\Phi$ given in Equation~\eqref{eq:r1_sm} is represented graphically in Figure~\ref{fig:rate}. In Figure~\ref{fig:rate}A, we set $\rho_{\text{R}} = 0.90$ and $\beta^{(\text{r})} = 0.16$, and we represent the rate of $\Phi$ as a function of the blocklength $n$ for different values of $\beta^{(\text{s})}$. Otherwise, in Figure~\ref{fig:rate}B, we evaluate the rate of $\Phi$ as a function of $n$ for different values of $\beta^{(\text{r})}$ when $\rho_{\text{R}} = 0.90$ and $\beta^{(\text{s})} = 0.30$. As mentioned in Section~\ref{sec:PCS_dbcnldls_encoder}, this~rate tends to be negligible for sufficiently large $n$. Moreover, according to the polar code construction proposed previously, for a fixed $n$, the cardinality of the set $\big( \mathcal{H}_{X|Y_1}^{(n)} \big)^{\text{C}} \cap \big( \mathcal{L}_{X|Y_1}^{(n)} \big)^{\text{C}}$ will be higher for larger values of $(\beta^{(\text{r})},\beta^{(\text{s})})$, or equivalently, smaller values of $(\delta_n^{(\text{r})},\delta_n^{(\text{s})})$. Therefore, as can be seen in Figure~\ref{fig:rate}, it is clear that higher values of $(\beta^{(\text{r})},\beta^{(\text{s})})$ mean also higher rate of the additional secret~sequence.

 \begin{figure}[H]
 \centering
 \resizebox{0.65\linewidth}{!}{%
%
%
\begin{tikzpicture}

\begin{axis}[%
		width=3in,
	    height=2.2in,
		scale only axis,
		xmin=9,
		xmax=21,
		xlabel style={font=\color{white!15!black}},
		xlabel={$log_2 n$},
		ymode=log,
		ymin=1e-12,
		ymax=0.01,
		yminorticks=true,
		ylabel style={font=\color{white!15!black}},
		ylabel={Average bit error probability},
		axis background/.style={fill=white},
		xmajorgrids,
		ymajorgrids,
		yminorgrids,
		legend style = { column sep = 10pt, legend columns = -1, legend to name = grouplegend,}]
        ]
 \addplot [color=black, line width=1.0pt, mark size=2.8pt, mark=o, mark options={solid, black}]
  table[row sep=crcr]{%
8	0.00793327803170879\\
9	0.00572894894362756\\
10	0.00442328055394923\\
11	0.00313934424860633\\
12	0.00237560749694456\\
13	0.0016846251647066\\
14	0.00125044233860375\\
15	0.000902387100607056\\
16	0.000663531848143671\\
17	0.00048053607866167\\
18	0.000343817316518261\\
19	0.000246619412221892\\
20	0.000179198046041084\\
21	0.000118244344754574\\
22	8.02893529969574e-05\\
};
\addlegendentry{$\beta^{(\text{r})} = 0.08$}

\addplot [color=black, line width=1.0pt, mark size=2.8pt, mark=square, mark options={solid, black}]
  table[row sep=crcr]{%
8	0.0036992432371179\\
9	0.0024544263797993\\
10	0.00177216658542016\\
11	0.00109038162552659\\
12	0.000723373929196064\\
13	0.000407904724274458\\
14	0.000216710057998383\\
15	0.000121358605958118\\
16	6.51842146327535e-05\\
17	3.18838055632279e-05\\
18	1.59848153828603e-05\\
19	7.32688962277638e-06\\
20	3.18113267757438e-06\\
21	7.95537631855016e-07\\
22	2.02684612628652e-07\\
};
\addlegendentry{$\beta^{(\text{r})} = 0.16$}

\addplot [color=black, line width=1.0pt, mark size=2.8pt, mark=triangle, mark options={solid, black}]
  table[row sep=crcr]{%
8	0.00103568323207606\\
9	0.000543498346791514\\
10	0.000341805062356229\\
11	0.000135441436785102\\
12	5.10843787854046e-05\\
13	1.663581653856e-05\\
14	4.291617432542e-06\\
15	9.31646058121652e-07\\
16	1.58629947861627e-07\\
17	2.15187317444772e-08\\
18	1.88864370229715e-09\\
19	1.17083542472591e-10\\
20	4.84282603356101e-12\\
21	2.07571802675318e-15\\
22	5.45901277629049e-20\\
};
\addlegendentry{$\beta^{(\text{r})} = 0.26$}

\addplot [color=black, line width=1.0pt, mark size=2.8pt, mark=asterisk, mark options={solid, black}]
  table[row sep=crcr]{%
8	0.000156003441341452\\
9	5.97125995312579e-05\\
10	3.58655436170883e-05\\
11	4.66487262296268e-06\\
12	4.96784924298228e-07\\
13	3.80012511338222e-08\\
14	1.3214019296735e-09\\
15	1.73528002878106e-11\\
16	7.86354498947004e-14\\
17	1.00234970465034e-16\\
18	1.76064550121254e-20\\
19	3.30703106170565e-25\\
20	2.95710544512746e-31\\
21	1.07944136971605e-48\\
22	3.74714710568221e-77\\
};
\addlegendentry{$\beta^{(\text{r})} = 0.36$}

\end{axis}

\node at ($(3.5,-1.6)$) {\ref{grouplegend}};
\end{tikzpicture}%
 }
 \caption{Reliability performance of the polar coding scheme for DBC-NLD-LS over BE-BC as a~function of $n$ and $\beta^{(\text{r})}$, which defines $\delta_n^{(\text{r})}$ for each $n$, when we set $\beta^{(\text{s})}=0.30$ and $\rho_{\text{R}} = 0.90$. That is, the bound $P_{\text{b}}^{\text{ub}(1)}$ on the average bit error probability at the legitimate Receiver 1 is defined as in Equation~\eqref{eq:r1_Pb}.} \label{fig:rel}
 \end{figure}
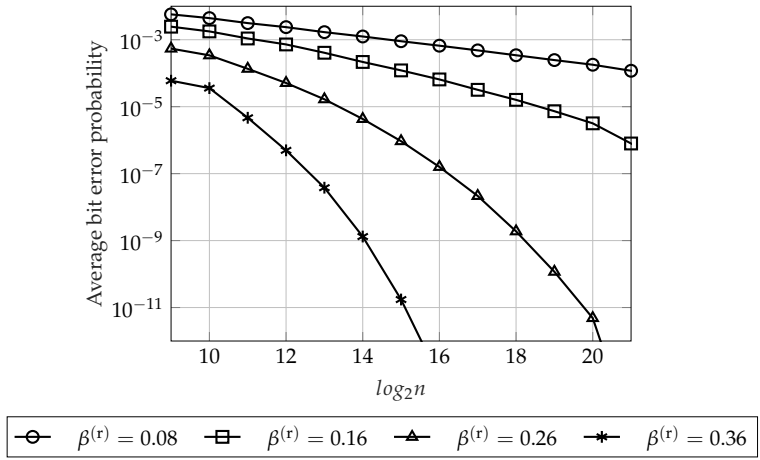\unskip

 \begin{figure}[H]
 \centering
 \resizebox{\linewidth}{!}{%
 \begin{tikzpicture}

\begin{groupplot}[
group style={
group name=myplot,
group size= 2 by 1,
horizontal sep=1in
},
]
        
\nextgroupplot[
		title={A},
        width=3in,
	    height=2.2in,
		scale only axis,
		xmin=9,
		xmax=21,
		xlabel style={font=\color{white!15!black}},
		xlabel={$log_2 n$},
		yminorticks=true,
		y tick label style={
        /pgf/number format/.cd,
            fixed,
            fixed zerofill,
            precision=2,
        /tikz/.cd
    	},
    	scaled y ticks=false,
		ylabel style={font=\color{white!15!black}},
		ylabel={Rate $\frac{1}{n} | \Phi | $},
		axis background/.style={fill=white},
		xmajorgrids,
		ymajorgrids,
		yminorgrids,
		legend style={legend cell align=left, align=left, draw=white!15!black}
        ]
       \addplot [color=black, line width=1.0pt, mark size=2.8pt, mark=o, mark options={solid, black}]
  	    table[row sep=crcr]{%
8	0.03125\\
9	0.0296223958333333\\
10	0.0279296875\\
11	0.0261962890625\\
12	0.02401123046875\\
13	0.0224853515625\\
14	0.0207244873046875\\
15	0.0186386108398438\\
16	0.016802978515625\\
17	0.0151203155517578\\
18	0.0135017395019531\\
19	0.012067699432373\\
20	0.0107521057128906\\
21	0.00948143005371094\\
22	0.00838327407836914\\
};
\addlegendentry{$\beta^{(\text{s})} = 0.10$}

\addplot [color=black, line width=1.0pt, mark size=2.8pt, mark=square, mark options={solid, black}]
  table[row sep=crcr]{%
8	0.04296875\\
9	0.0400390625\\
10	0.037646484375\\
11	0.0350830078125\\
12	0.03240966796875\\
13	0.0303466796875\\
14	0.0279937744140625\\
15	0.0256729125976563\\
16	0.0235191345214844\\
17	0.0215003967285156\\
18	0.0195175170898438\\
19	0.017710018157959\\
20	0.0160303592681885\\
21	0.0144046942392985\\
22	0.0129563808441162\\
};
\addlegendentry{$\beta^{(\text{s})} = 0.20$}

\addplot [color=black, line width=1.0pt, mark size=2.8pt, mark=triangle, mark options={solid, black}]
  table[row sep=crcr]{%
8	0.046875\\
9	0.0475260416666667\\
10	0.0453125\\
11	0.04375\\
12	0.04093017578125\\
13	0.03885498046875\\
14	0.0367767333984375\\
15	0.0344894409179687\\
16	0.0322975158691406\\
17	0.0301864624023437\\
18	0.0278848648071289\\
19	0.025479793548584\\
20	0.022950553894043\\
21	0.0201743443806966\\
22	0.0174212455749512\\
};
\addlegendentry{$\beta^{(\text{s})} = 0.30$}


\nextgroupplot[
		title ={B},
        width=3in,
		height=2.2in,
		scale only axis,
		xmin=9,
		xmax=21,
		xlabel style={font=\color{white!15!black}},
		xlabel={$log_2 n$},
		yminorticks=true,
		y tick label style={
        /pgf/number format/.cd,
            fixed,
            fixed zerofill,
            precision=2,
        /tikz/.cd
    	},
		ylabel style={font=\color{white!15!black}},
		ylabel={Rate $\frac{1}{n} | \Phi | $},
		axis background/.style={fill=white},
		xmajorgrids,
		ymajorgrids,
		yminorgrids,
		legend style={legend cell align=left, align=left, draw=white!15!black}
		]
\addplot [color=black, line width=1.0pt, mark size=2.8pt, mark=o, mark options={solid, black}]
  table[row sep=crcr]{%
8	0.03125\\
9	0.0338541666666667\\
10	0.0330078125\\
11	0.03271484375\\
12	0.030908203125\\
13	0.02955322265625\\
14	0.0280029296875\\
15	0.0265823364257813\\
16	0.0250625610351563\\
17	0.0236179351806641\\
18	0.0219841003417969\\
19	0.0201476097106934\\
20	0.0181309700012207\\
21	0.0158718427022298\\
22	0.013575553894043\\
};
\addlegendentry{$\beta^{(\text{r})} = 0.08$}

\addplot [color=black, line width=1pt, mark size=2.8pt, mark=square, mark options={solid, black}]
  table[row sep=crcr]{%
8	0.046875\\
9	0.0475260416666667\\
10	0.0453125\\
11	0.04375\\
12	0.04093017578125\\
13	0.03885498046875\\
14	0.0367767333984375\\
15	0.0344894409179687\\
16	0.0322975158691406\\
17	0.0301864624023437\\
18	0.0278848648071289\\
19	0.025479793548584\\
20	0.022950553894043\\
21	0.0201743443806966\\
22	0.0174212455749512\\
};
\addlegendentry{$\beta^{(\text{r})} = 0.16$}

\addplot [color=black, line width=1pt, mark size=2.8pt, mark=triangle, mark options={solid, black}]
  table[row sep=crcr]{%
8	0.0703125\\
9	0.0716145833333333\\
10	0.068798828125\\
11	0.0667724609375\\
12	0.06337890625\\
13	0.06005859375\\
14	0.05697021484375\\
15	0.0538314819335938\\
16	0.0506874084472656\\
17	0.0475334167480469\\
18	0.044281005859375\\
19	0.0409071922302246\\
20	0.0374085426330566\\
21	0.0337004661560059\\
22	0.0300455093383789\\
};
\addlegendentry{$\beta^{(\text{r})} = 0.26$}

\addplot [color=black, line width=1pt, mark size=2.8pt, mark=asterisk, mark options={solid, black}]
  table[row sep=crcr]{%
8	0.109375\\
9	0.112630208333333\\
10	0.111279296875\\
11	0.110595703125\\
12	0.10806884765625\\
13	0.1052734375\\
14	0.102349853515625\\
15	0.0991409301757813\\
16	0.0957374572753906\\
17	0.0922904968261719\\
18	0.0885183334350586\\
19	0.0846177101135254\\
20	0.0805491447448731\\
21	0.076146920522054\\
22	0.0717668533325195\\
};
\addlegendentry{$\beta^{(\text{r})} = 0.36$}

\end{groupplot}


\end{tikzpicture}
 }
 \caption{Rate of 
 the additional secret sequence $\Phi$ computed as in Equation~\eqref{eq:r1_sm} for DBC-NLD-LS over BE-BC as a function of the blocklength $n$ for different values of $(\beta^{(\text{r})},\beta^{(\text{s})})$, which defines $(\delta_n^{(\text{r})},\delta_n^{(\text{s})})$ for each $n$. {(\textbf{A})} 
Rate of $\Phi$ for different values of $\beta^{(\text{s})}$ when $\beta^{(\text{r})}=0.16$ and $\rho_{\text{R}} = 0.90$. {(\textbf{B})} Rate of $\Phi$ for different values of $\beta^{(\text{r})}$ when $\beta^{(\text{s})}=0.30$ and $\rho_{\text{R}} = 0.90$.} \label{fig:rate}
 \end{figure}
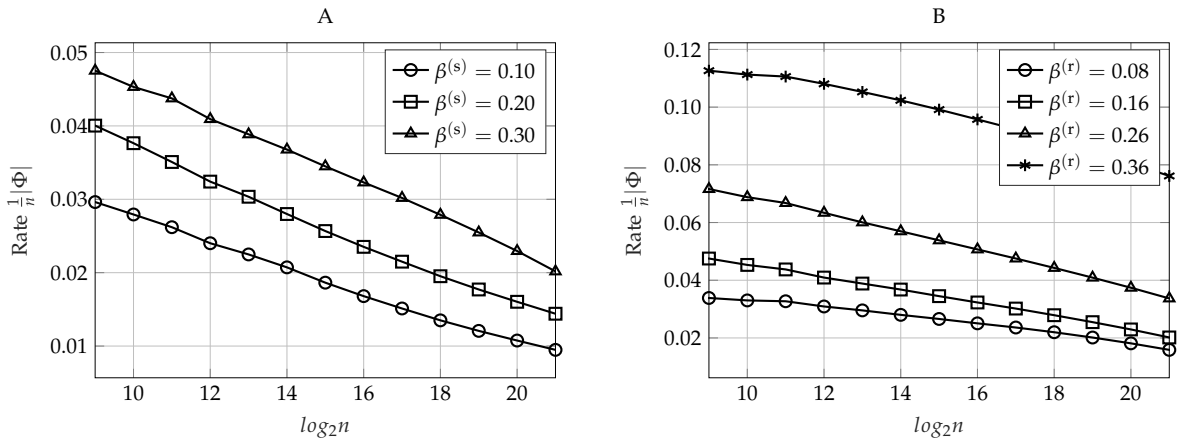

In conclusion, Figures~\ref{fig:sec1}--\ref{fig:rate} show that, for a particular value of the blocklength $n$, there is a trade-off between the reliability or the secrecy performance of the polar code and the length of the additional secret sequence $\Phi$, which can be controlled by the value of $\beta^{(\text{r})}$ or $\beta^{(\text{s})}$, respectively, in the polar code construction. Moreover, for sufficiently large $n$, the performance of the polar coding scheme always is improving as $n$ increases. Indeed, these figures show that we can transmit at rates very close to the capacity, providing good reliability and secrecy performance levels.

\subsection{DBC-LD-NLS}\label{sec:eval_dbcldnls}
For this model, we consider BS-BC with two legitimate receivers ($K=2$) and two eavesdroppers ($M=2$). Hence, each individual channel is a BSC where $\mathcal{X} = \mathcal{Y}_k = \mathcal{Z}_m = \{ 0 ,1 \}$, and~$k,m \in \{1,2 \}$. The individual channels are defined simply by their {crossover probability}, which is denoted by $\alpha_{Y_k}$ for the corresponding legitimate receiver $k$ ($\mathbb{P}[Y_k = 0 | X = 1] = \mathbb{P}[Y_k = 1 | X = 0] = \alpha_{Y_k}$) and $\alpha_{Z_m}$ for the corresponding eavesdropper $m$ ($\mathbb{P}[Z_m = 0 | X = 1] = \mathbb{P}[Z_m = 0 | X = 1] = \alpha_{Z_m}$). Due to the degradedness condition of the broadcast channel given in Equation~\eqref{eq:degch}, we have $\alpha_{Y_2} < \alpha_{Y_1} < \alpha_{Z_2} < \alpha_{Z_1}$. Due to the symmetry of the channel, it is easy to prove by using similar reasoning as in \cite{cover2012elements} (Ex.
~15.6.5) and by properly applying \cite{bloch2011physical} (Proposition 3.2) that the secrecy-capacity achieving distribution ${p}^{\star}_{V X}$ satisfies ${p}^{\star}_{V}(v) = {p}^{\star}_{X}(x) = \frac{1}{2}$ $\forall v,x \in \{0,1\}$, and consequently, $p^{\star}_{X|V}$ is symmetric. Thus, the distribution $p^{\star}_{X|V}$ can be characterized simply by the crossover probability $\alpha_{X|V} \triangleq p_{X|V}^{\star}(0 | 1) = p_{X|V}^{\star} (1 | 0)$, where $\alpha_{X|V} \in [0, \frac{1}{2}]$. Indeed, the overall rate in Proposition~\ref{prop:SCR_1} is maximized when $\alpha_{X|V} = \frac{1}{2}$, which implies that $R_1 = 0$. Then, by taking $\alpha_{X|V} < \frac{1}{2}$, we can {transfer} part of the rate associated with the message $W_2$ to the rate $R_1$, $R_2 = 0$ and $R_1$ being maximum if $\alpha_{X|V} = 0$. For~the simulations, we consider a BS-BC with $\alpha_{Y_2}=0.01$, $\alpha_{Y_1} = 0.04$, $\alpha_{Z_2}=0.2$ and $\alpha_{Z_1} = 0.35$. We set $\alpha_{X|V} = 0.1084$, which corresponds to the distribution that maximizes $\ln (R_1) + \ln (R_2)$ for this particular channel (proportional fair allocation). Thus, according to Corollary~\ref{coro:SCR_1}, the maximum achievable rates are $R_1^{\star} = 0.2507$ and $R_2^{\star} = 0.3254$. 

\subsubsection{Practical Polar Code Construction}\label{sec:pccm2}
Given the blocklength $n$ and the distribution $p^{\star}_{VXY_2Y_1Z_2Z_1} = p_{VX}^{\star} p_{Y_2Y_1Z_2Z_1|X}$, the goal of the polar code construction is to obtain the partition of the universal set $[n]$ defined in Equations~\eqref{eq:e1_setI}--\eqref{eq:e1_setT} and graphically represented in Figure~\ref{fig:pc_dbcldnls}. Hence, we need to define first the sets in Equations~\eqref{eq:pol1b1}--\eqref{eq:pol1c}, which means having to compute the entropy terms $\{H(U_1(j)|U_1^{1:j-1}) \}_{j=1}^n$, $\{H(U_1(j)|U_1^{1:j-1}, Y_1^n) \}_{j=1}^n$ and $\{H(U_1(j)|U_1^{1:j-1}, Z_2^n) \}_{j=1}^n$ associated with the polar transform $U_1^n = V^n G_n$ for the first superposition layer and $\{H(U_2(j)|U_2^{1:j-1}, V^n) \}_{j=1}^n$, $\{H(U_2(j)|U_2^{1:j-1}, V^n, Y_2^n) \}_{j=1}^n$ and $\{H(U_2(j)|U_2^{1:j-1}, V^n, Z_2^n) \}_{j=1}^n$ associated with the polar transform $U_2^n = X^n G_n$ for the second layer. In the following, we propose an adaptation of the Monte Carlo method \cite{vangala2015comparative} (PCC-1), which is based on the {butterfly algorithm} described in \cite{arikan2009channel} for SC decoding, to directly estimate these entropy terms. 

\vspace{0.3cm}
\noindent \textbf{Monte-Carlo method to estimate the entropy terms.} First, consider the entropy terms associated with to the first layer. As for the previous model, since ${p}^{\star}_{V}(v) = \frac{1}{2}$, we have $H(U_1(j)|U_1^{1:j-1}) = 1$ for all $j \in [n]$. In order to compute $\{H(U_1(j)|U_1^{1:j-1}, Y_k^n) \}_{j=1}^n$ and $\{H(U_1(j)|U_1^{1:j-1}, Z_m^n) \}_{j=1}^n$ for some $k,m \in \{1,2\}$, we run the Monte Carlo simulation as follows. First, due to the symmetry of the channel and the symmetry of $p^{\star}_{X|V}$, as in \cite{vangala2015comparative} (PCC-1), we can set $v^n = u_1^n = 0^n$ at each iteration. For the realization $\tau \in [1,N_{\tau}]$, $N_{\tau}$ being the number of realizations, we randomly generate $y_k^{n({\tau})}$ and $z_m^{n({\tau})}$ from $p^{\star}_{Y_k^n|V^n}$ and $p^{\star}_{Z_m^n|V^n}$, respectively ({by abuse of notation, we use $(\tau)$ in any sequence $a^{n({\tau})}$ to emphasize that it is generated at the iteration $\tau \in [1,N_{\tau}]$}). Next, we obtain the log-likelihood ratios $\{ L^{({\tau})}_{Y_k|V}(j) \}_{j=1}^n$ and $\{ L^{({\tau})}_{Z_m|V}(j) \}_{j=1}^n$ by using the algorithm \cite{vangala2015comparative} (PCC-1). For instance, consider $\{ L^{({\tau})}_{Y_k|V}(j) \}_{j=1}^n$. From the initial values $\{ p_{Y_k|V}^{\star}(y_k^{(\tau)}(j)|0) / p_{Y_k|V}^{\star}(y_k^{(\tau)}(j)|1) \}_{j=1}^n$, the algorithm recursively computes: 
\begin{align*}
L^{({\tau})}_{Y_k|V}(j) \triangleq \ln \frac{p^{\star}_{Y_k^n U_1^{1:j-1} |U_1(j)}(y_k^{n({\tau})}, 0^{j-1} | 0)}{p^{\star}_{Y_k^n U_1^{1:j-1}|U_1(j)}(y_k^{n({\tau})}, 0^{j-1} | 1)} \stackrel{(a)}{=} \frac{p^{\star}_{U_1(j) | U_1^{1:j-1} Y_k^n}(0 | 0^{j-1}, y_k^{n({\tau})})}{1 - p^{\star}_{U_1(j) | U_1^{1:j-1} Y_k^n}(0 | 0^{j-1} , y_k^{n({\tau})})} ,
\end{align*}
\textls[-15]{for all $j \in [n]$, where $(a)$ follows from the fact that $p_{U_1(j)}^{\star}(0) = p_{U_1(j)}^{\star}(1) = \frac{1}{2}$ because $H(U_1(j)|U_1^{1:j-1}) = 1$} for all $j \in [n]$. Hence, we can obtain $p^{\star}_{U_1(j) | U_1^{1:j-1} Y_k^n}(0 | 0^{j-1} , y_k^{n(\tau)})$ from $L^{({\tau})}_{Y_k|V}(j)$, and since: 
\begin{align*}
H(U_1(j) | U_1^{1:j-1}, Y_k^n) = \mathbb{E}_{U_1^{1:j-1} Y_k^n} \left[ h_2 \left( p^{\star}_{U_1(j) | U_1^{1:j-1} Y_k^n}(0 | u_1^{1:j-1}, y_k^n) \right) \right],
\end{align*}
after $N_{\tau}$ realizations, we can estimate $H(U_1(j) | U_1^{1:j-1}, Y_k^n)$ by computing the empirical mean, that is,
 \begin{align*}
H(U_1(j) | U_1^{1:j-1}, Y_k^n) \approx \frac{1}{N_{\text{r}}} \sum_{\tau = 1}^{N_{\tau}} h_2 \left( p^{\star}_{U_1(j) | U_1^{1:j-1} Y_k^n}(0 | 0^{j-1}, y_k^{n({\tau})}) \right).
\end{align*}

Now, consider the Monte Carlo method to estimate $\{H(U_2(j)|U_2^{1:j-1}, V^n) \}_{j=1}^n$, $\{H(U_2(j)|U_2^{1:j-1}, V^n, Y_k^n) \}_{j=1}^n$ and $\{H(U_2(j)|U_2^{1:j-1}, V^n, Z_m^n) \}_{j=1}^n$ for any $k,m \in \{1,2\}$ associated with the second layer. To obtain $\{H(U_2(j)|U_2^{1:j-1}, V^n) \}_{j=1}^n$, we can see $X$ and $V$ as the input and output random variables, respectively, of a symmetric channel with distribution $p^{\star}_{V|X}$. Now, although $p^{\star}_{X}$ is uniform and, consequently, $H(U_2(j)|U_2^{1:j-1}) = 1$ for all $j \in [n]$, notice that $\mathcal{H}_{X|V}^{(n)} \neq [n]$ and $\mathcal{T}_1^{(n)} \neq \emptyset$ because $\mathcal{H}_{X|V}^{(n)}$ and its complementary set depend on $p^{\star}_{X|V}$. On the other hand, to obtain $\{H(U_2(j)|U_2^{1:j-1}, V^n, Y_k^n) \}_{j=1}^n$ or $\{H(U_2(j)|U_2^{1:j-1}, V^n, Z_m^n) \}_{j=1}^n$, we can see $(V,Y_k)$ or $(V,Z_m)$ as the output of a symmetric channel with distribution $p^{\star}_{VY_k|X}$ or $p^{\star}_{VZ_m|X}$, respectively, where notice that $p^{\star}_{VY_k|X} = p^{\star}_{V|X}p^{\star}_{Y_k|X}$ and $p^{\star}_{VZ_m|X} = p^{\star}_{V|X}p^{\star}_{Z_m|X}$ because $V-X-Y_k - Z_m$ forms a Markov chain. Hence, due to the symmetry of the previous distributions, we can set $x^n = u_2^n = 0^n$ at each iteration. Then, for the realization $\tau \in [1,N_{\tau}]$, we draw $v^{n({\tau})}$, $y_k^{n({\tau})}$ and $z_m^{n({\tau})}$ from the distributions $p^{\star}_{V^n|X^n}$, $p_{Y_k^n|X^n}$ and $p_{Z_m^n|X^n}$, respectively. Next, we obtain the log-likelihood ratios $\{L_{V|X}^{(\tau)} (j) \}_{j=1}^n$, $\{L_{VY_k|X}^{(\tau)} (j) \}_{j=1}^n$ and $\{L_{VZ_m|X}^{(\tau)} (j) \}_{j=1}^n$ by using \cite{vangala2015comparative} (PCC-1). Since $H(U_2(j)|U_2^{1:j-1}) = 1$ for all $j \in [n]$, we have $p_{U_2(j)}^{\star}(u) = \frac{1}{2}$ for all $u \in \{0,1\}$, and we can compute $p^{\star}_{U_2(j) | U_2^{1:j-1} V^n}(0 | 0^{j-1} , v^{n({\tau})})$, $p^{\star}_{U_2(j) | U_2^{1:j-1} V^n Y_k^n}(0 | 0^{j-1} , v^{n({\tau})}, y_k^{n({\tau})})$ and $p^{\star}_{U_2(j) | U_2^{1:j-1} V^n Z_m^n}(0 | 0^{j-1} , v^{n({\tau})}, z_m^{n({\tau})})$ from the corresponding log-likelihood ratios. Finally, after $N_{\tau}$ realizations, we can estimate the corresponding entropy terms by computing the empirical mean.

\vspace{0.3cm}
\noindent \textbf{Partition of the universal set} $\bm{[} \mathbf{n} \bm{]}$. In order to provide more flexibility on the design, now we introduce $(\delta_n^{(1,\text{r})}, \delta_n^{(1,\text{s})})$ for the first layer, where $\delta_n^{(1,\text{r})} \triangleq 2^{-n^{\beta^{(1,\text{r})}}}$ and $\delta_n^{(1,\text{s})} \triangleq 2^{-n^{\beta^{(1,\text{s})}}}$ for some $\beta^{(1,\text{r})},\beta^{(1,\text{s})} \in (0,\frac{1}{2})$. For the second layer, we introduce $(\delta_n^{(2,\text{r})}, \delta_n^{(2,\text{s})})$ and $(\delta_n^{(2,\text{L})}, \delta_n^{(2,\text{H})})$, where $\delta_n^{(2,\text{r})} \triangleq 2^{-n^{\beta^{(2,\text{r})}}}$, $\delta_n^{(2,\text{s})} \triangleq 2^{-n^{\beta^{(2,\text{s})}}}$, $\delta_n^{(2,\text{L})} \triangleq 2^{-n^{\beta^{(2,\text{L})}}}$ and $\delta_n^{(2,\text{H})} \triangleq 2^{-n^{\beta^{(2,\text{H})}}}$ for some $\beta^{(2,\text{r})},\beta^{(2,\text{s})},\beta^{(2,\text{L})},\beta^{(2,\text{H})} \in (0,\frac{1}{2})$. 

\textls[-15]{Consider the partition of $[n]$ for the first layer ($\ell = 1$ in Equations~\eqref{eq:e1_setI}--\eqref{eq:e1_setT}). As mentioned previously, since ${p}^{\star}_{V}(v) = \frac{1}{2}$, we have $\mathcal{H}_V^{(n)} = [n]$ and $\mathcal{T}_1^{(n)} = \emptyset$. Let $R^{\prime}_{1} \in [0,R_{1}^{\star}]$ denote the target rate corresponding to the message $W_1$ that the polar coding scheme must approach. We obtain the partition in Equations~\eqref{eq:e1_setI}--\eqref{eq:e1_setT} as follows. First, we define $( \mathcal{H}_{V|Y_1}^{(n)} )^{\text{C}} \triangleq \{ j \in [n]: H ( U_1(j) | U_1^{1:j-1}, Y_1^n ) \leq 1 - \delta_n^{(1,\text{s})} \}$.} Then, we choose $\mathcal{I}_1^{(n)}$ by taking the $\ceil{n R_1^{\prime}}$ indices $j \in ( \mathcal{H}_{V|Y_1}^{(n)} )^{\text{C}}$ that correspond to the highest entropy terms $\{H(U_1(j)|U_1^{1:j-1}, Z_2^n) \}_{j=1}^n$ associated with Eavesdropper 2. Notice that $\delta_n^{(1,\text{s})}$ must guarantee $| ( \mathcal{H}_{V|Y_1}^{(n)} )^{\text{C}} | \leq R^{\prime}_{1}$. Finally, we obtain $\mathcal{C}_1^{(n)} = ( \mathcal{H}_{V|Y_1}^{(n)} )^{\text{C}} \setminus \mathcal{I}_1^{(n)}$ and $\mathcal{F}_1^{(n)} = \mathcal{H}_{V|Y_1}^{(n)}$. Furthermore, in order to evaluate the reliability performance, we define $\mathcal{L}_{V|Y_1}^{(n)} \triangleq \{ j \in [n]: H ( U_1(j) | U_1^{1:j-1}, Y_1^n ) \leq\delta_n^{(1,\text{r})} \}$. 

Consider the partition of $[n]$ for the second layer ($\ell = 2$ in Equations~\eqref{eq:e1_setI}--\eqref{eq:e1_setT}). Since~$\mathcal{H}_{X|V}^{(n)} \neq [n]$ and $\mathcal{T}_1^{(n)} \neq \emptyset$, we define $\mathcal{H}_{X|V}^{(n)} \triangleq \{ j \in [n]: H ( U_2(j) | U_2^{1:j-1}, V^n ) \geq 1- \delta_n^{(2,\text{H})} \}$ and $\mathcal{L}_{X|V}^{(n)} \triangleq \{ j \in [n]: H ( U_2(j) | U_2^{1:j-1}, V^n ) \leq \delta_n^{(2,\text{L})} \}$, where we have used $\delta_n^{(2,\text{H})}$ and $\delta_n^{(2,\text{L})}$, respectively. Let $R^{\prime}_{2} \in [0,R_{2}^{\star}]$ denote the target rate corresponding to $W_2$. We define $( \mathcal{H}_{X|VY_2}^{(n)} )^{\text{C}} \triangleq \{ j \in \mathcal{H}_{X|V}^{(n)}: H ( U_2(j) | U_2^{1:j-1}, V^n, Y_2^n ) \leq 1- \delta_n^{(2,\text{s})} \}$. Then, we choose $\mathcal{I}_2^{(n)}$ by taking the $\ceil{n R_2^{\prime}}$ indices $j \in ( \mathcal{H}_{X|VY_2}^{(n)} )^{\text{C}}$ that correspond to the highest entropy terms $\{ H ( U_2(j) | U_2^{1:j-1}, V^n, Z_2^n ) \}_{j=1}^n$ associated with Eavesdropper 2. Thus, notice that $\delta_n^{(2,\text{H})}$ and $\delta_n^{(2,\text{s})}$ must guarantee $| \mathcal{H}_{X|V}^{(n)} | \geq | ( \mathcal{H}_{X|VY_2}^{(n)} )^{\text{C}} | \geq R^{\prime}_{2}$. Then, we obtain $\mathcal{C}_2^{(n)} = ( \mathcal{H}_{X|VY_2}^{(n)} )^{\text{C}} \setminus \mathcal{I}_2^{(n)}$ and $\mathcal{F}_2^{(n)} = \mathcal{H}_{X|VY_2}^{(n)}$. Finally, in order to evaluate the reliability performance, we define $\mathcal{L}_{X|VY_2}^{(n)} \triangleq \{ j \in [n]: H ( U_2(j) | U_2^{1:j-1}, V^n, Y_2^n ) \leq \delta_n^{(2,\text{r})} \}$. 

\subsubsection{Performance Evaluation}
First, notice that the encoding at the first layer induces a distribution $\tilde{q}_{V^n} = p_{V^n}$.~For~the second layer, the entries $U[\mathcal{H}_{X|V}^{(n)}]$ of the original DMS only are {almost} independent of $V^n$ because $H ( U_2(j) | U_2^{1:j-1}, V^n ) \leq 1-\delta_n^{(2,\text{s})}$ for $j \in \mathcal{H}_{X|V}^{(n)}$.~Nevertheless, the encoding will construct $\tilde{U}_2[\mathcal{H}_{X|V}^{(n)}]$ by storing uniformly-distributed sequences that are totally independent of $V^n$.~On~the other hand, since $\mathcal{L}_{X|V}^{(n)} \subseteq \mathcal{T}_2^{(n)} \neq \emptyset$, the encoder will use the deterministic SC encoding in Equation~\eqref{eq:argmax1} to construct $\tilde{U}_2[\mathcal{L}_{X|V}^{(n)}]$.~Therefore, according to Lemma~\ref{lemma:distDMS_1} and Remark~\ref{remark:TV1}, we~will have $\mathbb{V}(\tilde{q}_{V^nX^nY_2^nY_1^nZ_2^nZ_1^n},p^{\star}_{V^nX^nY_2^nY_1^nZ_2^nZ_1^n}) \neq 0$ for finite $n$.~Since, as seen in Section~\ref{sec:performance1}, this~total variation distance impacts the performance, we obtain first an upper-bound $d_{\text{TV}}^{\text{ub}}$ on $\mathbb{V}(\tilde{q}_{V^nX^nY_2^nY_1^nZ_2^nZ_1^n},p^{\star}_{V^nX^nY_2^nY_1^nZ_2^nZ_1^n})$, which is defined as: 
\begin{align*}
d_{\text{TV}}^{\text{ub}} \triangleq d_{\text{TV}}^{\text{ub(L)}} + d_{\text{TV}}^{\text{ub(H)}},
\end{align*}
where $d_{\text{TV}}^{\text{ub(L)}}$ will measure the impact of using the deterministic SC encoding in Equation~\eqref{eq:argmax1} for the entries $\tilde{U}_2 \big[ \mathcal{L}_{X|V}^{(n)} \big]$, and $d_{\text{TV}}^{\text{ub(H)}}$ is the contribution on the total variation distance of storing uniformly-distributed random sequences into $\tilde{U}_2 \big[ \mathcal{H}_{X|V}^{(n)} \big]$ that are totally independent of $V^n$. 

Consider $d_{\text{TV}}^{\text{ub(L)}}$, which corresponds to the analytic bound found in Lemma~\ref{lemma:distU1cU2}.~For the simulations, we can use the Monte Carlo method to directly estimate Equation~\eqref{eq:comb1} by computing the empirical~mean, 
\begin{align}
d_{\text{TV}}^{\text{ub(L)}} \triangleq \frac{1}{N_{\tau^{\prime}}} \sum_{\tau^{\prime} = 1}^{N_{\tau^{\prime}}} \Bigg[ \sum_{j \in \mathcal{L}_{X|V}^{(n)}} \Bigg( 1 - p^{\star}_{U_{2}(j) | U_{2}^{1:j-1}V^n} \Big( u_{2}^{\ast}(j) \Big| \check{u}_{2}^{1:j-1(\tau^{\prime})}, \check{v}^{n({\tau^{\prime}})} \Big) \Bigg) \Bigg] , \label{eq:m1_TVL}
\end{align}
where $(\check{v}^{n(\tau^{\prime})}, \check{u}_{2}^{n(\tau^{\prime})})$ must be drawn at each iteration $\tau^{\prime} \in [1,N_{\tau^{\prime}}]$ according to Equation~\eqref{eq:dist_check}, $\mathcal{L}_{X|V}^{(n)}$ has been obtained previously in the polar code construction and, according to Equation~\eqref{eq:comb1}, $u_{2}^{\ast}(j) \triangleq \argmax_{u \in \{0,1\} } p^{\star}_{U_{2}(j) | U_{2}^{1:j-1}V^n} ( u | \check{u}_{2}^{1:j-1(\tau^{\prime})}, \check{v}^{n(\tau^{\prime})} )$. Due to the symmetry of $p^{\star}_{V|X}$, the probabilities $p^{\star}_{U_{2}(j) | U_{2}^{1:j-1}V^n}$ can be obtained with low complexity using the {butterfly} algorithm described in \cite{arikan2009channel}.

Consider now $d_{\text{TV}}^{\text{ub(H)}}$, which corresponds to the analytic bound found in Lemma~\ref{lemma:distUc1Uc2}.~We~can compute exactly the Kullback-Leibler divergence as in Equation~\eqref{eq:kldist} by using the corresponding entropy terms obtained in the polar code construction. Thus, by applying Pinsker's inequality, we~have:
\begin{align}
d_{\text{TV}}^{\text{ub}(\text{H})} \triangleq \Bigg( 2 \ln 2 \sum_{j \in \mathcal{H}_{X|V}^{(n)}} \Big( 1 - H \Big( U_{2}(j) \Big| U_{2}^{1:j-1}, V^{n} \Big) \Big) \Bigg)^{1/2}. \label{eq:m1_TVH}
\end{align}
According to the polar code construction, $| \mathcal{L}_{X|V}^{(n)} |$ and $| \mathcal{H}_{X|V}^{(n)} |$ will depend only on the values of $\delta_n^{(2,\text{L})}$ and $\delta_n^{(2,\text{H})}$, respectively, for a particular $n$. Hence, the value of $d_{\text{TV}}^{\text{ub}}$ can be controlled by adjusting $(\beta^{(2,\text{L})},\beta^{(2,\text{H})})$. It is clear that higher values of $(\beta^{(2,\text{L})},\beta^{(2,\text{H})})$ mean lower cardinalities of the sets $\mathcal{L}_{X|V}^{(n)}$ and $\mathcal{H}_{X|V}^{(n)}$ and, consequently, lower $d_{\text{TV}}^{\text{ub}}$. However, $| (\mathcal{H}_{X|V}^{(n)})^{\text{C}} \cap (\mathcal{L}_{X|V}^{(n)})^{\text{C}} |$ increases with $(\beta^{(2,\text{L})},\beta^{(2,\text{H})})$, and the encoder in Equation~\eqref{eq:distrenc2} requires more randomness to form \mbox{$\tilde{U}_2 [(\mathcal{H}_{X|V}^{(n)})^{\text{C}} \cap (\mathcal{L}_{X|V}^{(n)})^{\text{C}} ]$}.

To evaluate the reliability performance, we obtain the upper-bounds $P_{\text{b}}^{\text{ub}(1)}$ and $P_{\text{b}}^{\text{ub}(2)}$ on the average bit error probability at Receivers 1 and 2, respectively. From Equations~\eqref{eq:reliability1}~and~\eqref{eq:reliability1b} and by applying \cite{arikan2010source} (Proposition 2) to upper-bound the Bhattacharyya parameters from the entropy terms, we have:
\begin{align}
P_{\text{b}}^{\text{ub}(1)} & \triangleq d_{\text{TV}}^{\text{ub}} + \frac{1}{\big| \mathcal{L}_{V|Y_1}^{(n)} \big|} \sum_{j \in \mathcal{L}_{V|Y_1}^{(n)}} \! \! \! \! \sqrt{H \big({U}_1(j) \big| U_1^{1:j-1}, {Y}_1^n \big)}, \label{eq:2_r1_Pb_1} \\
P_{\text{b}}^{\text{ub}(2)} \! \! & \triangleq 2 d_{\text{TV}}^{\text{ub}} \! + \! \frac{2}{\big| \mathcal{L}_{V|Y_1}^{(n)} \big|} \! \sum_{j \in \mathcal{L}_{V|Y_1}^{(n)}} \! \! \! \! \! \! \sqrt{H \big({U}_1(j) \big| U_1^{1:j-1}, {Y}_2^n \big)} + \frac{1}{\big| \mathcal{L}_{X|VY_2}^{(n)} \big|} \! \sum_{j \in \mathcal{L}_{X|VY_2}^{(n)}} \! \! \! \! \! \! \sqrt{H \big({U}_2(j) \big| U_2^{1:j-1}, V^n, {Y}_2^n \big)}. \label{eq:2_r1_Pb_2} 
\end{align}

To evaluate the secrecy performance, we compute an upper-bound $I^{\text{ub}}(W_1, W_2; F_1,F_2, \tilde{Z}_{2}^n)$ on the information leakage $I(W_1,W_2; F_1,F_2,\tilde{Z}_2^n)$ for Eavesdropper 2. From Equation~\eqref{eq:leakage1c} we obtain:
\begin{align}
I^{\text{ub}}(W_1, W_2; F_1,F_2, \tilde{Z}_{2}^n) & \triangleq 4 n d_{\text{TV}}^{\text{ub}} - 2 d_{\text{TV}}^{\text{ub}} \log d_{\text{TV}}^{\text{ub}} + \sum_{\ell=1}^2 \big| \mathcal{I}_{\ell}^{(n)} \cup \mathcal{F}_{\ell}^{(n)} \big| \nonumber \\
& \quad - \! \! \! \! \! \! \sum_{j \in \mathcal{I}_{1}^{(n)} \cup \mathcal{F}_1^{(n)}} H \big({U}_1(j) \big| U_1^{1:j-1}, {Z}_2^n \big) - \! \! \! \! \! \! \sum_{j \in \mathcal{I}_{2}^{(n)} \cup \mathcal{F}_2^{(n)}} H \big({U}_2(j) \big| U_2^{1:j-1}, V^n , {Z}_2^n \big), \label{eq:2_r1_I} 
\end{align}
Due to the degradedness condition of BS-BC and, consequently, by Lemma~\ref{lemma:subsetproperty}, the information leakage at Eavesdropper 1 will be always less than the one at Eavesdropper 2. 

Finally, we evaluate the overall rate of the additional sequences $\{ \Phi_1, \Phi_2 \}$ by computing:
\begin{align}
\frac{1}{n} \big( |\Phi_1| + |\Phi_2| \big) = \frac{1}{n} \Big( \Big| \big( \mathcal{H}_{V|Y_1}^{(n)} \big)^{\text{C}} \cap \big( \mathcal{L}_{V|Y_1}^{(n)} \big)^{\text{C}} \Big| + \Big| \big( \mathcal{H}_{X|VY_2}^{(n)} \big)^{\text{C}} \cap \big( \mathcal{L}_{X|VY_2}^{(n)} \big)^{\text{C}} \Big| \Big). \label{eq:2_r1_R}
\end{align}

\begin{figure}[H]
 \centering
 \resizebox{\linewidth}{!}{%
 \begin{tikzpicture}

\begin{groupplot}[
group style={
group name=myplot,
group size= 2 by 2,
horizontal sep=1in,
vertical sep=1in
},
]
        
\nextgroupplot[
		title={A},
        width=3in,
	    height=2.2in,
		scale only axis,
		xmin=7,
		xmax=16,
		xlabel style={font=\color{white!15!black}},
		xlabel={$log_2 n$},
		ymode=log,
		ymin=1e-8,
		ymax=10,
		yminorticks=true,
		ylabel style={font=\color{white!15!black}},
		ylabel={$I_0^{\text{ub}}\left(W_1,W_2;F_1,F_2,Z_2^n \right)$},
		axis background/.style={fill=white},
		xmajorgrids,
		ymajorgrids,
		yminorgrids,
		legend style={legend cell align=left, align=left, draw=white!15!black, legend pos=south west}
        ]
       \addplot [color=black, line width=1.0pt, mark size=2.8pt, mark=o, mark options={solid, black}]
  	    table[row sep=crcr]{%
7	2.99659059044721\\
8	3.14279980792739\\
9	2.75018638414038\\
10	2.26411309081565\\
11	1.71028306394401\\
12	1.03095587135806\\
13	0.479198995301567\\
14	0.179517391850116\\
15	0.0525461031166939\\
16	0.010353\\
};
\addlegendentry{$\rho_{R} = 0.90$}

\addplot [color=black, line width=1.0pt, mark size=2.8pt, mark=square, mark options={solid, black}]
  table[row sep=crcr]{%
7	1.73667466556314\\
8	1.4652691232601\\
9	1.05430634667737\\
10	0.711864337973856\\
11	0.418273012922623\\
12	0.180220618881935\\
13	0.0399606312608285\\
14	0.00508463359296911\\
15	5.53736826987006e-05\\
16	2.96682992484421e-06\\
};
\addlegendentry{$\rho_{R} = 0.80$}

\addplot [color=black, line width=1.0pt, mark size=2.8pt, mark=triangle, mark options={solid, black}]
  table[row sep=crcr]{%
7	1.01664470582199\\
8	0.649286183315525\\
9	0.418156198872777\\
10	0.214996119675328\\
11	0.10054088986949\\
12	0.0287545673182194\\
13	0.00331630707810291\\
14	0.000170079570534654\\
15	5.83304426982068e-08\\
16	0\\
};
\addlegendentry{$\rho_{R} = 0.70$}


\nextgroupplot[
		title ={B},
        width=3in,
		height=2.2in,
		scale only axis,
		xmin=7,
		xmax=16,
		xlabel style={font=\color{white!15!black}},
		xlabel={$log_2 n$},
		ymode=log,
		ymin=3e-7,
		ymax=0.05,
		ylabel style={font=\color{white!15!black}},
		ylabel={Average bit error probability},
		axis background/.style={fill=white},
		xmajorgrids,
		ymajorgrids,
		]
\addplot [color=black, line width=1pt, mark size=2.8pt, mark=o, mark options={solid, black}]
  table[row sep=crcr]{%
7	0.0315715982604217\\
8	0.0198724929329397\\
9	0.013115954353868\\
10	0.00690355555846413\\
11	0.00311594351221345\\
12	0.00132915397933704\\
13	0.000458843065443816\\
14	0.000137947806745977\\
15	1.29424686683532e-05\\
16	6.62070131746584e-07\\
};
\addlegendentry{$P_{\text{b},0}^{\text{ub}(1)}$}

\addplot [color=black, line width=1pt, mark size=2.8pt, mark=square, mark options={solid, black}]
  table[row sep=crcr]{%
7	0.028389628517896\\
8	0.0173420855221836\\
9	0.011861210574974\\
10	0.00635328083638573\\
11	0.00350625997982382\\
12	0.00175591822825694\\
13	0.000718257280773492\\
14	0.000318937913868682\\
15	9.14676105834786e-05\\
16	2.54017970890577e-05\\
};
\addlegendentry{$P_{\text{b},0}^{\text{ub}(2)}$}


\nextgroupplot[
		title ={C},
        width=3in,
		height=2.2in,
		scale only axis,
		xmin=7,
		xmax=16,
		xlabel style={font=\color{white!15!black}},
		xlabel={$log_2 n$},
		ymin=0.06,
		ymax=0.16,
		ylabel style={font=\color{white!15!black}},
		ylabel={Overall rate $\frac{1}{n} ( | \Phi_{1} | + | \Phi_{2} | )$},
		axis background/.style={fill=white},
		xmajorgrids,
		ymajorgrids,
		tick label style={/pgf/number format/fixed}
		]
\addplot [color=black, line width=1.0pt, mark size=2.8pt, mark=o, mark options={solid, black}]
  table[row sep=crcr]{%
7	0.1484375\\
8	0.136067708333333\\
9	0.12978515625\\
10	0.12236328125\\
11	0.11455078125\\
12	0.10528564453125\\
13	0.09527587890625\\
14	0.0853668212890625\\
15	0.0752207438151042\\
16	0.0671539306640625\\
};


\nextgroupplot[
		title ={D},
        width=3in,
		height=2.2in,
		scale only axis,
		xmin=7,
		xmax=16,
		xlabel style={font=\color{white!15!black}},
		xlabel={$log_2 n$},
		ymode=log,
		ymin=1e-8,
		ymax=0.5,
		ylabel style={font=\color{white!15!black}},
		ylabel={Total variation distance},
		axis background/.style={fill=white},
		xmajorgrids,
		ymajorgrids,
		]
\addplot [color=black, line width=1.0pt, mark size=2.8pt, mark=o, mark options={solid, black}]
  table[row sep=crcr]{%
7	0.346577971941243\\
8	0.224680381833073\\
9	0.151637191535308\\
10	0.0835620088167886\\
11	0.0435896815369012\\
12	0.0182805633223594\\
13	0.00579650015562058\\
14	0.00145228791220794\\
15	3.09747719479062e-05\\
16	0\\
};
\addlegendentry{$d_{\text{TV}}^{\text{ub(H)}}$}

\addplot [color=black, line width=1.0pt, mark size=2.8pt, mark=square, mark options={solid, black}]
  table[row sep=crcr]{%
7	0.00737759901875788\\
8	0.00850779253961201\\
9	0.00614393068882758\\
10	0.00466844454967863\\
11	0.00248525601109265\\
12	0.00103929236013702\\
13	3.03072394869763e-05\\
14	3.71950766461707e-08\\
15	0\\
16	0\\
};
\addlegendentry{$d_{\text{TV}}^{\text{ub(L)}}$}

\end{groupplot}


\end{tikzpicture}
 }
 \caption{Performance 
{of the polar coding scheme for DBC-LD-NLS over BS-BC as a function of the blocklength $n$ when $\beta^{(1,\text{r})}=\beta^{(2,\text{r})} = 0.24$, $\beta^{(1,\text{s})} = 0.30$, $\beta^{(2,\text{s})} = 0.36$ and \mbox{$\beta^{(2,\text{H})} = \beta^{(2,\text{H})} = 0.36$.}} {(\textbf{A})} Upper-bound on the information about $(W_1,W_2)$ leaked to Eavesdropper 2 defined as in Equation~\eqref{eq:2_r1_I} for different normalized target rates $\rho_{\text{R}}$ when we set $d_{\text{TV}}^{\text{ub}} = 0$. {(\textbf{B})} Upper-bounds on the average error probability at legitimate Receivers 1 and 2 defined as in Equations~\eqref{eq:2_r1_Pb_1}~and~\eqref{eq:2_r1_Pb_2}, respectively, when $d_{\text{TV}}^{\text{ub}}=0$. {(\textbf{C})} Overall rate of the sequences $\{\Phi_1,\Phi_2\}$ computed as in Equation~\eqref{eq:2_r1_R}. {(\textbf{D})}~terms $d_{\text{TV}}^{\text{ub(H)}}$ and $d_{\text{TV}}^{\text{ub(L)}}$ that contribute to the bound on the total variation distance $d_{\text{TV}}^{\text{ub}}$ defined as in Equations~\eqref{eq:m1_TVL}~and~\eqref{eq:m1_TVH}, respectively.}\label{fig:mod2}
 \end{figure}
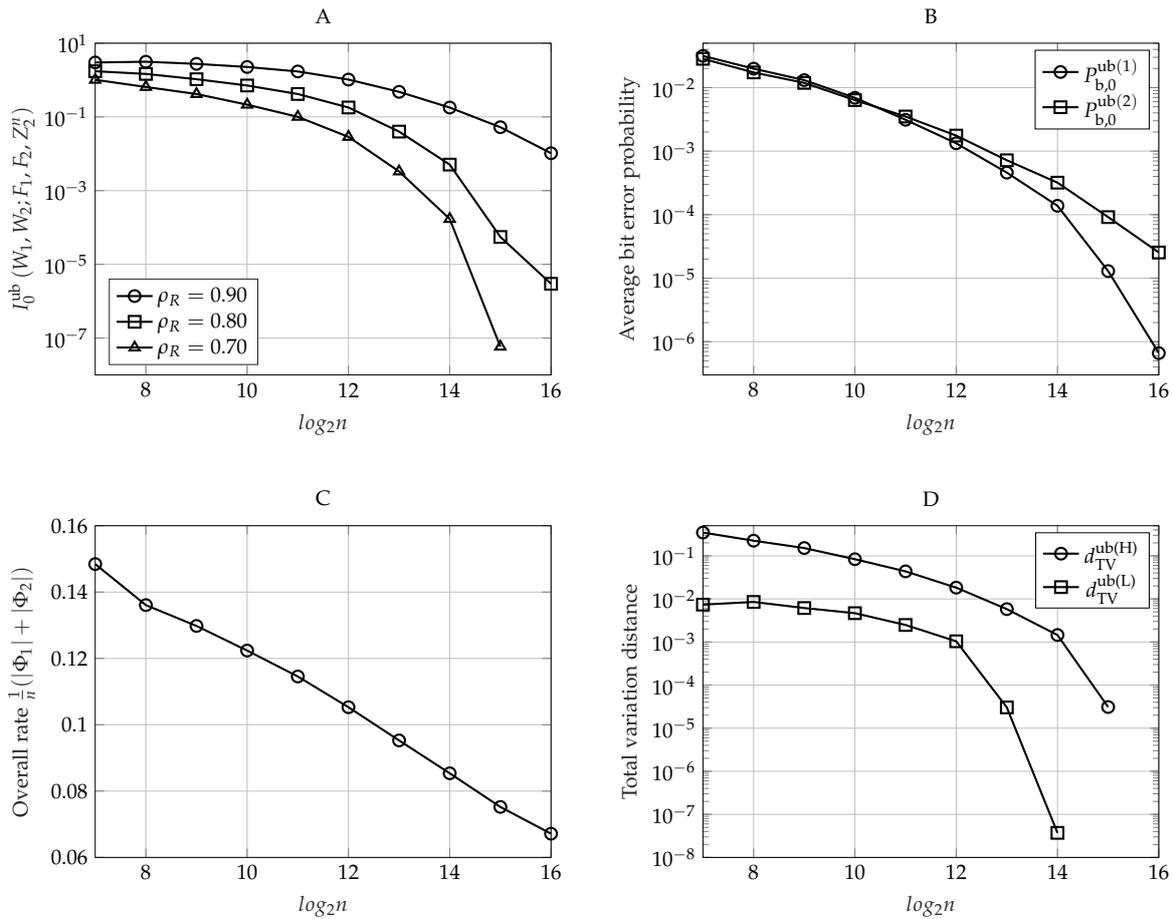

The performance of the polar coding scheme is graphically shown in Figure~\ref{fig:mod2}.~As for the previous model, let $\rho_{\text{R}}$ be the normalized target rate in which the polar coding scheme operates, that is $\rho_{\text{R}} \triangleq \frac{R_1^{\prime}}{R_1^{\star}} = \frac{R_2^{\prime}}{R_2^{\star}}$. In Figure~\ref{fig:mod2}A, we evaluate the upper-bound $I_0^{\text{ub}}(W_1, W_2; F_1,F_2, {Z}_{2}^n)$, which corresponds to the upper-bound on the information leakage defined in Equation~\eqref{eq:2_r1_I} when we consider $d_{\text{TV}}^{\text{ub}}=0$, as a function of the blocklength $n$ for different values of $\rho_{\text{R}}$.~For this plot, we set $\beta^{(1,\text{s})} = 0.30$ and $\beta^{(2,\text{s})} = 0.36$. Notice that $(\beta^{(1,\text{r})}, \beta^{(2,\text{r})})$ and $(\beta^{(2,\text{L})},\beta^{(2,\text{H})})$ if we set $d_{\text{TV}}^{\text{ub}}=0$ will not impact the information leakage. As we have proven in Section~\ref{sec:m1_sec}, the secrecy performance is improving as $n$ increases. Moreover, to satisfy a particular secrecy performance level, the polar code will need higher values of $n$ as the target rates approach the capacity.

In Figure~\ref{fig:mod2}B, we evaluate the upper-bounds $P_{\text{b},0}^{\text{ub}(1)}$ and $P_{\text{b},0}^{\text{ub}(2)}$, which correspond to the bounds on the average bit error probability at the legitimate Receivers 1 and 2, respectively, when we set $d_{\text{TV}}^{\text{ub}}=0$, as a function of the blocklength $n$.~For this plot, we set $\beta^{(1,\text{r})}=\beta^{(2,\text{r})}=0.24$ and notice that the reliability performance will not depend on the values of $(\beta^{(1,\text{s})},\beta^{(2,\text{s})})$ and $\rho_{\text{R}}$. If we set $d_{\text{TV}}^{\text{ub}}=0$, then~it is clear that it will not depend on $(\beta^{(2,\text{L})},\beta^{(2,\text{H})})$ either. As shown theoretically in Section~\ref{sec:m1_rel}, the~error probability becomes lower as the blocklength $n$ increases.

Figure~\ref{fig:mod2}C plots the overall rate of the additional secret sequences computed as in Equation~\eqref{eq:2_r1_R} when we set $\beta^{(1,\text{r})}=\beta^{(2,\text{r})}=0.24$, $\beta^{(1,\text{s})}=0.30$ and $\beta^{(2,\text{s})}=0.36$. As mentioned in Section~\ref{sec:PCS_dbcldnls_encoder}, we can see that this rate tends to be negligible for $n$ sufficiently large.

Finally, Figure~\ref{fig:mod2}D plots the upper-bounds $d_{\text{TV}}^{\text{ub(L)}}$ and $d_{\text{TV}}^{\text{ub(H)}}$ defined in Equations~\eqref{eq:m1_TVL}~and~\eqref{eq:m1_TVH}, respectively, when we set $\beta^{(2,\text{L})} = \beta^{(2,\text{H})} = 0.36$. As we have proven theoretically in Lemma~\ref{lemma:distDMS_1}, notice that the total variation distance decays with the blocklength $n$. Precisely, notice that $d_{\text{TV}}^{\text{ub(L)}}$ is lower than $d_{\text{TV}}^{\text{ub(H)}}$, and therefore, the bound on the total variation distance is practically governed by $d_{\text{TV}}^{\text{ub(H)}}$ ($d_{\text{TV}}^{\text{ub}} \approx d_{\text{TV}}^{\text{ub(H)}}$). This happens because although we can compute exactly the Kullback--Leibler divergence as in Equation~\eqref{eq:kldist} from the entropy terms estimated in the polar code construction,~Pinsker's inequality to obtain $d_{\text{TV}}^{\text{ub(H)}}$ as in Equation~\eqref{eq:m1_TVH} can be too loose for $n$ not sufficiently large. Consider the impact of $d_{\text{TV}}^{\text{ub}}$ on the reliability performance of the code. The average error probability bounds in Equations~\eqref{eq:2_r1_Pb_1}~and~\eqref{eq:2_r1_Pb_2} are modeled as the sum of two terms, one depending directly on $d_{\text{TV}}^{\text{ub}}$ and the other depending on the polar construction (which has been plotted in Figure~\ref{fig:mod2}B). Since~$d_{\text{TV}}^{\text{ub}(H)}$ is too loose, what we obtain is that the reliability performance of the code will be governed practically by the bound $d_{\text{TV}}^{\text{ub}}$ for small values of the blocklength $n$. Now, consider the impact of $d_{\text{TV}}^{\text{ub}}$ on the secrecy performance of the code. The bound on the information leakage in Equation~\eqref{eq:2_r1_I} is modeled as the sum of two terms, one also depending only on the polar code construction (which has been plotted in Figure~\ref{fig:mod2}A) and the other depending on $d_{\text{TV}}^{\text{ub}}$. However, in this situation, $d_{\text{TV}}^{\text{ub}}$ impacts the information leakage approximately as $n \cdot d_{\text{TV}}^{\text{ub}}$, which means that this term will totally govern the secrecy performance. Recall that this term follows from Equation~\eqref{eq:entropy1bound}, which bounds the impact of the encoding in Equation~\eqref{eq:distrenc2} on the conditional entropy term of the information leakage as a function of the total variation distance. Hence, we can conclude that this bound, which follows from applying \cite{csiszar2011information} (Lemma~2.9), can be too loose for $n$ not sufficiently large.


\section{Conclusions}\label{sec:conclusions}
We have described two polar coding schemes for two different models over the degraded broadcast channel: DBC-NLD-LS and DBC-LD-NLS. For both models, we have proven that the proposed polar coding schemes are asymptotically secrecy-capacity achieving, providing reliability and strong secrecy simultaneously. Then, we have discussed how to construct these polar codes in practice, and we have evaluated their performance for a finite blocklength by means of simulations. Although several polar code constructions methods have been proposed in the literature, this paper, as~far as we know, is the first to discuss practical constructions when the polar code must satisfy both reliability and secrecy constraints.~In addition, we have evaluated the secrecy performance of the polar code in terms of the strong secrecy performance, which has been possible by obtaining an upper-bound on the corresponding information leakage at the eavesdroppers. Indeed, we have shown that the proposed polar coding schemes can perform well in practice for a finite blocklength.

The criteria we have chosen for designing the polar codes are: to provide reliability and strong secrecy in one block of size $n$ by using only a secret key that is negligible in terms of rate and to minimize the amount of random decisions for the SC encoding.~For the first purpose, we~have introduced the source of common randomness, and we have avoided the use of the chaining construction given in \cite{6620400} (which is possible due to the degraded nature of the broadcast channel); for the second one, we have adapted the deterministic SC encoding given in \cite{7447169}. These two types of randomness have different implications on the practical design: while the common randomness is uniformly distributed and can be provided by the communication system, the~randomness for SC encoding is not and must be drawn by the encoder. In communication scenarios requiring several transmissions of size $n$, we have shown that one realization of the common randomness can be reused without worsening the performance.

Despite the good performance of the polar coding schemes, some issues still persist. How to avoid the transmissions of the additional secret sequences is a problem that remains open. Despite the length of the required secret key being asymptotically negligible in terms of rate, these additional transmissions can be problematic in practical scenarios. As pointed out in Remark~\ref{remarkA}, one can adopt the chaining construction in \cite{6620400} to further reduce the length of these sequences, but this requires the transmission to take place over several blocks of size $n$ and a very large memory capacity at the transmitter or receiver side.~Furthermore, despite the rate of the amount of randomness required for SC encoding being negligible, how to replace the random decisions entirely by deterministic ones is a problem that still remains unsolved. Another problem that remains open is how to avoid the use of the common randomness, which allows {keyless} secret communication over a single block of size $n$ (keyless in the sense that the rate of the required secret key is negligible).~Finally, to design polar codes based on the proposed performance evaluation, it seems necessary to find tighter upper-bounds on the total variation distance between the distribution induced by the encoder and the original distribution used in the code construction, particularly for the term that models the impact of storing uniformly-distributed sequences. Also, for the secrecy performance, it would be interesting to find a tighter upper-bound to evaluate the impact of the total variation distance on the information leakage.

Lastly, it is worth mentioning that having to know the statistics of the eavesdropper channels for the polar code construction may seem problematic. Nevertheless, for the polar code construction, one can consider virtual eavesdroppers with some target channel qualities. For DBC-LD-NLS, we can design a polar code according to the statistics of this virtual eavesdropper, and due to the degradedness condition of the channel, this code will perform well if the real eavesdroppers have worse channel quality (worst-case design). On the other hand, for the DBC-NLD-LS, one can simply consider different levels of secrecy depending on different target channel qualities. Depending on the channel quality of the real eavesdropper with respect to the virtual ones considered for the design, the polar coding scheme will provide a particular secrecy performance level.

\vspace{6pt} 

\supplementary{The MATLAB code used in this paper for Section~\ref{sec:results} is available at \linksupplementary{?}.}

\authorcontributions{Conceptualization, J.d.O.A. and J.R.F. Formal analysis, J.d.O.A. Funding acquisition, J.R.F. Investigation, J.d.O.A. and J.R.F. Methodology, J.d.O.A. and J.R.F. Software, J.d.O.A. Supervision, J.R.F. Validation, J.R.F. Writing, original draft, J.d.O.A.}

\funding{This work is supported by the ``Ministerio de Ciencia, Innovaci{\'o}n y Universidades'' and the ``Agencia Estatal de Investigaci{\'o}n'' of the Spanish Government, ERDF funds (TEC2013-41315-R, TEC2015-69648-REDC, TEC2016-75067-C4-2-R) and the Catalan Government (2017 SGR 578 AGAUR).}


\conflictsofinterest{The authors declare no conflict of interest.} 

\abbreviations{The following abbreviations are used in this manuscript:\\

\noindent 
\begin{tabular}{@{}ll}
DBC & Degraded Broadcast Channel \\
DBC-NLD-LS & Degraded Broadcast Channel with Non-Layered Decoding and Layered Secrecy \\
DBC-LD-NLS & Degraded Broadcast Channel with Layered Decoding and Non-Layered Secrecy \\
SC & Successive Cancellation \\
DMS & Discrete Memoryless Source \\
BEC & Binary Erasure Channel \\
BSC & Binary Symmetric Channel \\
BE-BC & Binary Erasure Broadcast Channel \\
BS-BC & Binary Symmetric Broadcast Channel 
\end{tabular}}

\appendixtitles{yes} 
\appendixsections{multiple} 
\appendix

\section{Proof of Lemmas \ref{lemma:distDMS_2} and \ref{lemma:distDMS_1}}\label{app:distributionDMS}
Consider a DMS $( \mathcal{V}_{1} \times \cdots \times \mathcal{V}_{L} \times \mathcal{Y}_{K} \times \cdots \times \mathcal{Y}_{1} \times \mathcal{Z}_{M} \times \cdots \times \mathcal{Z}_{1}, p_{V_1\dots V_L Y_K \dots Y_1 Z_M \dots Z_1} )$, the~joint distribution of which satisfies the Markov chain condition $V_1 - \cdots - V_L - Y_K - \cdots - Y_1 - Z_M - \cdots - Z_1$. Consider an i.i.d. $n$-sequence $( V_1^n,\dots,V_L^n, Y_K^n, \dots, Y_1^n, Z_M^n, \dots, Z_1^n )$ of this DMS, $n$ being any power of two. We define the polar transforms $( U_1^n,\dots,U_L^n )$, where $U_{\ell}^n \triangleq V_{\ell}^n G_n$ for each $\ell \in [1,L]$, with~joint distribution $p_{U_1^n \dots U_L^n }$. Then, define $\mathcal{H}_{V_{\ell}|V_{{\ell}-1}}^{(n)}$ and $\mathcal{L}_{V_{\ell}|V_{{\ell}-1}}^{(n)}$ as in Equations~\eqref{eq:pol1b1}~and~\eqref{eq:pol1b2}, \mbox{where $V_{0} = U_{0} \triangleq \varnothing$}. Let $V_L \triangleq X$; if $L \triangleq 1$, notice that this DMS is the one considered for the code construction of DBC-NLD-LS. Otherwise, if $L \triangleq K$, it is the one considered for DBC-LD-NLS. 

Now, consider the polar encoding procedures described for both models in Sections~\ref{sec:PCS_dbcnldls_encoder}~and~\ref{sec:PCS_dbcldnls_encoder}. Let $\tilde{q}_{U_{1}^n \dots U_{L}^n}$ be the joint distribution of $( \tilde{U}_1^n,\dots,\tilde{U}_L^n )$ after the encoding. For both models, we have:
\begin{align}
\tilde{q}_{U_{1}^n \dots U_{L}^n}(\tilde{u}_{1}^n, \dots, \tilde{u}_{L}^n) & = \prod_{\ell = 1}^L \prod_{j =1}^n \tilde{q}_{U_{\ell}(j) | U_{\ell}^{1:j-1} V_{\ell-1}^n} \big( \tilde{u}_{\ell}(j) \big| \tilde{u}_{\ell}^{1:j-1}, \tilde{u}_{\ell-1}^n G_n \big), \nonumber
\end{align}
where, for all $\ell \in [1,L]$, 
\begin{align}
& \tilde{q}_{U_{\ell}(j) | U_{\ell}^{1:j-1} V_{{\ell}-1}^n} \! \big( \tilde{u}_{\ell}(j) \big| \tilde{u}_{\ell}^{1:j-1}, \tilde{v}_{{\ell}-1}^n \big)
 \nonumber \\
 & \quad = \left\{
\begin{array}{ll}
\frac{1}{2} & \text{if } j \in \mathcal{H}_{V_{\ell}|V_{{\ell}-1}}^{(n)}, \\
p_{U_{\ell}(j) | U_{\ell}^{1:j-1} V_{{\ell}-1}^n} \big( \tilde{u}_{\ell}(j) \big| \tilde{u}_{\ell}^{1:j-1}, \tilde{v}_{{\ell}-1}^n \big) & \text{if } j \in \big( \mathcal{H}_{V_{\ell}|V_{\ell-1}}^{(n)} \big)^{\text{C}} \cap \big( \mathcal{L}_{V_{\ell}|V_{\ell-1}}^{(n)} \big)^{\text{C}}, \\
 \mathds{1} \big\{ \tilde{u}_{\ell}(j) = \xi^{(j)} \big( \tilde{u}_{\ell}^{1:j-1},\tilde{v}_{\ell-1}^{n} \big) \big\} & \text{if } j \in \mathcal{L}_{V_{\ell}|V_{{\ell}-1}}^{(n)} , 
\end{array} 
\right. \label{eq:dist_tilde}
\end{align}
$p_{U_{\ell}(j) | U_{\ell}^{1:j-1} V_{\ell-1}^n}$ being the distribution induced by the original DMS and $\xi^{(j)}$ being the deterministic $\argmax$ function given in Equation~\eqref{eq:argm2a} for DBC-NLD-LS or given in Equation~\eqref{eq:argmax1} for DBC-LD-NLS.

Additionally, consider another encoding process that constructs $(\check{U}_1^n, \dots, \check{U}_L^n)$ by omitting the use of the deterministic $\argmax$ function, but samples $\check{U}_1 (j)$ from the distribution:
\begin{align}
& \check{q}_{U_{\ell}(j) | U_{\ell}^{1:j-1} V_{{\ell}-1}^n} \! \big( \check{u}_{\ell}(j) \big| \check{u}_{\ell}^{1:j-1}, \check{v}_{{\ell}-1}^n \big) = \! \left\{
\begin{array}{ll}
\frac{1}{2} & \text{if } j \in \mathcal{H}_{V_{\ell}|V_{{\ell}-1}}^{(n)}, \\
p_{U_{\ell}(j) | U_{\ell}^{1:j-1} V_{{\ell}-1}^n} \! \big( \check{u}_{\ell}(j) \big| \check{u}_{\ell}^{1:j-1}, \check{v}_{{\ell}-1}^n \big) & \text{if } j \in \big( \mathcal{H}_{V_{\ell}|V_{\ell-1}}^{(n)} \big)^{\text{C}}.
\end{array}
\right. \label{eq:dist_check}
\end{align}

First, the following lemma shows that the joint distributions $p_{U_{1}^n \dots U_{L}^n}$ and $\check{q}_{U_{1}^n \dots U_{L}^n}$ are nearly statistically indistinguishable for sufficiently large $n$.

\begin{Lemma}\label{lemma:distUc1Uc2}
Let $\delta_n = 2^{-n^{\beta}}$ for some $\beta \in (0, \frac{1}{2})$, and define ${\delta}^{(1)}_n \triangleq \sqrt{2 n \delta_n \ln 2}$. Then, 
\begin{align*}
\mathbb{V} (\check{q}_{U_{1}^n \dots U_{L}^n}, p_{U_{1}^n \dots U_{L}^n}) \leq \sqrt{L}{\delta}^{(1)}_n.
\end{align*}
\end{Lemma}

\begin{proof}
The Kullback-Leibler distance between $p_{U_{1}^n \dots U_{L}^n}$ and $\check{q}_{U_{1}^n \dots U_{L}^n}$ is:
\begin{align} 
\mathbb{D} \big( p_{U_{1}^n \dots U_{L}^n} \big\| \check{q}_{U_{1}^n \dots U_{L}^n} \big) & \stackrel{(a)}{=} \sum_{\ell = 1}^L \sum_{j=1}^n \mathbb{E}_{p_{U_{\ell}^{1:j-1} V_{\ell-1}^n} } \Big[ \mathbb{D} \Big( p_{U_{\ell}(j)|U_{\ell}^{1:j-1} V_{\ell-1}^n } \Big\| \check{q}_{U_{\ell}(j)|U_{\ell}^{1:j-1} V_{\ell-1}^n } \Big) \Big] \nonumber \\
& \stackrel{(b)}{=} \sum_{\ell = 1}^L \sum_{j \in \mathcal{H}_{V_{\ell}|V_{\ell-1}}^{(n)}} \Big( 1 - H \Big( U_{\ell}(j) \Big| U_{\ell}^{1:j-1}, V_{\ell-1}^{n} \Big) \Big) \nonumber \\
& \stackrel{(c)}{\leq} L \delta_n \big| \mathcal{H}_{V_{\ell}|V_{\ell-1}}^{(n)} \big|, \label{eq:kldist}
\end{align}
where $(a)$ holds by the chain rule, the invertibility of $G_n$ and the fact that $U_{1}^n - U_{2}^n - \dots - U_{L}$ (and $\check{U}_{1}^n - \check{U}_{2}^n - \dots - \check{U}_{L}$) forms a Markov chain, $(b)$ follows from Equation~\eqref{eq:dist_check} and by applying \cite{6975233} (Lemma~10), and $(c)$ holds by the definition of $\mathcal{H}_{V_{\ell}|V_{\ell-1}}^{(n)}$ in Equation~\eqref{eq:pol1b1}.~Finally, since~$| \mathcal{H}_{V_{\ell}|V_{\ell-1}}^{(n)} | \leq n$ and by using Pinsker's inequality, we~obtain \mbox{$\mathbb{V} (\check{q}_{U_{1}^n \dots U_{L}^n}, p_{U_{1}^n \dots U_{L}^n}) \leq \sqrt{2 L n \delta_n \ln 2}$}.
\end{proof}

Now, we show that $\check{q}_{U_1^n \dots U_L^n}$ and $\tilde{q}_{U_1^n \dots U_L^n}$ are nearly indistinguishable for $n$ large enough.

\begin{Lemma}\label{lemma:distU1cU2}
Let $\delta_n = 2^{-n^{\beta}}$ for some $\beta \in (0, \frac{1}{2})$. Then, 
\begin{align*}
\mathbb{V} (\tilde{q}_{U_1^n \dots U_L^n}, \check{q}_{U_1^n \dots U_L^n}) \leq {\delta}^{(2)}_n,
\end{align*}
where ${\delta}^{(2)}_n \triangleq L n \sqrt{ 2 \sqrt{2} \delta_n^{(1)} \big( 2n - \log \sqrt{2} \delta_n^{(1)} \big) + \delta_n}$ and ${\delta}^{(1)}_n$ defined as in Lemma~\ref{lemma:distUc1Uc2}. 
\end{Lemma}

\begin{proof}
The proof follows similar reasoning as the one for \cite {7447169} (Lemma~2). Hence, define a coupling \cite{levin2009markov} for $( \check{U}_{1}^n , \dots , \check{U}_{L}^n )$ and $( \tilde{U}_{1}^n , \dots, \tilde{U}_{L}^n )$ such that $\check{U}_{\ell} [ ( \mathcal{L}_{V_{\ell}|V_{\ell-1}}^{(n)} )^{\text{C}} ] = \tilde{U}_{\ell}[ ( \mathcal{L}_{V_{\ell}|V_{\ell-1}}^{(n)} )^{\text{C}} ]$. Thus, we have:
\begin{align}
 \mathbb{V} (\tilde{q}_{U_1^n \dots U_L^n}, \check{q}_{U_1^n \dots U_L^n}) 
& \stackrel{(a)}{\leq} \mathbb{P} \Big[ \big( \tilde{U}_{1}^n , \dots, \tilde{U}_{L}^n \big) \neq \big( \check{U}_{1}^n , \dots , \check{U}_{L}^n \big)\Big] \nonumber \\
& \stackrel{(b)}{\leq} \sum_{\ell=1}^L \mathbb{P} \Big[ \tilde{U}_{\ell}^n \neq \check{U}_{\ell}^n \Big| \tilde{V}_{\ell-1}^n = \check{V}_{\ell-1}^n \Big] \nonumber \\
& \stackrel{(c)}{\leq} \sum_{\ell=1}^L \sum_{j=1}^n \mathbb{P} \Big[ \tilde{U}_{\ell}(j) \neq \check{U}_{\ell}(j) \Big| \tilde{U}_{\ell}^{1:j-1} = \check{U}_{\ell}^{1:j-1} , \tilde{V}_{\ell-1}^n = \check{V}_{\ell-1}^n \Big] \nonumber \\
 & \stackrel{(d)}{=} \sum_{\ell=1}^L \sum_{j \in \mathcal{L}_{V_{\ell}|V_{\ell-1}}^{(n)}} \! \! \! \! \mathbb{E}_{\big( \check{U}_{\ell}^{1:j-1}, \check{V}_{\ell-1}^n \big)} \Bigg[ \Bigg( 1 - p_{U_{\ell}(j) | U_{\ell}^{1:j-1}V_{\ell-1}^n} \Big( u_{\ell}^{\ast}(j) \Big| \check{U}_{\ell}^{1:j-1}, \check{V}_{\ell-1}^n \Big) \Bigg) \Bigg] , \label{eq:comb1}
\end{align}
where $(a)$ follows from the coupling lemma \cite{levin2009markov} (Proposition~4.7), $(b)$ holds by the union bound, the invertibility of $G_n$ and the fact that $\tilde{U}_{1}^n - \tilde{U}_{2}^n - \dots - \tilde{U}_{L}$ (and $\check{U}_{1}^n - \check{U}_{2}^n - \dots - \check{U}_{L}$) forms a Markov chain, $(c)$ also holds by the union bound and $(d)$ follows from Equations~\eqref{eq:dist_tilde}~and~\eqref{eq:dist_check} given that $\check{U}_{\ell} [ ( \mathcal{L}_{V_{\ell}|V_{\ell-1}}^{(n)} )^{\text{C}} ] = \tilde{U}_{\ell} [ ( \mathcal{L}_{V_{\ell}|V_{\ell-1}}^{(n)} )^{\text{C}} ]$ and from defining $u_{\ell}^{\ast}(j) \triangleq \argmax_{u \in \{0,1\} } p_{U_{\ell}(j) | U_{\ell}^{1:j-1}V_{\ell-1}^n} ( u \big| \check{U}_{\ell}^{1:j-1}, \check{V}_{\ell-1}^n )$. 

Next, for any $\ell \in [1,L]$ and $j \in [n]$, for sufficiently large $n$, we have:
\begin{align}
& \Big| H\big( U_{\ell} (j) \big| U_{\ell}^{1:j-1}, V_{\ell-1}^n \big) - H\big( U_{\ell} (j) \big| \check{U}_{\ell}^{1:j-1}, \check{V}_{\ell-1}^n \big) \Big| \nonumber \\
& \quad \stackrel{(a)}{\leq} \Big| H\big( U_{\ell}^{1:j-1}, V_{\ell-1}^n \big) - H\big( \check{U}_{\ell}^{1:j-1}, \check{V}_{\ell-1}^n \big) \Big| + \Big| H\big( U_{\ell}^{1:j}, V_{\ell-1}^n \big) - H\big( U_{\ell}(j), \check{U}_{\ell}^{1:j-1}, \check{V}_{\ell-1}^n \big) \Big| \nonumber \\
& \quad \stackrel{(b)}{\leq} 2 \mathbb{V} \big( \check{q}_{U_{\ell}^{1:j-1} U_{\ell-1}^n}, p_{U_{\ell}^{1:j-1} U_{\ell-1}^n} \big) \log \frac{2^n}{\mathbb{V} \big( \check{q}_{U_{\ell}^{1:j-1} U_{\ell-1}^n}, p_{U_{\ell}^{1:j-1} U_{\ell-1}^n} \big)} \nonumber \\
& \quad \stackrel{(c)}{\leq} 2 \sqrt{2} \delta_n^{(1)} \big( 2 n - \log \sqrt{2} \delta_n^{(1)} \big) , \label{eq:absentr}
\end{align}
where $(a)$ holds by the chain rule of entropy and the triangle inequality, $(b)$ follows from applying~\cite{csiszar2011information} (Lemma~2.9), the invertibility of $G_n$ and because $\mathbb{V} ( p_{U_{\ell}(j) | U_{\ell}^{1:j-1} U_{\ell-1}^n} \check{q}_{U_{\ell}^{1:j-1} U_{\ell-1}^n}, p_{U_{\ell}^{1:j} U_{\ell-1}^n} ) = \mathbb{V} ( \check{q}_{U_{\ell}^{1:j-1} U_{\ell-1}^n}, p_{U_{\ell}^{1:j-1} U_{\ell-1}^n} )$, and $(c)$ holds because $\mathbb{V} \big( \check{q}_{U_{\ell}^{1:j-1} U_{\ell-1}^n}, p_{U_{\ell}^{1:j-1} U_{\ell-1}^n} \big) \leq \mathbb{V} ( \check{q}_{U_{\ell-1}^{n} U_{\ell}^n}, p_{U_{\ell-1}^{n} U_{\ell}^n} ) \leq \sqrt{2} \delta_n^{(1)}$ (by using Lemma~\ref{lemma:distUc1Uc2} and taking $L\triangleq 2$) and because the function $x \mapsto x \log x$ is monotonically decreasing for $x > 0$ small enough.

Thus, for any $\ell \in [1,L]$ and $j \in \mathcal{L}_{V_{\ell}|V_{\ell-1}}^{(n)}$, we have:
\begin{align}
& 2 \sqrt{2} \delta_n^{(1)} \big( 2n - \log \sqrt{2} \delta_n^{(1)} \big) + \delta_n \nonumber \\
& \quad \stackrel{(a)}{\geq} 2 \sqrt{2} \delta_n^{(1)} \big( 2 n - \log \sqrt{2} \delta_n^{(1)} \big) + H \Big( U_{\ell}(j)| U_{\ell}^{1:j-1} , V_{\ell-1}^n \Big) \nonumber \\
& \quad \stackrel{(b)}{\geq} H \Big( U_{\ell}(j)| \check{U}_{\ell}^{1:j-1} , \check{V}_{\ell-1}^n \Big) \nonumber \\
& \quad = \mathbb{E}_{\big( \check{U}_{\ell}^{1:j-1}, \check{U}_{\ell-1}^n \big)} \left[ h_2 \left( p_{U_{\ell}(j) | U_{\ell}^{1:j-1}V_{\ell-1}^n}\Big( u_{\ell}^{\star}(j) \Big| \check{U}_{\ell}^{1:j-1}, \check{V}_{\ell-1}^n \Big) \right)
 \right] \nonumber \\
& \quad \geq \mathbb{E}_{\big( \check{U}_{\ell}^{1:j-1}, \check{U}_{\ell-1}^n \big)} \left[ - \left(1 - p_{U_{\ell}(j) | U_{\ell}^{1:j-1}V_{\ell-1}^n}\Big( u_{\ell}^{\star}(j) \Big| \check{U}_{\ell}^{1:j-1}, \check{V}_{\ell-1}^n \Big) \right) \right. \nonumber \\
& \qquad \cdot \left. \log \left( 1 - p_{U_{\ell}(j) | U_{\ell}^{1:j-1}V_{\ell-1}^n}\Big( u_{\ell}^{\star}(j) \Big| \check{U}_{\ell}^{1:j-1}, \check{V}_{\ell-1}^n \Big) \right) \right] \nonumber \\
& \quad \stackrel{(c)}{\geq} \mathbb{E}_{\big( \check{U}_{\ell}^{1:j-1}, \check{U}_{\ell-1}^n \big)} \left[ \left(1 - p_{U_{\ell}(j) | U_{\ell}^{1:j-1}V_{\ell-1}^n}\Big( u_{\ell}^{\star}(j) \Big| \check{U}_{\ell}^{1:j-1}, \check{V}_{\ell-1}^n \Big) \right)^2 \right] \nonumber \\
& \quad \stackrel{(d)}{\geq} \left( \mathbb{E}_{\big( \check{U}_{\ell}^{1:j-1}, \check{U}_{\ell-1}^n \big)} \left[ \left(1 - p_{U_{\ell}(j) | U_{\ell}^{1:j-1}V_{\ell-1}^n}\Big( u_{\ell}^{\star}(j) \Big| \check{U}_{\ell}^{1:j-1}, \check{V}_{\ell-1}^n \Big) \right) \right] \right)^2 , \label{eq:comb2}
\end{align}
where $(a)$ holds because, by definition, $H \Big( U_{\ell}(j)| U_{\ell}^{1:j-1} , V_{\ell-1}^n \Big) \leq \delta_n$ if $j \in \mathcal{L}_{V_{\ell}|V_{\ell-1}}^{(n)}$, $(b)$ holds by Equation~\eqref{eq:absentr}, $(c)$ holds because $p_{U_{\ell}(j) | U_{\ell}^{1:j-1}V_{\ell-1}^n}\big( u_{\ell}^{\star}(j) \big| \check{U}_{\ell}^{1:j-1}, \check{V}_{\ell-1}^n \big) \geq 1/2$ and $\log (x) < -x$ if $x \in [0, 1/2 )$ and $(d)$ follows from Jensen's inequality. 

Finally, by combining Equations~\eqref{eq:comb1} and~\eqref{eq:comb2} and because $| \mathcal{L}_{V_{\ell}|V_{\ell-1}}^{(n)} | \leq n$, we have $\mathbb{V} (\tilde{q}_{U_1^n \dots U_L^n}, \check{q}_{U_1^n \dots U_L^n}) \leq L n \sqrt{ 2 \sqrt{2} \delta_n^{(1)} ( 2n - \log \sqrt{2} \delta_n^{(1)} ) + \delta_n}$.
\end{proof}

Hence, by Lemma~\ref{lemma:distUc1Uc2}, Lemma~\ref{lemma:distU1cU2} and by applying the triangle inequality, we obtain:
\begin{align}
\mathbb{V} (\tilde{q}_{U_1^n \dots U_L^n}, p_{U_1^n \dots U_L^n}) & \leq \mathbb{V} (\tilde{q}_{U_1^n \dots U_L^n}, \check{q}_{U_1^n \dots U_L^n}) + \mathbb{V} (\check{q}_{U_1^n \dots U_L^n}, p_{U_1^n \dots U_L^n}) \nonumber \\
& \leq L n \sqrt{ 2 \sqrt{2} \delta_n^{(1)} \big( 2n - \log \sqrt{2} \delta_n^{(1)} \big) + \delta_n} + \sqrt{L}{\delta}^{(1)}_n.
\end{align}
Consequently, since $\tilde{q}_{Y_K^n \dots Y_1^n Z_M^n \dots Z_1^n | V_1^n \dots V_L^n} = p_{Y_K^n \dots Y_1^n Z_M^n \dots Z_1^n | V_1^n\dots V_L^n}$ and the invertibility of $G_n$, we obtain $\mathbb{V} (\tilde{q}_{V_1^n\dots V_L^n Y_K^n \dots Y_1^n Z_M^n \dots Z_1^n}, p_{V_1^n \dots V_L^n Y_K^n \dots Y_1^n Z_M^n \dots Z_1^n}) = \mathbb{V} (\tilde{q}_{U_1^n\dots U_L^n}, p_{U_1^n \dots U_L^n})$, and this concludes the proof. 



\reftitle{References and Notes}




\end{document}